\documentclass[letterpaper,11pt]{article} 

\usepackage{amsmath}
\usepackage{amssymb}
\usepackage[comma,longnamesfirst]{natbib}
\usepackage[dvips]{epsfig}
\usepackage{dcolumn}
\usepackage{enumerate}
\usepackage{hhline}
\usepackage{dsfont}
\usepackage{afterpage}
\usepackage{arydshln}
\usepackage{graphicx}
\usepackage{color}
\usepackage[usenames,dvipsnames]{xcolor}
\usepackage{rotating}
\usepackage[breaklinks]{hyperref}
\usepackage{breakurl} 
\usepackage{xr}
\usepackage[percent]{overpic}
\usepackage[]{caption}
\usepackage{algorithmicx}
\usepackage[]{algpseudocode}
\usepackage{algorithm}
\usepackage{diagbox}
\usepackage{graphicx}
\usepackage{wrapfig}
\usepackage{lscape}
\usepackage{fontenc}
\usepackage{setspace}
\usepackage{bm}
\usepackage{slashbox}
\usepackage{lscape}
\usepackage{breakurl} 
\usepackage{multirow,booktabs}
\usepackage{eurosym}
\usepackage{titlesec}
\usepackage{sectsty}
\usepackage{placeins}
\usepackage{subcaption}
\usepackage[thinc]{esdiff}
\usepackage{derivative}
\usepackage[flushleft]{threeparttable}
\usepackage{adjustbox}

\epsfverbosetrue
\def\theequation{\thesection.\arabic{equation}}  
\def\abstract{\if@twocolumn
\section*{Abstract}
\else \normalsize 
\begin{center}
{\bf Summary\vspace{-.5em}\vspace{0pt}} 
\end{center}
\quotation 
\fi}
\def\endabstract{\if@twocolumn\else\endquotation\fi}

\makeatletter
\newcommand{\myappendix}[1]{
	\setcounter{section}{1}
        \renewcommand{\thesection}{A\arabic{section}}}

\DeclareMathOperator{\tr }{tr}

\DeclareMathOperator{\D   }{D}

\DeclareMathOperator{\diag}{diag}
\DeclareMathOperator{\Diag}{Diag}

\DeclareMathOperator{\vect}{vec}
\DeclareMathOperator{\dd}{d}

\def \calL {\mathcal L}

\def \calT {\mathcal T}

\def \lvec {\text{\boldmath$l$}}

\def \uvec {\text{\boldmath$u$}}    
    
\def \wvec {\text{\boldmath$w$}}    
\def \xvec {\text{\boldmath$x$}}    
\def \yvec {\text{\boldmath$y$}}    
\def \zvec {\text{\boldmath$z$}}

\def \ltildevec {\text{\boldmath$\tilde l$}}

\def \alphavec        {\text{\boldmath$\alpha$}}

\def \deltavec        {\text{\boldmath$\delta$}}

\def \varepsilonvec   {\text{\boldmath$\varepsilon$}}

\def \thetavec        {\text{\boldmath$\theta$}}

\def \lambdavec       {\text{\boldmath$\lambda$}}
\def \muvec           {\text{\boldmath$\mu$}}

\def \psivec          {\text{\boldmath$\psi$}}

\usepackage{color}
\usepackage{colordvi}
\newlength{\breite}
\breite\textwidth
\newcounter{aufg}[section]
  {\refstepcounter{aufg}\noindent\textbf{Exercise \arabic{aufg}:}
   \\*[1ex]\noindent}{\vspace{.5cm}}
   
 \newcounter{notes}[section]
  {\refstepcounter{aufg}\noindent\textbf{}
   \\*[1ex]\noindent}{\vspace{.5cm}}
   
\usepackage{amsthm}  

\theoremstyle{definition}

\newtheorem*{beisp*}{Example}
\newtheorem{Proof}{Proof}


\newtheoremstyle{break}
  {}
  {}
  {}
  {}
  {\bfseries}
  {.}
  {\newline}
  {}
  
\theoremstyle{break}

\newcommand{\head}[2]%
 {\hrule \vspace{.15cm} {\sfbold Advanced Statistical Inference, Summer Term 2012, Georg-August-University G\"ottingen}\hfill
{\sfbold Sheet #1}\\
{\sfbold Prof. Dr. Thomas Kneib, Nadja Klein}\hfill {\sfbold #2}

\vspace{.2cm}
\hrule

\vspace{1cm}

}

\newcounter{auf}
{\refstepcounter{auf}
\begin{center}
\fcolorbox[gray]{0}{.95}{
\makebox[\breite]{
\textbf{Exercise \arabic{auf}}
}}\\*[1ex]\noindent
\end{center}
}{\vspace{.5cm}}

\newcounter{loes}[section]
{\stepcounter{loes}
\begin{center}
\fcolorbox[gray]{0}{.95}{
\makebox[\breite]{
\textbf{L"osung \arabic{loes}}
}}\\*[1ex]\noindent
\end{center}
}{}

{\begin{center}
\fcolorbox[gray]{0}{.95}{
\makebox[\breite]{
\textbf{Zu Aufgabe #1}
}}\\*[1ex]\noindent
\end{center}\vspace{1cm}
}{\vspace{1cm}}

\newcounter{ka}
{\refstepcounter{ka}
\begin{center}
\framebox[\textwidth]{
\textbf{Aufgabe \arabic{ka}} \hfill #1 Punkte
}\\*[1ex]\noindent
\end{center}
}{\vspace{1cm}}

\newcounter{lka}
{\refstepcounter{lka}
\begin{center}
\framebox[\textwidth]{
\textbf{L\"osung \arabic{lka}} \hfill #1 Punkte
}\\*[1ex]\noindent
\end{center}
}{\vspace{1cm}}

\usepackage[margin=1in]{geometry}
\titlespacing*\section{0pt}{0pt plus 4pt minus 2pt}{0pt plus 2pt minus 2pt}
\titlespacing*\subsection{0pt}{0pt plus 4pt minus 2pt}{0pt plus 2pt minus 2pt}
\titlespacing*\subsubsection{0pt}{0pt plus 4pt minus 2pt}{0pt plus 2pt minus 2pt}

\definecolor{myblue}{RGB}{0,73,114}

\newcounter{mynotation}

\usepackage{paralist}

\renewenvironment{itemize}[1]{\begin{compactitem}#1}{\end{compactitem}}

\usepackage[skins,breakable]{tcolorbox}
\newtcolorbox{cvbox}[2][]{%
	blanker,
	after skip=8mm,
	title=#2,
	coltitle=cyan,
	#1
}

\makeatletter
\def\@seccntformat#1{\@ifundefined{#1@cntformat}%
	{\csname the#1\endcsname\quad}  
	{\csname #1@cntformat\endcsname}
}
\let\oldappendix\appendix 
\renewcommand\appendix{%
	\oldappendix
	\newcommand{\section@cntformat}{\appendixname~\thesection\quad}
}
\makeatother

\usepackage{titlesec}

\begin{document}
\setlength{\abovedisplayskip}{0.15cm}
\setlength{\belowdisplayskip}{0.15cm}
\pagestyle{empty}
\begin{titlepage}

\title{\bfseries\sffamily\color{myblue}  
Large Skew-t Copula Models and Asymmetric Dependence in
Intraday Equity Returns}
\author{Lin Deng, Michael Stanley Smith and Worapree Maneesoonthorn}
\date{\today}
\maketitle
\noindent
{\small Lin Deng is a PhD student and Michael Smith is Professor of Management (Econometrics), both at the Melbourne Business School, University of Melbourne, Australia.
Worapree Maneesoothorn is Associate Professor of Statistics and Econometrics at Monash University, Australia.  Correspondence should be directed to Michael Smith at {\tt mikes70au@gmail.com}. MATLAB
code to implement the method is available at https://github.com/lindenglab/ACCopVB-MATLAB. 
\\

\noindent \textbf{Acknowledgments:} This research was supported by The University of Melbourne’s Research Computing Services and the Petascale Campus Initiative. 
A/Prof. Maneesoonthorn has been supported by the Australian Research Council (ARC) Discovery Project Grant (DP200101414). We thank an Associate Editor and two referees for comments that have helped improve the paper, and the Co-Editor Ivan A. Canay for his guidance.}\\

\newpage
\begin{center}
\mbox{}\vspace{2cm}\\
{\LARGE \title{\bfseries\sffamily\color{myblue} Large Skew-t Copula Models and Asymmetric Dependence in Intraday Equity Returns}
}\\
\vspace{1cm}
{\Large Abstract}
\end{center}
\vspace{-1pt}
\onehalfspacing
\noindent
Skew-t copula models are attractive for the modeling of financial
data because they allow for asymmetric and extreme tail dependence. 
We show that the copula implicit
in the skew-t distribution of~\cite{Azzalini_Capitanio_2003} allows for a higher
level of pairwise asymmetric dependence than two popular alternative skew-t copulas. Estimation of this copula in high dimensions is challenging, and we propose
a fast and accurate Bayesian variational inference (VI) approach to do so. The method
uses a generative representation of the skew-t distribution to define an augmented
posterior that can be approximated accurately. A stochastic gradient ascent algorithm is used to solve the variational optimization. 
The methodology is used to estimate skew-t factor copula models with up to 15 factors for intraday returns
from 2017 to 2021 on 93 U.S. equities. The copula
captures substantial heterogeneity in asymmetric dependence over equity pairs, in addition to the
variability in pairwise correlations. In a moving window study we show that the asymmetric dependencies also vary over time, and that intraday predictive densities from the skew-t copula are more accurate than those from benchmark copula models. Portfolio selection strategies based on 
the estimated pairwise asymmetric dependencies improve performance
relative to the index.
\vspace{20pt}
 
\noindent
{\bf Keywords}: Asymmetric Dependence, Bayesian Data Augmentation, Factor Copula, Generative Representation, Intraday Equity Returns, Variational Inference
\end{titlepage}

\newpage
\pagestyle{plain}
\setcounter{equation}{0}
\renewcommand{\theequation}{\arabic{equation}}
\section{Introduction}\label{sec:intro}
Copula models are widely employed for multivariate data because they separate the selection of marginals from the copula function.
When modeling financial data, the copula should possess two features. 
First, it should allow for asymmetric dependence, where the level of dependence varies for different quantiles of the variables. Second, it should allow for high tail dependence, where extreme values of both variables are positively or negatively dependent.
Skew-t copulas can capture both features in high dimensions. 
However, there exist multiple variants of skew-t copulas, with no consensus on the most effective choice,
and their estimation in high dimensions is challenging. 
This paper addresses both issues by showing that the skew-t copula implicit in the~\cite{Azzalini_Capitanio_2003} distribution, combined with Bayesian variational inference, 
offers an attractive solution. We employ the methodology in a 
study using 15 minute intraday returns on 93 U.S. equities over five years, and show that pairwise asymmetric dependence varies
in a complex fashion over both equity pairs and time.  

A skew-t copula is that implicit in a parametric skew-t distribution,
and three types have been used previously. The first was
proposed
by~\cite{demarta_mcneil_2005} and is based on the generalized hyperbolic (GH)
skew-t distribution. It is the most popular of the three in financial studies,
with applications by~\cite{Christoffersen_Errunza_Jacobs_Langlois_2012,creal2015high,oh2017,lucas2017}
and~\cite{oh_patton_2021}. The second was suggested by~\cite{smith_gan_kohn_2010} and
is based on the distribution of~\cite{sahu_dey_branco_2003}, while
the third proposed by~\cite{kollo2010} and~\cite{Yoshiba_2018} is based on the distribution of~\cite{Azzalini_Capitanio_2003}, which we label the ``AC skew-t copula''.
All three skew-t distributions
have latent  generative representations for which factor models can be used, and their implicit copulas are called ``factor copulas'' by~\cite{oh2017}.\footnote{This type of factor copula is not to be confused with the similarly named vine-based factor copulas of~\cite{krupskii2013} and others, which are not the implicit copulas of high-dimensional
	skew-t distributions studied here.} 
In this paper we show the AC skew-t factor copula allows
for higher levels of asymmetry in pairwise dependence than the other two copulas,
making it an attractive choice for financial data. 

\cite{Yoshiba_2018} studies maximum likelihood estimation of the AC skew-t copula in low dimensions. However, the likelihood (and therefore also posterior) has a complex geometry and its direct optimization
is hard in high dimensions, which has precluded its use with large financial panels. 
Here we propose a scalable Bayesian solution that
uses a conditionally Gaussian generative representation of the AC skew-t distribution. 
Latent variables are introduced and augmented with the copula parameters to 
provide a tractable joint (or ``augmented'') posterior.
However, evaluation of this augmented posterior using Markov chain Monte Carlo (MCMC) methods 
is prohibitively slow in high dimensions. 
To overcome this we develop a new variational inference estimator
for evaluating the proposed augmented posterior of the AC skew-t copula in high dimensions.

Variational inference (VI) approximates
a target density with a tractable density called a variational approximation (VA).
The VA is calibrated by solving an optimization problem where
the Kullback-Leibler divergence between it and the target density is minimized.
Well designed VI methods are effective for estimating models with large numbers of parameters
and/or big datasets; see~\cite{Blei_Kucukelbir_McAuliffe_2017} for an introduction.
But here, the complex geometry of the posterior of the AC skew-t copula parameters is difficult to approximate directly using standard VI methods. Therefore, we instead propose using the augmented posterior as the target
density, and then adopt a flexible VA of the form suggested by~\cite{Loaiza-Maya_Smith_Nott_Danaher_2021}. 
To solve the optimization problem stochastic
gradient methods~\citep{ranganath2014} are used with re-parameterization gradients~\citep{Kingma_2014}. To increase the speed of the method
we derive these gradients analytically.
The speed and accuracy of the resulting VI method is demonstrated using simulated data.

We use our new copula methodology
in two studies using 15 minute intraday financial data.
The first illustrates the flexibility of the AC skew-t copula by modeling
returns on two equities and the VIX index during two 40 day periods: a low market volatility period, and the high market volatility period of the COVID-19 pandemic crash. 
We find strong asymmetries in the pairwise dependencies, which 
change in direction and scale from the low to high market volatility periods; asymmetries that are not identified
using the two other variants of skew-t copula.

The second example is our main study, and uses an AC skew-t factor copula to 
capture dependence between returns on 93 equities. 
Using a rolling window of width 40 trading days from January 2017 to 
December 2021, 
we make three findings. First, we show that one-step-ahead forecast densities are more accurate than those from benchmark copula models. 
Forecast 
accuracy is maximized with 10 to 15 factors, suggesting a rich
factor structure is worthwhile. 
This is consistent with~\cite{oh2017,opschoor2021} and~\cite{oh_patton_2021}, who
find that industry or other group-based factors increase accuracy.
Second, a major finding
is that pairwise dependence asymmetries can differ substantially over equity pairs in
addition to overall correlations. 
Third, we show that (absent of trading costs) 
portfolio selection strategies based on pairwise quantile dependencies from the copula model
can improve performance relative to index returns. Our findings are 
consistent with evidence of cross-sectional heterogeneity in
asymmetric dependence of equity returns~\citep{bollerslev2022realized,ando2022quantile}
and its importance in 
risk management~\citep{patton2004out,harvey2010,giacometti2021}.

Skew-t factor copulas have been used previously to capture dependence in high dimensional 
financial data. \cite{ohpatton2013} use the implicit copula of the GH distribution, and of a mixture of GH and t distributions. \cite{creal2015high} and \cite{oh_patton_2021} consider the implicit copula of the GH distribution with a single global factor, which the latter authors
augment with latent group specific factors. \cite{lucas2017} considers
the same copula with a 
block equicorrelation structure.
In comparison, ours is the first application of which we are aware of the AC skew-t copula to a large financial panel. We adopt an unrestricted specification
for the skew parameters of the skew-t distribution, allowing for greater flexibility in its implicit copula. 
Previous studies consider dynamic latent factor specifications for fitting
daily or weekly financial time series. In contrast, we employ intraday data at the 15 minute frequency
in a rolling window study using a static factor copula model with up to 15 global factors. 
Our objective
is to allow for a high degree of heterogeneity in the level of
asymmetric dependence over equity pairs, rather than capturing time variation at
a lower data frequency.

Our paper also contributes to the literature on Bayesian estimation of skew-t copulas. 
\cite{creal2015high} use
MCMC to estimate their skew-t factor copula, where a particle filter is used to evaluate
the intractable likelihood. 
\cite{smith_gan_kohn_2010} use MCMC with data augmentation, but for 
the skew-t copula
implicit in the distribution of~\cite{sahu_dey_branco_2003}.
Both approaches can also be adopted to estimate the AC skew-t copula, but they incur
a much higher computational burden than the VI approach suggested here.
A few recent studies have used
VI methods to estimate other high-dimensional copula models. 
\cite{loaiza2019variational} use VI to estimate vine copulas with discrete margins, while 
\cite{nguyen2020variational} use VI to estimate factor copulas based on pair-copula constructions. However, both studies employ simple mean field or other VAs
that are ineffective approximations for the complex geometry of the posterior
of the AC skew-t copula.  

The rest of the paper is organized as follows. Section~\ref{sec:copula} outlines
the three skew-t copulas. Section~\ref{sec:est} develops the variational inference method for 
estimating the AC skew-t factor copula, and its efficacy is demonstrated using simulated data in Section~\ref{sec:sim}. Sections~\ref{sec:vixeg} and~\ref{sec:risk} contain the two empirical studies
of intraday equity returns, while Section~\ref{sec:conc} concludes.

\section{Skew-t Copulas}\label{sec:copula}
\subsection{Copula models}
A copula model~\citep{Nelsen_1999,Joe_2014} expresses the joint distribution
function
of a continuous-valued random vector $\bm{Y}=(Y_1,\ldots,Y_d)^\top$ as
\begin{equation}\label{eq:copula}
	F_{Y}(\yvec) = C(F_{Y_1}(y_1),...,F_{Y_d}(y_d))\,. 
\end{equation}
Here, $\yvec=(y_1,\ldots,y_d)^\top$, $F_{Y_j}$ is the 
marginal distribution function of $Y_j$, and 
$C:[0,1]^d \rightarrow \mathbb{R}$ is a copula function that captures
the dependence structure. Differentiating~\eqref{eq:copula} gives
the joint density 
\begin{equation}\label{eq:copula_pdf}
	f_Y(\yvec) =
	\frac{\partial}{\partial \yvec}
	F_Y(\yvec)=c(F_{Y_1}(y_1),...,F_{Y_d}(y_d)) \prod_{j=1}^d f_{Y_j}(y_j)\,,
\end{equation}
where $f_{Y_j}=\frac{\partial}{\partial y_j} F_{Y_j}(y_j)$,
$c(\uvec)=\frac{\partial }{\partial \uvec}C(\uvec)$ is widely 
called the ``copula density'', and $\uvec=(u_1,\ldots,u_d)^\top$.\footnote{The notations $C(u_1,\ldots,u_d)$ and $C(\uvec)$ are used interchangeably, as are the notations $c(u_1,\ldots,u_d)$ and $c(\uvec)$.} 

The main
advantage of using a copula model is that the marginals and the copula
function can be modeled separately.
When the elements of $\bm{Y}$ are asset 
returns, their
distribution features asymmetric and high tail dependence; e.g. see~\cite{Patton_2006} and \cite{Christoffersen_Errunza_Jacobs_Langlois_2012}. A skew-t copula is one of only a few 
copulas that can account for these features in high dimensions.

\subsection{Three skew-t copulas}\label{sec:skewtdist}
Let the continuous random vector $\bm{Z}=(Z_1,\ldots,Z_d)^\top\sim F_Z$, with marginals $F_{Z_1},\ldots,F_{Z_d}$. Then if 
$U_j=F_{Z_j}(Z_j)$, the distribution function of $\bm{U}=(U_1,\ldots,U_d)^\top$ is
	$C(\uvec) = F_Z(F_{Z_1}^{-1}(u_1),...,F_{Z_d}^{-1}(u_d))$.
This is called an implicit copula, and the elements of $\bm{Z}$ are called pseudo or auxiliary variables because their values
are unobserved; see~\cite{smith2021implicit} for an introduction.
The copula density is
\begin{equation}
c(\uvec)=\frac{\partial}{\partial \uvec}C(\uvec)=f_Z(\zvec)/\prod_{j=1}^d f_{Z_j}(z_j)\,,\label{eq:coppdf}
\end{equation}
where $\zvec=(z_1,\ldots,z_d)^\top$, $f_Z(\zvec)=\frac{\partial}{\partial \zvec}F_Z(\zvec)$ and 
$f_{Z_j}=\frac{\partial}{\partial z_j}F_{Z_j}(z_j)$.
A skew-t copula is where $F_Z$ is a multivariate skew-t distribution. Because there are multiple skew-t distributions, there are also multiple copulas. Three types have been used
previously, which we briefly outline below. 
The marginal moments of $\bm{Z}$ are unidentified in $C$, so that
location parameters are unnecessary and the scale matrices are 
restricted to be a correlation matrix $\bar\Omega$ for all three skew-t copulas.

\subsubsection{GH skew-t copula}
\cite{demarta_mcneil_2005} construct the implicit copula of the generalized hyperbolic (GH) skew-t distribution~\citep{barndorff1977}. This distribution
 has the generative representation 
\begin{equation*}
    \bm{Z}_{\mbox{\tiny GH}} = \deltavec W +W^{-1/2}\bm{X}\,,\label{eq:ghgen}
\end{equation*}
where $\bm{X} \sim N_d(\bm{0}, \bar{\Omega}), W \sim \textrm{Gamma}(\nu/2, \nu/2)$, and $\deltavec \in \mathbb{R}^d$ is the skewness parameter. The joint density of $\bm{Z}_{\mbox{\tiny GH}}$ is available in closed form (see Part~A of the Web Appendix). However, the marginal distributions and quantile functions are not and are difficult to compute using numerical methods, so that they are usually 
evaluated by Monte Carlo simulation from the generative representation. For high $d$ this is 
slow. \cite{oh_patton_2021} note that if $\deltavec=\delta (1,1,\ldots,1)^\top$ with $\delta$ a scalar, then
the $d$ marginals of the GH distribution are the 
same. This speeds computation greatly, but at the cost of restricting the flexibility of the pairwise asymmetric dependencies in its implicit copula.

\subsubsection{SDB skew-t copula}
\cite{smith_gan_kohn_2010} construct the implicit copula of
the skew-t distribution proposed by~\cite{sahu_dey_branco_2003}.
This distribution can be constructed from a t-distribution by hidden conditioning as follows. 
Let $\bm{X}$ and $\bm{L}$
be $d$-dimensional random variables jointly distributed  
\begin{equation*}
\left(\begin{array}{l}\bm{X} \\\bm{L}\end{array}\right) \sim t_{2 d}\left(\bm{0} , \Omega, \nu\right),\,\;\;\Omega=\left(\begin{array}{cc} \bar\Omega +D^{2} & D \\D & I\end{array}\right)\,,\label{eq:sdbgen}
\end{equation*}
with $D = \operatorname{diag}(\deltavec), \deltavec \in \mathbb{R}^d$, $\bar\Omega$ a positive definite matrix, and $t_{2d}(\bm{0},\Omega,\nu)$
denoting a t-distribution with mean zero, scale matrix $\Omega$ and $\nu$
degrees of freedom. Then,
$\bm{Z}_{\mbox{\tiny SDB}} \equiv (\bm{X}|\bm{L}>\bm{0})$ \footnote{The notation 
	$\bm{X}|\bm{L}>\bm{0}$ corresponds to the conditional distribution of $\bm{X}$ given that all elements of $\bm{L}$ are positive. It does not denote the 
conditional distribution of $\bm{X}$ given a specific value for $\bm{L}$. This notational convention is used throughout the paper.}
is a 
 skew-$t$ distribution with skewness parameters $\deltavec$ and 
joint density 
        $f_{Z_{\mbox{\tiny SDB}}}(\zvec ; \bar\Omega, \deltavec, \nu)= 2^d f_t(\zvec;\bm{0},\bar\Omega+D^{2}, \nu) \mbox{Pr}(\bm{V}>\bm{0};\zvec)$
where $f_t(\zvec;\bm{0},\bar{\Omega},\nu)$ is the density of a $t_{d}(\bm{0},\bar{\Omega},\nu)$ distribution
evaluated at $\zvec$, and $\bm{V}$ follows a $d$-dimensional t-distribution
with parameters that are functions of $\zvec$. 
\cite{sahu_dey_branco_2003} show that 
the $j$th marginal has density 
$f_{Z_{\mbox{\tiny SDB}}}(z_j;1,\delta_j,\nu)$ with distribution and quantile functions
that are computed numerically.
Evaluation of the 
copula density at~\eqref{eq:coppdf} for large $d$ is difficult because computation of the term $\mbox{Pr}(\bm{V}>\bm{0};\zvec)$ is also. However,~\cite{smith_gan_kohn_2010}
show how to estimate the distribution and its implicit copula
using Bayesian data augmentation.

\subsubsection{AC skew-t copula}\label{sec:ac_cop}
\cite{kollo2010} and~\cite{Yoshiba_2018} construct the implicit copula 
of the skew-t distribution proposed by~\cite{Azzalini_Capitanio_2003}. This distribution
is formed via hidden conditioning as follows. Let $\bm{X}$ be
a $d$-dimensional random vector and $L$ a random variable with joint
distribution
\begin{equation}
	\left(\begin{array}{l} \bm{X} \\ L \end{array}\right) \sim t_{d+1}(\bm{0}, \Omega,\nu)\,, \;\; \Omega = \left(\begin{array}{ll} \bar{\Omega} & \deltavec \\\deltavec^{\top} & 1 \end{array}\right)\,, 
	\label{eq:ac_jointnorm}
\end{equation}
where $\deltavec = (\delta_1,\dots, \delta_d)^\top$. Then, 
$\bm{Z}_{\mbox{\tiny AC}} \equiv (\bm{X}|L > 0)$ is a skew-{\em t}
distribution with joint density
\begin{equation}\label{eq:fAC}
		f_{Z_{\mbox{\tiny AC}}}(\zvec;\bar{\Omega},\deltavec,\nu)=2 f_{t}(\zvec ; \bm{0},\bar{\Omega}, \nu) T\left(\alphavec^{\top} \zvec \sqrt{\frac{\nu+d}{\nu+{\cal M}(z)}} ; \nu+d\right)\,,
\end{equation}
where ${\cal M}(\zvec)=\zvec^{\top} \bar{\Omega}^{-1} \zvec$, and  $T(x;\nu)$ is the distribution function of a univariate student-t with 
degrees of freedom $\nu$, evaluated at $x$. 
There is a one-to-one relationship between $\alphavec$ and 
$\deltavec$, given by
\[
        \alphavec = (1 - \deltavec^{\top} \bar{\Omega}^{-1} \deltavec)^{-1 / 2} \bar{\Omega}^{-1} \deltavec\,, \;\mbox{ and }\;\; \deltavec = (1+\alphavec^{\top} \bar{\Omega} \alphavec)^{-1 / 2} \bar{\Omega} \alphavec\,.
\]
It is more convenient to employ $\alphavec\in \mathbb{R}^d$ as the skewness parameter, rather than $\deltavec$, because it is unconstrained. The $j$th marginal of~\eqref{eq:fAC} has
density
 $f_{Z_{\mbox{\tiny AC}}}(z_j; 1,\delta_j, \nu)$, and the distribution function
 $F_{j}(z_j;\delta_j, \nu)=\int_{-\infty}^{z_j} f_{Z_{\mbox{\tiny AC}}}(\xi; 1,\delta_j, \nu)d\xi$ is computed using numerical integration.

A draw from the AC skew-t distribution can be obtained by
first generating $W\sim \mbox{Gamma}(\nu/2,\nu/2)$, followed by $\tilde L\sim N(0,1)$ constrained so that $\tilde L>0$, and then from
\begin{equation}
\bm{Z}_{\mbox{\tiny AC}}|\tilde L,W \sim N_d\left(\frac{\tilde L }{\sqrt{W}}\deltavec,\frac{1}{W}
\left(\bar\Omega - \deltavec\deltavec^\top\right)\right)\,,
\label{eq:ACgen}
\end{equation}
where $\tilde L = W^{1/2}L$, and $L$ is defined at~\eqref{eq:ac_jointnorm}.
An alternative generative representation for the AC skew-t distribution is given in 
Appendix~\ref{app:gr}. To convert a draw $\bm{Z}_{\mbox{\tiny AC}}=(Z_1,\ldots,Z_d)^\top$ from the skew-t distribution to a draw $\bm{U}\sim C$ from its implicit copula, set 
$U_j=F_j(Z_j;\delta_j, \nu)$ for all $j$.
 
\subsection{Asymmetric dependence} \label{sec:properties}
The motivation for using any skew-t copula over a t-copula is to 
capture asymmetric dependence. To measure this here
we compute the four pairwise 
quantile dependence metrics, and
define measures of asymmetry in the major and minor diagonals as follows. Let $(Y_1,Y_2)$ follow a bivariate copula model\footnote{Because all three skew-t distributions are closed under marginalization, so are the skew-t copulas. Therefore, 
	 maximum asymmetric tail dependence in the bivariate case is equal to that for variable pairs in higher dimensions.}
with copula function $C(u_1,u_2)=\mbox{Pr}(U_1\leq u_1,U_2 \leq u_2)$. Then, for $0<u<0.5$, we define the four quantile dependencies as: lower left
		$\lambda_{\mbox{\tiny LL}}(u) = P(U_2\leq u|U_1\leq u)$, upper right
	$\lambda_{\mbox{\tiny UR}}(u) = P(U_2>1-u|U_1>1-u)$, lower right
	$\lambda_{\mbox{\tiny LR}}(u) = P(U_2 \leq u| U_1 > 1-u)$,
	and upper left
	$\lambda_{\mbox{\tiny UL}}(u) = P(U_2 > 1 - u| U_1 \leq u)$.
They are computed for the AC and SDB skew-t copulas using numerical integration, and by Monte Carlo simulation for the GH skew-t copula; see Appendix~\ref{app:qdep} for more details.
For a given quantile value $u$, the metrics 
\[
\Delta_{\mbox{\tiny Major}}(u)\equiv \lambda_{\mbox{\tiny UR}}(u)-\lambda_{\mbox{\tiny LL}}(u)\,,\;\;\; \mbox{ and }\;\;\;
\Delta_{\mbox{\tiny Minor}}(u)\equiv \lambda_{\mbox{\tiny UL}}(u)-\lambda_{\mbox{\tiny LR}}(u)\,,
\]
 measure asymmetry 
in the major and minor diagonals, respectively. \cite{ando2022quantile} suggest similar quantile metrics, although based on a network model, rather than a copula model.

\begin{figure}[htp]
	\centering
	\includegraphics[width=1\textwidth]{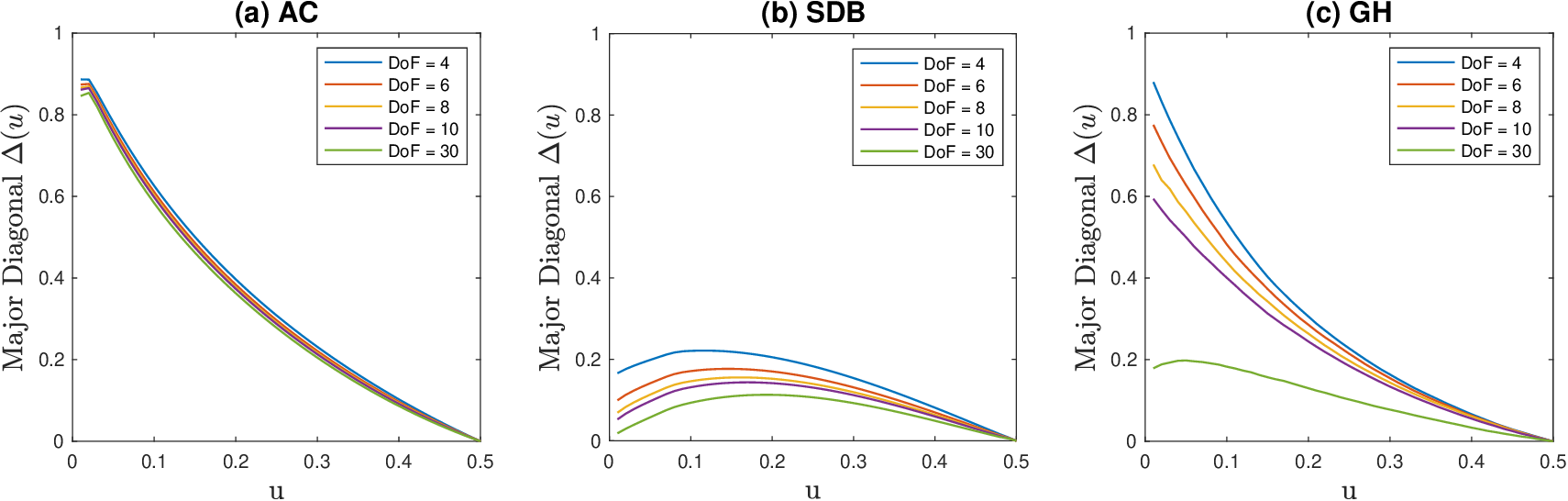}
	\caption{Maximum asymmetric dependence $\max_{\deltavec,\rho} \left\{\Delta(u) \right\}$ of the three skew-t copulas (a)~AC, (b)~SDB, and~(c) GH. 
		Results are given for degrees of freedom $\nu = 4, 6, 8, 10, 30$, and plotted as a function of $u$.}
	\label{fig:max_asym}
\end{figure} 
The maximum asymmetry for each of the
three copulas is obtained by solving the optimization
\[
\max_{\deltavec,\rho} \left\{\Delta(u) \right\}
\]
for given values of $0<u<0.5$ and 
$\nu$. The solution
is the same for $\Delta_{\mbox{\tiny Major}}(u)$ and $\Delta_{\mbox{\tiny Minor}}(u)$, and optimization is with respect to
$\deltavec=(\delta_1,\delta_2)$ and the off-diagonal element $\rho$
of the $(2\times 2)$ correlation matrix $\bar \Omega$. 
Figure~\ref{fig:max_asym} plots the maximums against $u$. 
At every quantile the GH skew-t copula
has a higher maximum asymmetric dependence than the SDB skew-t copula. 
However, the AC skew-t copula has a
 higher maximum level of asymmetric dependence than both alternatives for all but the lowest values of $\nu$, at which
$\Delta(u)$ is almost equal for the AC and GH skew-t copulas.
Figure~A1 
in the Web Appendix plots contours of
the bivariate densities for
the three copulas with maximal $\Delta_{\mbox{\tiny Major}}(0.01)$, further illustrating
the strong differences between the skew-t copulas




In summary, the AC skew-t copula allows for a higher level of asymmetric dependence, making it
an attractive choice. In addition, the computational demands of evaluating the marginals of the GH skew-t 
distribution complicates the use of its implicit copula in higher dimensions. A comparison 
of the differing skew-t copulas is undertaken later in our study of intraday data. 

 
\subsection{Factor copula}\label{sec:factor}
For large $d$ an unrestricted correlation matrix $\bar \Omega$ is difficult to estimate and one
solution is adopt a factor model for the auxiliary variables $\bm{Z}$.
Following~\cite{Murray_Dunson_Carin_Lucas_2013},
a static factor copula is used here, which corresponds 
to adopting the factorization
\begin{equation*}
\bar\Omega = V_1 V V_1 = V_1(GG^\top + D)V_1. 
\end{equation*}
 Here, $G=\{g_{ij}\}$ is an $(d \times k)$ loadings matrix with $k<d$, zero upper triangle
and positive leading diagonal elements (i.e. $g_{ii}>0$), and to identify the parameters $D=I_d$.
The diagonal matrix 
 $V_1=\mbox{diag}(V)^{-1/2}$ normalizes $V$ to a correlation matrix. If $\widetilde{G}$ equals $G$ with the leading diagonal elements
 replaced by their logarithmic values, then $\bar \Omega$ is parameterized by
 $\mbox{vech}(\widetilde{G})$. We adopt this transformation because our VI method 
 is applied to unconstrained real-valued model parameters.
 
\section{Variational Inference for the AC Skew-t Copula}\label{sec:est}
\subsection{Likelihood and Extended Likelihood}\label{sec:like}
Consider observations $\yvec_i=(y_{i1},\ldots,y_{id})^\top$, for $i=1,\ldots,n$, drawn independently  
from the copula model at~\eqref{eq:copula} with copula parameters $\thetavec$.
The pseudo variables $\zvec_i=(z_{i1},\ldots,z_{id})^\top$
are given by 
\begin{equation}\label{eq:transform}
z_{ij}=F_{Z_j}^{-1}\left(F_{Y_j}\left(y_{ij}\right);\thetavec_j\right)\,,
\end{equation}
where $F_{Z_j}$ is a function of $\thetavec_j \subseteq \thetavec$. 
If $\yvec=\{\yvec_1,\ldots,\yvec_n\}$, then from~\eqref{eq:copula_pdf} and~\eqref{eq:coppdf}, the likelihood is
\[
p(\yvec|\thetavec)=\prod_{i=1}^n \left\{p(\zvec_i|\thetavec)\prod_{j=1}^d \frac{ f_{Y_j}(y_{ij})}{f_{Z_j}(z_{ij};\thetavec_j)} \right\}\,.
\]
For the AC skew-t copula, this likelihood has a complex geometry, making  
direct maximization challenging for large $d$; see Part D.2 of the Web Appendix for an illustration. 
However, a 
more tractable extended likelihood of the AC skew-t copula model can be obtained from the generative 
representation for the AC skew-t distribution given in Section~\ref{sec:ac_cop} as follows.

Let $\wvec=(w_1,\ldots,w_n)^\top$ be a vector $n$ draws of $W$, and $\tilde\lvec=(\tilde l_1,\ldots,\tilde l_n)^\top$ be a vector of $n$ draws of $\tilde L$, 
then 
\begin{equation}
	p(\yvec,\tilde \lvec,\wvec|\thetavec) =\prod_{i=1}^n p(\yvec_i ,\tilde l_i,w_i|\thetavec)
	= \prod_{i=1}^n \left\{p(\zvec_i,\tilde l_i,w_i|\thetavec)
	\prod_{j=1}^d \frac{ f_{Y_j}(y_{ij})}{f_{Z_j}(z_{ij};\thetavec_j)}\right\}\,,\label{eq:elike1}
\end{equation}
is the extended likelihood.
The product over $j$ in~\eqref{eq:elike1} is the Jacobian of the transformation from $\zvec$ to $\yvec$ at~\eqref{eq:transform}. 
From the generative representation of the AC skew-t distribution, the joint density
\begin{equation}\label{eq:elike2}
	p(\zvec_i,\tilde l_i,w_i|\thetavec)=p(\zvec_i|\tilde l_i,w_i,\thetavec)p(\tilde l_i|w_i)p(w_i|\nu)\,,
\end{equation}
where $p(w_i|\nu)$ is the density of a Gamma($\nu/2,\nu/2)$ distribution, $p(\tilde l_i|w_i)= 2\phi_1(\tilde l_i;0,1)\mathds{1}(l_i>0)$ is the density of a constrained standard normal, and 
$p(\zvec_i|\tilde l_i,w_i,\thetavec)=\phi_d(\zvec_i;\deltavec \tilde l_i w_i^{-1/2},w_i^{-1}(\bar\Omega - \deltavec\deltavec^\top))$.
Here, $\phi_m(\cdot;\muvec,\Sigma)$ denotes the density
of a $N_m(\muvec,\Sigma)$ distribution, and the indicator function $\mathds{1}(X)=1$ if $X$ is true, and zero otherwise.
The copula model likelihood can be recovered by integrating over the two latent vectors;
i.e. $p(\yvec|\thetavec)=\int p(\yvec,\tilde \lvec,\wvec|\thetavec)\mbox{d}(\tilde \lvec,\wvec)$.

%


This extended likelihood is tractable and fast to compute. Evaluation of $F_{Z_1}^{-1},\ldots,F_{Z_d}^{-1}$ 
at~\eqref{eq:transform} is undertaken 
using the spline interpolation method outlined by~\cite{Smith_Maneesoonthorn_2018} and~\cite{Yoshiba_2018} that
uses the closed form densities $f_{Z_1},\ldots,f_{Z_d}$. These authors show this approach is
highly accurate and is scalable with respect to $n$. An alternative extended likelihood
can be specified using
the other generative representation given in Appendix~\ref{app:gr}. However, we found
it to be less effective than that adopted here; see Parts~C.3 and~D.2 of the Web Appendix.

%
%

\subsection{Priors and Posterior}
The copula
parameters with a factor decomposition for $\bar \Omega$ are  $\thetavec=\{\mbox{vech}(\widetilde{G}),\alphavec,\nu\}$, where
$\alphavec = (\alpha_1,...,\alpha_d)^\top$ is a one-to-one function of $\deltavec$.
 Bayesian estimation uses
the posterior density $p(\thetavec|\yvec)$, which is the marginal in $\thetavec$ of the augmented posterior 
\begin{equation}
	p(\thetavec,\tilde \lvec,\wvec|\yvec)\propto p(\yvec,\tilde \lvec,\wvec|\thetavec)p(\thetavec)\equiv h(\psivec)\,,\label{eq:augpost}
\end{equation}
where $\psivec\equiv\{\thetavec,\tilde \lvec,\wvec\}$. The augmented posterior is the product of the 
extended likelihood at~\eqref{eq:elike1} and the prior
 $p(\thetavec)=p(\mbox{vech}(\widetilde{G}))p(\alphavec)p(\nu)$.
We adopt the
generalized double Pareto distribution prior for each element of $\mbox{vech}(\widetilde{G})$ suggested by \cite{Murray_Dunson_Carin_Lucas_2013}.
The prior $\alpha_j \sim N(0, 5^2)$, which places
99\%  mass on $\delta_j\in(-0.997,0.997)$.
The prior for $\nu$ is constrained so that $\nu>2$, with $(\nu-2) \sim\operatorname{Gamma}(3, 0.2)$. 
This places 99\%  mass on the range $\nu\in (3.69,48.47)$. 

An MCMC scheme that produces draws from the augmented posterior of the 
AC skew-t copula is outlined in Appendix~\ref{app:mcmcscheme}.
However, it is slow and for high $d$ (including when $d=93$ as in our equity application in Section~\ref{sec:risk}) it does not evaluate the augmented posterior in reasonable computing time. 
Variational inference  
is a faster and scalable alternative, as we now discuss.

\subsection{Hybrid variational inference}
 VI approximates the augmented posterior using a density $q(\psivec)\in {\cal Q}$ called
the ``variational approximation'' (VA) from a family of tractable approximations ${\cal Q}$.
Typically, the VA is obtained by minimizing the Kullback-Leibler divergence (KLD) between the target density $p(\psivec|\yvec)\propto p(\yvec|\psivec)p(\psivec)=h(\psivec)$ and its approximation $q(\psivec)$. It is easily shown that this is equivalent to  maximizing the Evidence Lower Bound (ELBO)  
\begin{align*}
\calL = E_{q}\left[\log h(\psivec) - \log q(\psivec)\right] 
\end{align*}
over $q\in {\cal Q}$, where the expectation is with respect to $\psivec\sim q$.

The selection of families ${\cal Q}$ that balance accuracy of the VA and the time required to maximize the 
ELBO function, is the topic of much current research. For models with 
a large number of latent variables, such as the augmented
posterior at~\eqref{eq:augpost}, \cite{Loaiza-Maya_Smith_Nott_Danaher_2021} 
point out that assuming simple fixed form approximations (such as the widely used
mean field approximation) can be very inaccurate. Instead, they 
suggest using a VA family of the form
\begin{align}
q_{\lambda}(\psivec) = p(\tilde \lvec,\wvec|\thetavec,\yvec) q^0_{\lambda}(\thetavec) \,,
\label{eq:hybrid_q}
\end{align}
where 
$p(\tilde \lvec,\wvec|\thetavec,\yvec) $ is the conditional posterior of the latent variables, and $q^0_{\lambda}(\thetavec)$ is the density of a fixed form  VA with 
parameters $\lambdavec$ which we discuss further below.\footnote{In the machine learning literature a parametric density $q_\lambda$ with parameters $\lambdavec$ is often called a fixed form density.}
These authors show that when using the VA at~\eqref{eq:hybrid_q},
the ELBO of the target density $p(\psivec|\yvec)$ is exactly equal to the ELBO for
the target density $p(\thetavec|\yvec)$; that is,
\[
\calL = E_{q_\lambda}\left[\log h(\psivec) - \log q_\lambda(\psivec)\right] =
E_{q^0}\left[\log\left(p(\yvec|\thetavec)p(\thetavec)\right) - \log q^0_\lambda(\thetavec)\right]\,.
\] 
Thus, maximizing $\cal L$ is equivalent
to solving the variational optimization for the parameter posterior with the latent variables $\{\tilde \lvec,\wvec\}$ integrated out exactly, which makes~\eqref{eq:hybrid_q} more accurate than other choices of VA for this target density.

To maximize ${\cal L}$ with respect to 
$\lambdavec$,
stochastic gradient ascent (SGA) is the most frequently used algorithm~\citep{ranganath2014}. From an initial
value $\lambdavec^{(0)}$, SGA recursively updates 
\begin{equation}\label{eq:Update}
	\bm{\lambda}^{(t+1)} = \bm{\lambda}^{(t)}+\bm{\rho}^{(t)}\odot\left.\widehat{\nabla_\lambda\mathcal{L}}\right|_{\lambdavec=\lambdavec^{(t)}},\; t=0,1,\ldots
\end{equation}
until reaching convergence. Here, $\bm{\rho}^{(t)}$ is a vector of adaptive learning rates set using the momentum method of~\cite{zeiler},
`$\odot$' is the Hadamard product,
and $\widehat{\nabla_\lambda\mathcal{L}}$ is an unbiased 
estimator of the gradient, 
which is 
evaluated at $\lambdavec=\lambdavec^{(t)}$. 
The key to fast convergence and computational 
efficiency of this SGA algorithm is that $\widehat{\nabla_\lambda\mathcal{L}}$
has low variability and is fast to evaluate.
One of the most effective ways to achieve this is to use the re-parameterization 
trick of~\cite{Kingma_2014}. For the VA at~\eqref{eq:hybrid_q}, 
the re-parameterization is a transformation from $\thetavec$ to  
$\varepsilonvec\sim f_\varepsilon$, where 
 $\thetavec=\tau(\varepsilonvec,\lambdavec)$ and $\tau$ is a 
 deterministic function. 
In this case, \cite{Loaiza-Maya_Smith_Nott_Danaher_2021} show that 
\begin{equation}
	\nabla_\lambda\mathcal{L} = E_{f_{\varepsilon,\tilde l,w}}\left[
	\frac{\partial \thetavec^\top}{\partial \lambdavec}
	\left(
	\nabla_\theta \log h(\psivec) - \nabla_\theta \log q_\lambda^0(\thetavec) 
	\right)
	\right]\,,
	\label{eq:repargrad}
\end{equation}
where the expectation is with respect to the density 
$f_{\varepsilon,\tilde l,w}(\varepsilonvec,\tilde \lvec,\wvec)=f_\varepsilon(\varepsilonvec)
p(\tilde \lvec,\wvec|\thetavec,y)$. 
Typically only a single 
draw from $f_{\varepsilon,\tilde l,w}$ is required to obtain
a low variance estimate of the expectation in~\eqref{eq:repargrad}, resulting
in a major computational advantage.


For $q^0_\lambda$ we use a Gaussian density with a factor model 
covariance matrix as suggested by~\cite{miller2017} and \cite{Ong_Nott_Smith_2018}
with $r$ factors. This is not to 
be confused with the factor model used to define the copula.
\cite{Ong_Nott_Smith_2018} give the transformation $\tau$ required for the re-parametrization trick, fast to compute closed form expressions for 
the derivatives 
$\frac{\partial \thetavec^\top}{\partial \lambdavec}$ and 
$\nabla_\theta \log q_\lambda^0(\thetavec)$ in~\eqref{eq:repargrad}, along with MATLAB routines for their evaluation.
The derivative 
\[
\nabla_\theta \log h(\psivec)=( \nabla_{{\rm vech}(\widetilde G)}^\top \log h(\psivec), \nabla_{\alpha}^\top \log h(\psivec), \nabla_\nu^\top  \log h(\psivec) )^\top\,,
\]
is specific
to the target density $p(\psivec|\yvec)$, which is the augmented posterior at~\eqref{eq:augpost} here. 
Table~\ref{tab:grad} provides this gradient, which is evaluated recursively from the bottom of the columns upwards. Its derivation is given in Part~B of the Web Appendix, and uses the trace operator to express the gradient in a computationally efficient directional
derivative form. 
The gradient can also be computed using 
automatic differentiation, although this is much slower. 

\cite{Loaiza-Maya_Smith_Nott_Danaher_2021} call an approach that combines the VA at~\eqref{eq:hybrid_q} with SGA optimization 
and the re-parameterized gradient at~\eqref{eq:repargrad}, a ``hybrid VI'' method. This is because it nests
an MCMC step within a well-defined stochastic optimization algorithm for solving 
the VI problem.  
Algorithm~\ref{alg:hvi} details our proposed hybrid VI algorithm for estimating
the AC skew-t copula. 
At Step~(b)
a small number of Gibbs steps that first draw from $\tilde \lvec|\wvec,\thetavec,\yvec$, and then from $\wvec|\tilde \lvec,\thetavec,\yvec$, are used as outlined in Appendix~\ref{app:mcmcscheme}.   
We demonstrate that this works well here, although
other methods, such as the Hamiltonian Monte Carlo sampler of~\cite{hoffman2014no}, can also be used at this step. The output of the algorithm is 
$q^0_{\widehat{\lambda}}(\thetavec)$, which is usually called the ``variational posterior'', 
because
it is the optimal VA to the parameter posterior $p(\thetavec|\yvec)$.

\begin{algorithm}[ht!]
	\begin{algorithmic}
		\State Initiate $\lambdavec^{(0)}$ and set $t = 0$
		\Repeat
		\State (a) Generate $\thetavec^{(t)}$ using its re-parametrized representation
		\State (b)  Generate $\{\tilde \lvec^{(t)},\wvec^{(t)}\} \sim p(\tilde \lvec,\wvec| {\thetavec}^{(t)},\yvec)$
		\State (c) Compute the gradient 
		$\widehat{\nabla_\lambda\mathcal{L}}$ using~\eqref{eq:repargrad}
		evaluated at the values $\{\lambdavec^{(t)},\thetavec^{(t)},\tilde \lvec^{(t)},\wvec^{(t)}\}$
		\State (d) Compute step size $\boldsymbol{\rho}^{(t)}$ using an automatic adaptive method (e.g. an ADA method)
		\State (e) Set $\lambdavec^{(t+1)} = \lambdavec^{(t)} + \boldsymbol{\rho}^{(t)} \circ \widehat{\nabla_{\lambda}\mathcal{L}}(\lambdavec^{(t)})$
		\State (f) Set $t = t+1$
		\Until{either a stopping rule is satisfied or a fixed number of steps is taken}
		\State Set $\widehat{\lambdavec}=\frac{1}{500}\sum_{s=1}^{500}\lambdavec^{(t+1-s)}$, and the variational parameter posterior to $q^0_{\widehat{\lambda}}(\thetavec)$
	\end{algorithmic}
	\caption{Hybrid VI for AC Skew-t Copula}\label{alg:hvi}
\end{algorithm}

\section{Simulation}\label{sec:sim}
We first demonstrate the efficacy of the VI method using simulated data from lower dimensional examples where
the exact posterior can be calculated using MCMC to measure accuracy.

\subsection{Design}
A sample of size $n=1024$ is generated from each of two AC skew-t factor copulas. The first is a low-dimensional single factor model
($d=5,k=1$; Case~1), and the second is higher dimensional five factor model ($d=30,k=5$; Case~2). The copula parameters were obtained from fitting these models 
to returns data;
see Part~C of the Web Appendix for details of the data generating processes. 

\subsection{Approximation accuracy}
For both examples, the variational posterior is compared to the exact posterior $p(\thetavec|\yvec)$
computed using the (slower) MCMC scheme
in Appendix~\ref{app:mcmcscheme}. MCMC evaluates
the posterior up to an arbitrary error, which we made small 
using a large Monte Carlo sample size.
A skew-t copula with very different parameter values can have similar densities $c(\uvec)$ and thus dependence structures. Therefore, 
estimation 
accuracy is best measured using pairwise dependence
metrics of the copula for all possible variable pairs, rather than parameter values. The metrics computed here include the
quantile dependencies computed at the 1\% and 5\% quantiles, and the
Spearman correlation. For  the pair $(Y_i,Y_j)$, the latter is  $\rho^S_{ij}= 12\int \int C_{ij}(u_i',u_j')du_i'du_j'-3$, with $C_{ij}$ the bivariate marginal copula for $(Y_i,Y_j)$ from the skew-t copula
evaluated as in Appendix~\ref{app:qdep}. 

\begin{figure}[tbhp]
	\centering
	\includegraphics[width=0.6\textwidth]{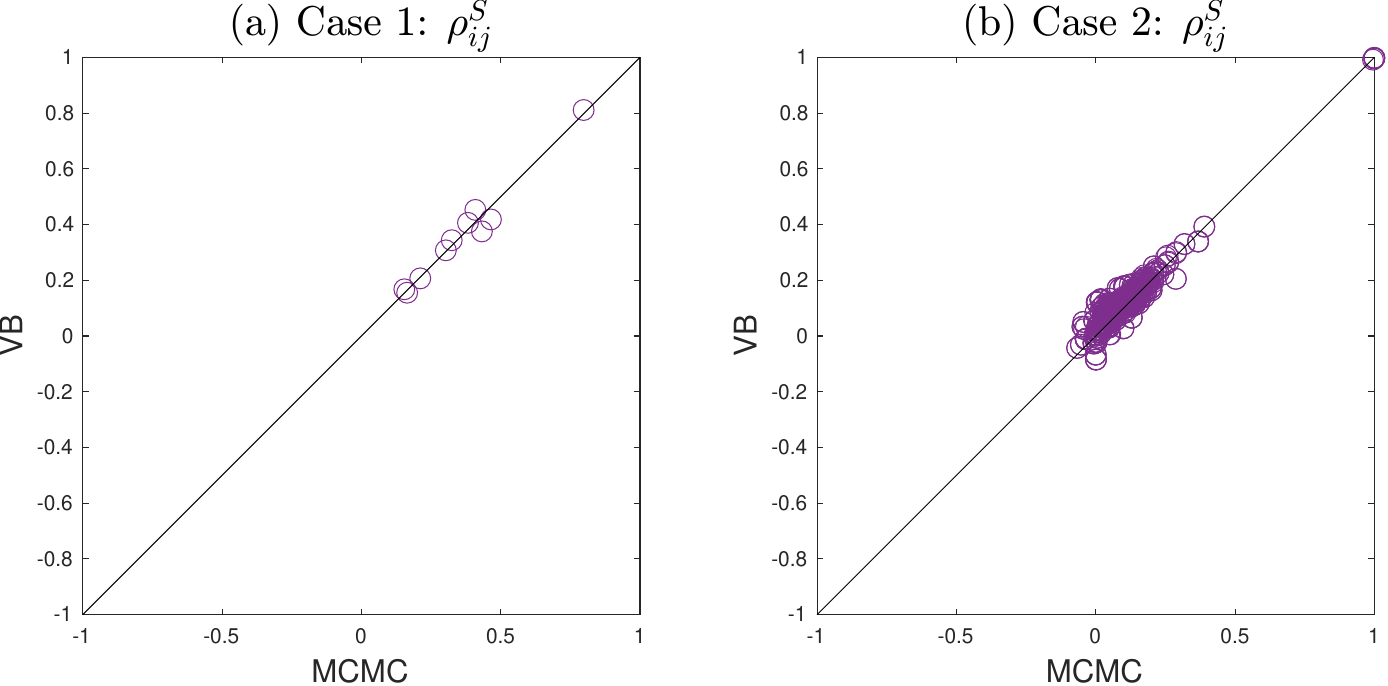}
	\caption{Comparison of the posterior mean estimates of the pairwise Spearman correlations $\rho^S_{ij}$ computed exactly
		using MCMC (horizontal axis) and approximately using VI (vertical axis). Panel~(a) plots these for the 5-dimensional skew-t copula in Case 1, and panel~(b) for the 30-dimensional skew-t copula in Case~2.}
	\label{fig:mcmc_vs_vb_spearmanrho_mh25_mean}
\end{figure}
\begin{figure}[htbp] 
	\centering
	\includegraphics[width=0.6\textwidth]{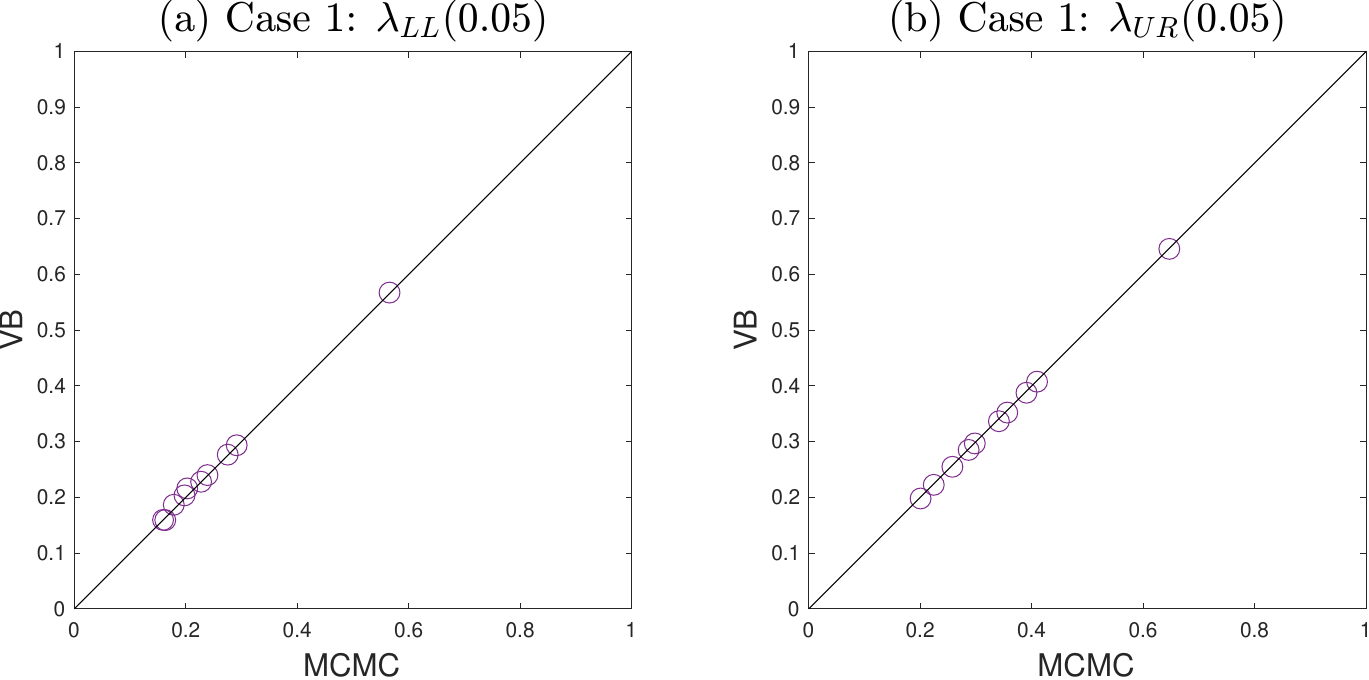}
	\caption{Comparison of the posterior mean estimates of the pairwise 5\% quantile dependence metrics $\lambda_{LL}(0.05)$ and $\lambda_{UR}(0.05)$ for Case~1 computed exactly
		using MCMC (horizontal axis) and approximately using VI (vertical axis).} 
	\label{fig:mcmc_vs_vb_5D_qd_05_major_mean}
\end{figure}
\begin{figure}[htbp] 
	\centering
	\includegraphics[width=0.6\textwidth]{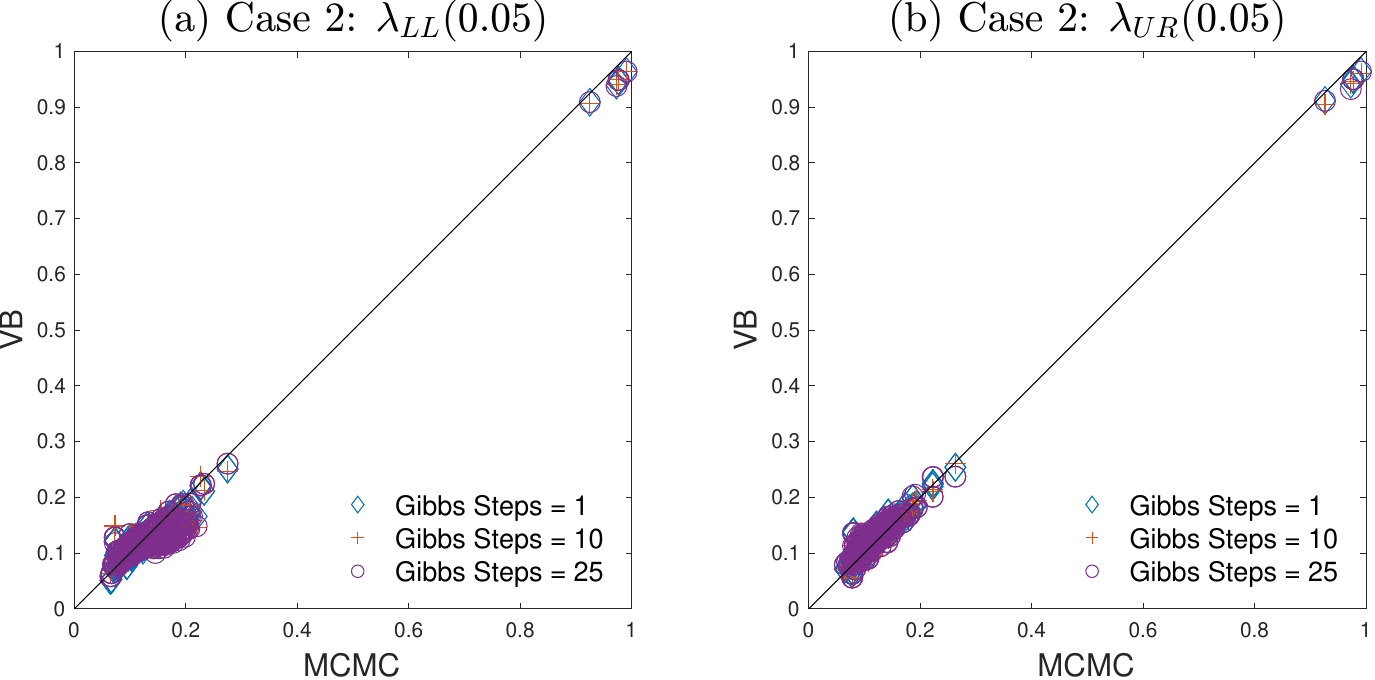}
	\caption{Comparison of the posterior mean estimates of  pairwise metrics $\lambda_{LL}(0.05)$ and $\lambda_{UR}(0.05)$ 
		for Case~2 computed exactly
		using MCMC (horizontal axis) and approximately using VI (vertical axis). Results are given for three VI 
		implementations where 1, 10 and 25 random walk Metropolis-Hastings draws were used to obtain a draw
		at step~(b) of Algorithm~\ref{alg:hvi}, with almost identical values.}
	\label{fig:mcmc_vs_vb_30D_iter5k_qd05_major_mean}
\end{figure}

The exact and approximate posterior means of the metrics are evaluated by averaging their values computed at Monte Carlo draws from $p(\thetavec|\yvec)$ and $q_{\widehat{\lambda}}^0(\thetavec)$, respectively. 
Figure~\ref{fig:mcmc_vs_vb_spearmanrho_mh25_mean} plots the exact and variational posterior means of 
the pairwise Spearman correlations. Figures~\ref{fig:mcmc_vs_vb_5D_qd_05_major_mean} 
and~\ref{fig:mcmc_vs_vb_30D_iter5k_qd05_major_mean} do the same for the quantile dependence metrics at the 5\% level. They show that VI produces a copula estimate
with a dependence structure that is close to that of the exact posterior; 
further results are given in 
Part~C of the Web Appendix. Variational inference is an approximate method
and it typically gives
estimates of the second moments of the Bayesian posterior that are less accurate
than the posterior mean, which we also observe here, although this has limited
impact on prediction; see~\cite{frazier2023variational} for a discussion.  

\subsection{Approximation speed}
In general, we implement Algorithm~\ref{alg:hvi} using 25 Gibbs draws at step~(b).
To illustrate that the algorithm is robust to the number of draws, 
Figure~\ref{fig:mcmc_vs_vb_30D_iter5k_qd05_major_mean} plots estimates for three implementations
with 1, 10 and 25 Gibbs draws. The VI estimates are very similar, 
which is consistent with the finding of~\cite{Loaiza-Maya_Smith_Nott_Danaher_2021} for some other
models.

Table~\ref{tab:computation_time} compares the computation time for MCMC and VI (with 25 Gibbs draws) using a standard laptop.
The diagnostic of \cite{Geweke_1991} was used to determine the number of MCMC sweeps required.
The number of steps in the SGA algorithm was 5000, which is a conservative number as judged by the 
stability of the optimal VA; for example, in Case~1 the VA based on only 2500 steps is very similar. The  VI estimator is faster,
and its computational advantage grows with $d$ and $k$. Finally, in the empirical 
work in Section~\ref{sec:risk} where $d=93$,
we found the VI estimator stable,
whereas the MCMC algorithm in Appendix~\ref{app:mcmcscheme} exhibited poor mixing and 
was unable to estimate the model in reasonable time.

  
\begin{table}[htbp]
	\centering
	\caption{Computation times to estimate skew-t copulas for simulated datasets}
	  \begin{tabular}{cccc}
	  \toprule
	  \toprule
			& \multicolumn{3}{c}{Case 1  (Dimension $d=5$ with $k=1$ Factor)} \\
  \cmidrule{2-4}          & Time/Step (seconds) & Number of Steps  & Total Time (hours)  \\
	  MCMC  & 0.207 & 27000 & 1.553 \\
	  VB    & 0.268 & 5000  & 0.372 \\
  \cmidrule{1-4} 
			& \multicolumn{3}{c}{Case 2 (Dimension $d=30$ with $k=5$ Factors)} \\
  \cmidrule{2-4}    MCMC  & 1.580 & 33000 & 14.480 \\
	  VB    & 1.446 & 5000  & 2.008 \\
	  \bottomrule
	  \bottomrule
	  \end{tabular}%
  \caption*{Note: Computation times in MATLAB for a 2018 MacBook Pro (with 2.3GHz Intel i5 CPU) laptop.}
	\label{tab:computation_time}%
  \end{table}%

\subsection{Choice of data augmentation}
Our approach uses the generative representation at~\eqref{eq:ACgen} (labelled `GR2')
to construct a variational
data augmentation method. One drawback is that a nested (albeit fast) MCMC sampler to draw from
$p(\tilde{\lvec},\wvec|\thetavec,\yvec)$ is required at step~(b) of  Algorithm~\ref{alg:hvi}. 
An alternative generative representation `GR1' in Appendix~\ref{app:gr} with latent vector $\lvec=(l_1,\dots,l_n)^\top$ can also be used to construct another hybrid VI algorithm 
that targets $p(\lvec,\thetavec|\yvec)$, where at step~(b) direct generation from $p(\lvec|\thetavec,\yvec)$ is possible
without the use of a nested MCMC sampler. We implemented this alternative hybrid VI algorithm based on GR1, but found it
to be less efficient than that based on GR2; see Part~C.3 of the Web Appendix for more details. Additional discussion of why GR2 is an effective target 
distribution is also given in Part~D.2 of the Web Appendix.

%
%
\section{Asymmetric Dependence Between Equity Returns and the VIX}\label{sec:vixeg}
The first example applies the AC skew-t copula model, with $k=1$ factor and fitted using the proposed VI method, to two equity returns series and the VIX. 
The latter
is an index of overall market volatility published by the Chicago Board of Exchange~\citep{britten2000option}. The objective is to illustrate the 
flexibility of the dependence structure of the AC skew-t copula.


\subsection{Data Description and Marginal Models}\label{sec:data}
Trade prices for Bank of America (BAC) and JP Morgan (JPM) were
obtained from Revinitiv DataScope Select database.
For each equity, 
the $N=26$ fifteen minute returns between 09:30 and 16:00 EDT of each trading day are constructed as follows.
If $P_{\tau,t}$ denotes the last trade price of a given equity in the $\tau$th 15 minute interval on trading day $t$, 
then the return for that interval is  $r_{\tau,t}=\log(P_{\tau,t})-\log(P_{\tau-1,t})$, with $P_{0,t}\equiv P_{N,t-1}$  
the last trade price on the previous day. 
Days where trading is suspended are removed.

We follow~\cite{Patton_2006} and many others by using the copula
to capture cross-sectional dependence in the conditional distribution of the financial variables.
The intraday GARCH model of~\cite{Engle_Sokalska_2012} is used as a marginal model for the returns on each equity, which decomposes the return into components
\begin{equation*}
	r_{\tau,t} = (h_t s_\tau)^{1/2} \sigma_{\tau,t} \varepsilon_{\tau,t} \label{eq:rti}\,.
\end{equation*}
Here, $h_t = \sum^N_{\tau=1}(r_{\tau,t})^2$ is the realized variance of day $t$, 
$s_\tau= \frac{1}{T}\sum_{t=1}^T(r_{\tau,t})^2 / h_t$ is an estimate of the diurnal pattern in volatility for the $\tau$th  interval, and the conditional volatility movement
\begin{equation*}
	\sigma_{\tau,t}^2 = \beta_0 + \beta_1 \epsilon^2_{\tau-1,t} + \beta_2\sigma^2_{\tau-1,t}\,, \label{eq:sigti}
\end{equation*}
with $\epsilon_{\tau,t}=r_{\tau,t}/(h_t s_\tau)^{1/2}$, $\epsilon_{0,t}=\epsilon_{N,t-1}$ and $\sigma_{0,t}=\sigma_{N,t-1}$. In our analysis, we assume that the innovation $\varepsilon_{\tau,t} \sim t_1(0,1,\tilde{\nu})$, and
estimate the parameters $(\beta_0,\beta_1,\beta_2,\tilde \nu)$ for each equity using maximum likelihood. 
The copula data is computed
as $T(\varepsilon_{\tau,t};\tilde{\nu})$ at the model 
estimates.\footnote{To match this with the copula model notation in Section~\ref{sec:like}, note that if $i=(N-1)t+\tau$ (so that $i=1,2,\ldots,NT$) and the equity return is the $j$th marginal, then $y_{ij}=r_{\tau,t}$
	and $F_{Y_j}(y_{ij})=T(\varepsilon_{\tau,t};\tilde \nu)$ in~\eqref{eq:transform}.}

Fifteen minute observations of the VIX were also obtained from the Revinitiv DataScope Select database. The VIX marginal model is
a first order autoregression with a nonparametric disturbance estimated using a  kernel density estimator.

\subsection{Empirical Results}
The copula model was fit to data from two periods of 40 trading days: a low volatility 
period between 24 Oct. and 22 Dec. 2017, and a high volatility period between 11 Feb. and 15 Apr. 2020.\footnote{We select these because the first period contains the lowest VIX value from 2017 to 2021, whereas the second contains the highest VIX value that corresponds to the COVID-19 crash.}
Table~\ref{tab:vix_pair_corr_theta} reports the estimates of the copula parameters, correlations and  asymmetry measures. The two equity returns are positively correlated, whereas they are both negatively correlated with the VIX. 
There are strong asymmetries in the quantile dependencies, which change in direction and scale from
the low to high volatility periods. For example, between the two equity returns, 
$\Delta_\text{Major}(0.05)$ changes from $0.169$ to $-0.279$. This suggests that during the
period of high market volatility, negative returns are more dependent than positive returns. Whereas, between the equity returns and the VIX, 
$\Delta_\text{Minor}(0.05)$ changes from $-0.059, -0.061$ to $0.240, 0.211$. 
This suggests that during the period of high market volatility there is an increase
in the dependence between 
high VIX values and negative equity returns.

  \begin{table}[htbp]
	\begin{center}
	\caption{Estimated AC Skew-t Copula with $d=3$ for the BAC-JPM-VIX Example}
	\resizebox{1.0\textwidth}{!}{
      \begin{tabular}{ccccccccccc}
      \toprule
      \toprule
            &       & \multicolumn{9}{c}{Low Volatility Period: From 24 October 2007 to 22 December 2017} \\
  \cmidrule{3-11}          &       & \multicolumn{2}{c}{Correlation} &       & \multicolumn{2}{c}{Tail Asymmetry} &       & \multicolumn{3}{c}{$\thetavec$} \\
  \cmidrule{3-4}\cmidrule{6-7}\cmidrule{9-11}    Pair  &       & Spearman & Kendall &       & $\Delta_\text{Major}(0.05)$ & $\Delta_\text{Minor}(0.05)$ &       & $\rho$ & $\deltavec^\top$ & $\nu$ \\
      \multirow{2}[0]{*}{BAC -- VIX} &       & -0.322 & -0.219 &       & 0.003 & -0.059 &       & -0.459 & 0.987, -0.422  &  \\
            &       & (0.029) & (-0.021) &       & (0.003) & (0.023) &       & (0.036) & (0.041, 0.006) &  \\
      \multirow{2}[0]{*}{JPM -- VIX} &       & -0.355 & -0.242 &       & 0.000 & -0.061 &       & -0.484 & 0.829,-0.422 & 25.767 \\
            &       & (0.041) & (0.029) &       & (0.001) & (0.013) &       & (0.037) & (0.034, 0.006) & (4.758) \\
      \multirow{2}[1]{*}{JPM -- BAC} &       & 0.759 & 0.566 &       & 0.169 & 0.000 &       & 0.881 & 0.829, 0.987 &  \\
            &       & (0.028) & (0.025) &       & (0.041) & (0.000) &       & (0.024) & (0.034, 0.041) &  \\
      \midrule
      \midrule
            &       & \multicolumn{9}{c}{High Volatility Period: From 11 February 2020 to 15 April 2020} \\
  \cmidrule{3-11}          &       & \multicolumn{2}{c}{Correlation} &       & \multicolumn{2}{c}{Tail Asymmetry} &       & \multicolumn{3}{c}{$\thetavec$} \\
  \cmidrule{3-4}\cmidrule{6-7}\cmidrule{9-11}    Pair  &       & Spearman & Kendall &       & $\Delta_\text{Major}(0.05)$ & $\Delta_\text{Minor}(0.05)$ &       & $\rho$ & $\deltavec^\top$ & $\nu$ \\
      \multirow{2}[0]{*}{BAC -- VIX} &       & -0.536 & -0.381 &       & -0.019 & 0.240 &       & -0.697 & -0.987, 0.649 &  \\
            &       & (0.024) & (0.019) &       & (0.004) & (0.030) &       & (0.023) & (0.004, 0.030) &  \\
      \multirow{2}[0]{*}{JPM -- VIX} &       & -0.547 & -0.393 &       & -0.008 & 0.211 &       & -0.708 & -0.865, 0.649 & 4.108 \\
            &       & (0.028) & (0.023) &       & (0.002) & (0.019) &       & (0.023) & (0.025, 0.030) & (0.339) \\
      \multirow{2}[1]{*}{JPM -- BAC} &       & 0.805 & 0.620 &       & -0.279 & -0.002 &       & 0.911 & -0.865, -0.987 &  \\
            &       & (0.023) & (0.024) &       & (0.038) & (0.001) &       & (0.017) & (0.025, 0.004) &  \\
      \bottomrule
      \bottomrule
      \end{tabular}%
      }%
    \end{center}
      Note: The variational posterior means, along with standard deviations in parentheses below, of the correlation coefficients (Spearman, and Kendall), 5\% quantile asymmetry metrics defined in the text, and copula parameters are reported. Results are given for two windows of 40 days of 15 minute data that correspond
	  	to low and high market volatility periods.
    \label{tab:vix_pair_corr_theta}%
\end{table}%

\begin{figure}[p]
	\centering
	\includegraphics[width=1\textwidth]{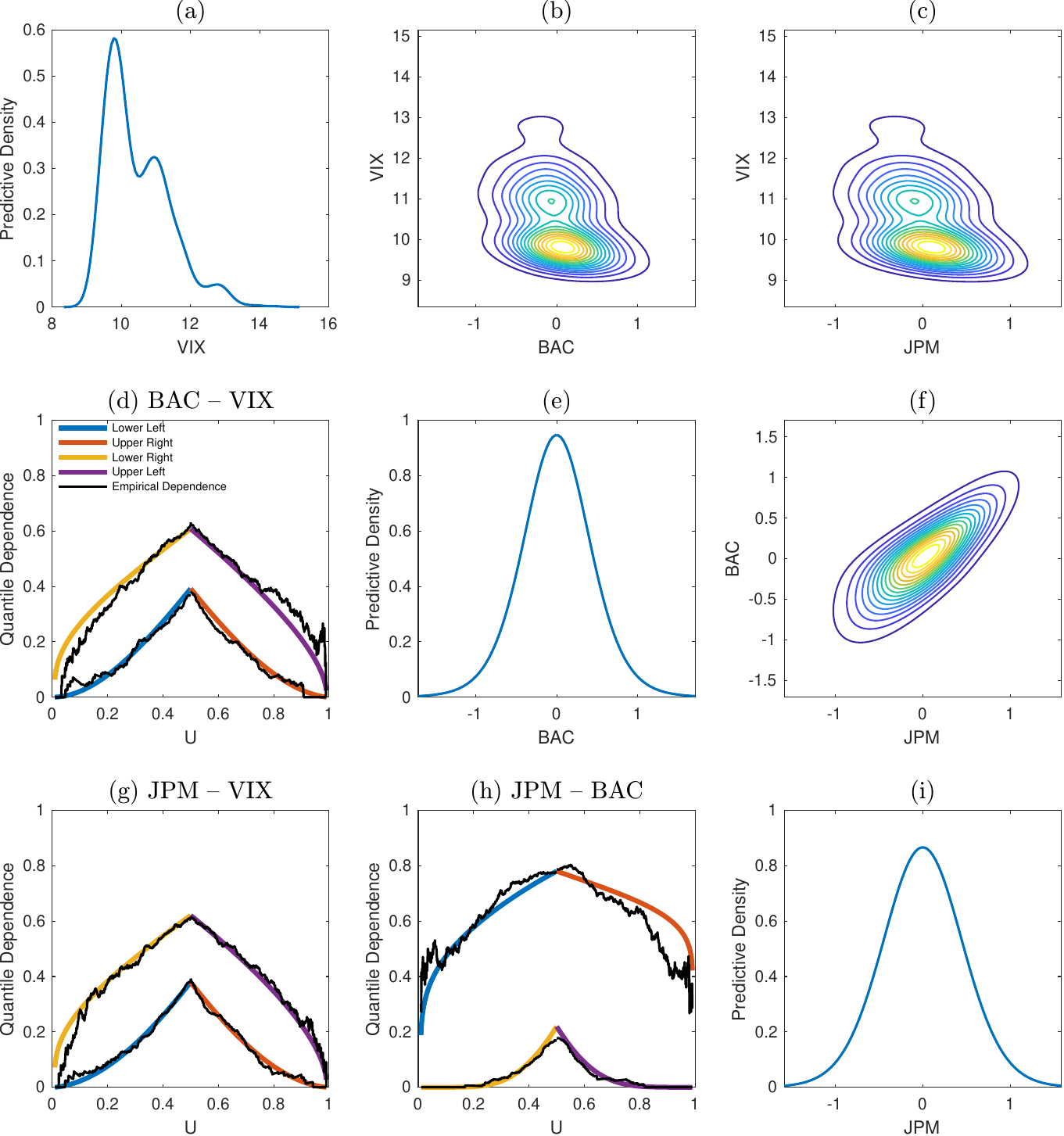}
	\caption{{\bf Low Volatility Period} results from the AC Skew-t copula model for variables (VIX, BAC, JPM). Panels~(a,e,g) give marginal predictive distributions for the series, and panels~(b,c,f) give bivariate predictive distributions for the variable pairs, all on 23 Dec. 2017 at 09:45. Estimated quantile dependence plots for the pairs are given in panels~(d) BAC--VIX, (g) JPM--VIX, and (h) JPM--BAC. Each quantile dependence plot visualizes asymmetry along the major diagonal by plotting $\lambda_{\tiny LL}(u)$ \& $\lambda_{\mbox{\tiny UR}}(1-u)$ against $u$ (blue \& red lines), and along the minor diagonal by plotting $\lambda_{\tiny LR}(u)$ \& $\lambda_{\mbox{\tiny UL}}(1-u)$ versus $u$ (yellow \& purple lines). Empirical quantile dependence plots are given for comparison (black lines).}
	\label{fig:app_lowVIX}
\end{figure}

\begin{figure}[p]
	\centering
	\includegraphics[width=1\textwidth]{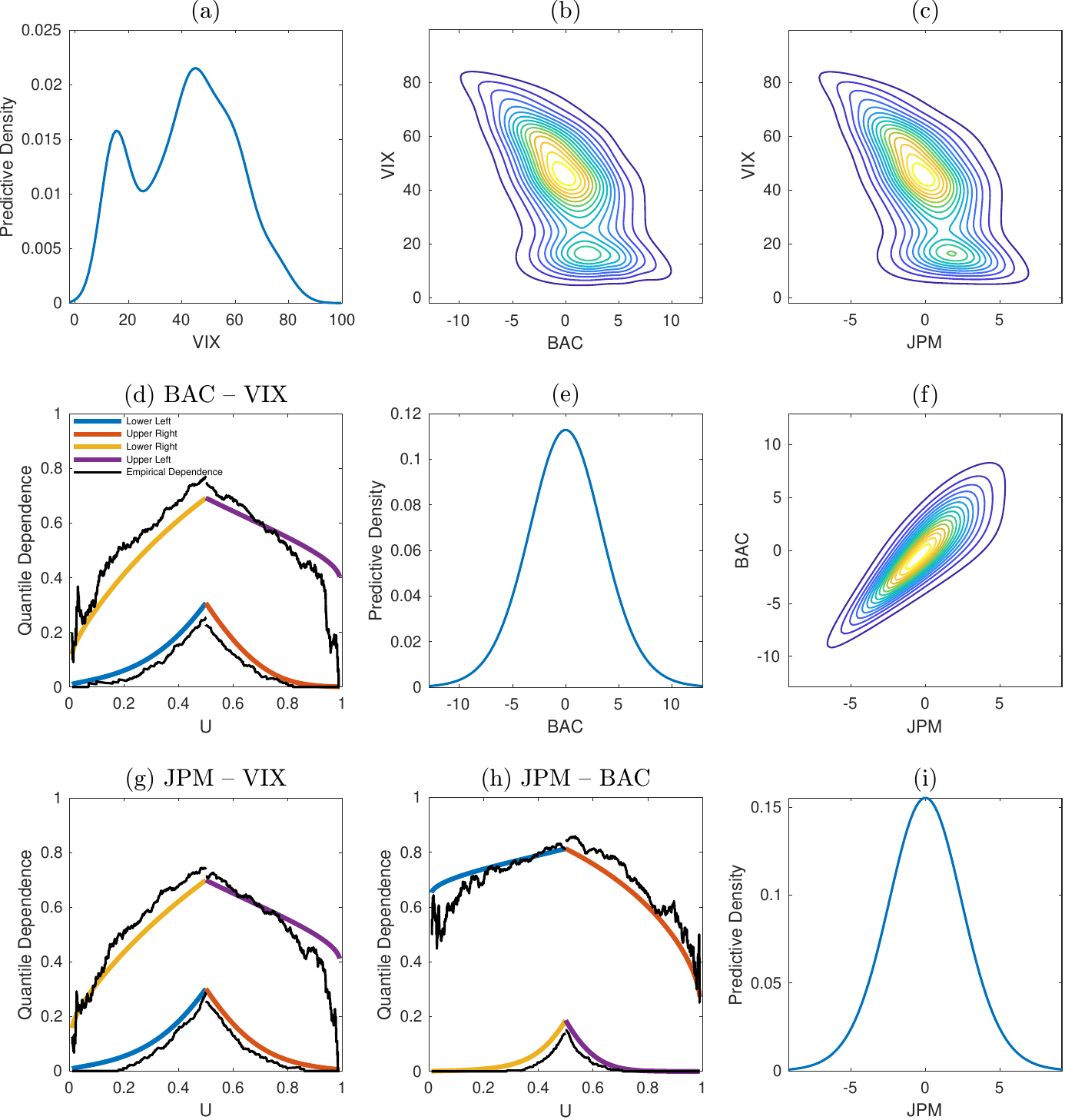}
	\caption{{\bf High Volatility Period} results from the AC Skew-t copula model for variables (JPM, BAC, VIX). Panels~(a,e,g) give marginal predictive distributions for the series, and panels~(b,c,f) give bivariate predictive distributions for the variable pairs, all on 16 Apr. 2020 at 09:45. Estimated quantile dependence plots for the pairs are given in panels~(d) BAC--VIX, (g) JPM--VIX, and (h) JPM--BAC. Each quantile dependence plot visualizes asymmetry along the major diagonal by plotting $\lambda_{\tiny LL}(u)$ \& $\lambda_{\mbox{\tiny UR}}(1-u)$ versus $u$ (blue \& red lines), and along the minor diagonal by plotting $\lambda_{\tiny LR}(u)$ \& $\lambda_{\mbox{\tiny UL}}(1-u)$ versus $u$ (yellow \& purple lines). Empirical quantile dependence plots are given for comparison (black lines).}
	\label{fig:app_hiVIX}
\end{figure}

Figures~\ref{fig:app_lowVIX} and \ref{fig:app_hiVIX} further summarize 
the dependence structure
of the estimated copula for the low and high market volatility periods,
respectively. Following~\cite{ohpatton2013}, panels~(d,g,h)
give ``quantile dependence plots'' for the pairs BAC-VIX, JPM-VIX and JPM-BAC, respectively. 
These visualize asymmetric dependence along the major diagonal by plotting the 
quantile dependencies $\lambda_{\mbox{\tiny LL}}(u)$ \& $\lambda_{\mbox{\tiny UR}}(1-u)$ against $u$ (blue \& red lines), and along the minor diagonal by plotting $\lambda_{\mbox{\tiny LR}}(u)$ \& $\lambda_{\mbox{\tiny UL}}(1-u)$ versus $u$ (yellow \& purple lines).
The dependence between the two equity returns in the major diagonal  is positively asymmetric during the low volatility
period in Fig.~\ref{fig:app_lowVIX}(h), and negatively asymmetric during the high volatility period in Fig.~\ref{fig:app_hiVIX}(h). 
In contrast, dependence in the minor diagonal between the returns on both equities and the VIX is close to
symmetric in the low volatility period in Fig.~\ref{fig:app_lowVIX}(d,g), but
positively asymmetric during the high volatility period in Fig.~\ref{fig:app_hiVIX}(d,g), as also indicated by the metrics in Table~\ref{tab:vix_pair_corr_theta}.

To visualize the flexibility of the predictive distributions from the copula model, we compute these for the first 15-minute trading interval of the following day of each sample, 
which are 23 Dec. 2017 (low volatility period) and 16 Apr. 2020 (high volatility period). Figures~\ref{fig:app_lowVIX} and~\ref{fig:app_hiVIX} plot the marginal
predictive densities in panels~(a,e,i), and the bivariate slices of the 
joint predictive density in
panels~(b,c,f). In the high volatility period both the tails and asymmetric dependence of the 
distributions are greatly accentuated, compared to the low volatility period.

For comparison, we also fit the SDB and GH skew-t copula models with the same marginals. To fit the former we used the MCMC algorithm
outlined by~\cite{smith_gan_kohn_2010}, and for the latter we used the code provided by~\cite{oh_patton_2021}. For SDB and AC we use $k=1$ factors
for $\bar{\Omega}$, and for the GH we use 1 global factor plus 3 group-specific factors. 
Part~D.1 of the Web Appendix reports the dependence metrics and parameter estimates
for both copulas.
We observe 
that all copulas have similar correlation estimates, but both SDB and GH exhibit much lower asymmetry in dependence compared to AC. Moreover, during the second period
 $\hat\nu =4.11$ for AC, whereas $\hat \nu=20.5$ and $21.2$ for SDB and GH, so that 
 the AC skew-t copula also captures higher extremal tail dependence.

\section{S\&P100 Portfolio} \label{sec:risk}
We now apply the AC skew-t copula model to the 15 minute returns on the
constituents of the S\&P100 index
for the period 1 Jan. 2017 to 31 Dec. 2021, which is our main application.
Only the $d=93$ equities that are listed over the entire period are used in our empirical analysis. The returns were constructed as in the previous example,
and the same marginal models used.
Rolling windows of width $T=40$ trading days are employed, and the copula model is estimated in each window using $n=26\times 40=1040$ intraday returns for each equity. 
The model is re-fit every 20 trading days, resulting in 60 overlapping windows.
Our empirical study focuses on three research questions. 
The first 
is whether adopting the AC skew-t copula improves the accuracy of 15 minute ahead density forecasts of portfolio returns, relative to benchmarks. The 
second is what is the degree of asymmetric dependence over the five year period of our data. 
The third question 
is whether, or not, exploiting asymmetry in dependence when forming investment portfolios can improve returns.

\subsection{Density forecasting} \label{sec:density}
We consider 15 minute (i.e. one-step-ahead) density forecasts of the return on a market value
weighted portfolio
of the equities. The weights are re-calculated at the same frequency as the model estimation; i.e. every 20 trading days.
The forecasts are
Bayesian posterior predictive distributions 
for every 15 minute period from 3 Mar. 2017 to 25 Jan. 2022,
giving a total of $1200 \times 26= 31,200$ forecasts. Each of these is evaluated by
drawing 10,000 iterates from the posterior predictive joint distribution of returns (see Part~E.1 of the Web Appendix), from which draws of the portfolio return are obtained. Kernel density estimates of these
are the density forecasts of the portfolio return.

\begin{table}[htbp]
	\centering
	\caption{Mean log-score and CRPS of the equally-weighted S\&P100 portfolio return forecasts}
	\begin{tabular}{ccccccccccc}
		\toprule
		\toprule
		\cmidrule{1-6}\cmidrule{8-11}    \multirow{2}[4]{*}{\# Copula Factors $k$} &       & \multicolumn{4}{c}{Panel A: Mean Log-Score} &       & \multicolumn{4}{c}{Panel B: Mean CRPS} \\
		\cmidrule{3-6}\cmidrule{8-11}          &       & \multicolumn{4}{c}{\# Variational Factors $r$} &       & \multicolumn{4}{c}{\# Variational Factor $r$} \\
		\cmidrule{1-1}\cmidrule{3-6}\cmidrule{8-11}          &       & 0     & 3     & 5     & 10    &       & 0     & 3     & 5     & 10 \\
		1     &       & 6.097 & 6.179 & 6.111 & 6.325 &       & 8.468 & 8.462 & 8.456 & 8.404 \\
		3     &       & 6.856 & 6.744 & 6.853 & 6.890 &       & 8.334 & 8.344 & 8.345 & 8.307 \\
		5     &       & 7.106 & 7.131 & 7.117 & 7.106 &       & 8.277 & 8.276 & 8.254 & 8.285 \\
		10    &       & 7.156 & \textbf{7.185} & 7.132 & 7.167 &       & 8.251 & \textbf{8.246} & 8.252 & 8.288 \\
		15    &       & 7.132 & 7.174 & 7.143 & 7.193 &       & 8.246 & 8.246 & 8.246 & 8.248 \\
		\bottomrule
		\bottomrule
	\end{tabular}
	\caption*{Note: LS values are multiplied by 10, and CRPS values are multiplied by 100, for presentation. The means are computed over all 15 minute density forecasts between
		03/03/2017 and 25/01/2020.}
	\label{tab:crps_ls}%
\end{table}%

Two density forecasting metrics are calculated:
the log-score (LS) and the continuous ranked probability score (CRPS) of \cite{Gneiting_Balabdaoui_Raftery_2007}.
Higher values of the LS and lower values of the CRPS correspond to increased accuracy.
Table~\ref{tab:crps_ls} reports the 
mean of these over all portfolio return density forecasts.
Results are given for different numbers of factors $k$ in the copula model, and estimated using 
VI where the covariance matrix of the Gaussian VA $q^0_\lambda$ has different numbers of factors $r$. Note that while setting
$r=0$ corresponds to a fully factorized VA for $q^0_\lambda$, we stress that the
VA $q_\lambda$ at~\eqref{eq:hybrid_q} for the target density $p(\psivec|\yvec)$ is not of a mean field type. We make two observations. First, using higher values of $k$ (i.e. $k=10$ and $k=15$) increases accuracy. 
This result is consistent with~\cite{oh_patton_2021}, who 
found latent group factors, over-and-above a single global factor, improve
the dependence structure in their skew-t copula model.
Second, the results are similar for $r=3,5,10$, which is consistent with~\cite{Ong_Nott_Smith_2018} and subsequent authors who find lower
values of $r$ typically work well. 
We focus on results for $k=10$ factors (which corresponds to a total of 979 copula parameters in $\thetavec$) with $r=3$ for the VA.
Estimation for a single window takes
29.7 hours using 20,000 steps of the SGA algorithm on a 2018 MacBook Pro.

\begin{figure}[htbp]
	\centering
	\includegraphics[width=\textwidth]{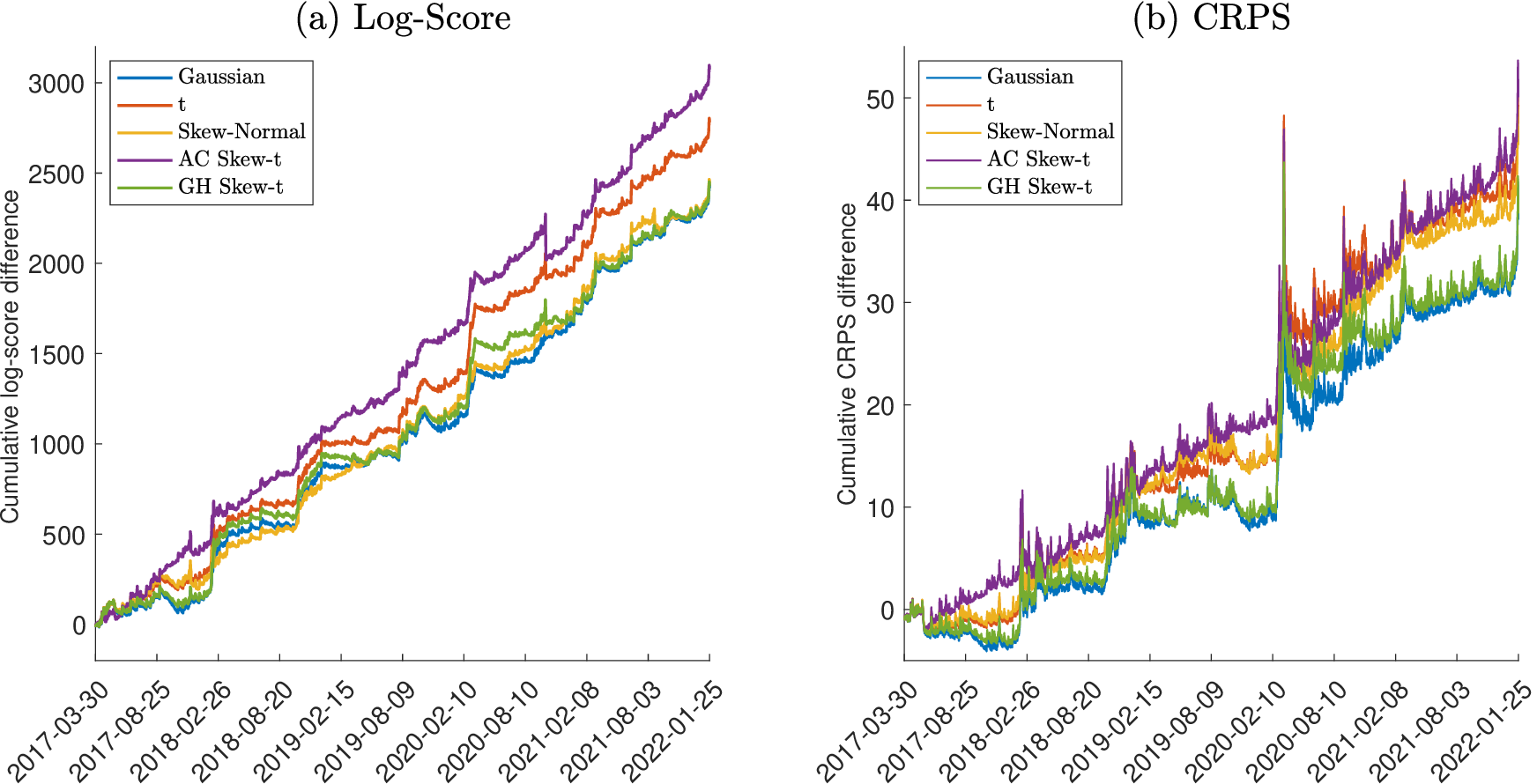}
	\caption{Cumulative difference in CRPS and LS scores between each of (i) Gaussian copula, (ii) t-copula, (iii) AC Skew-Normal copula, (iv) AC Skew-t copula, and (v) static GH Skew-t coupla, and the baseline CCC-GARCH model. All copulas are factor copulas as described in the text.}
	\label{fig:cumsumdiff}%
\end{figure} 
   
The predictive accuracy of the AC skew-t copula model is compared with five benchmarks models. The first four are copula models with the same marginals. Three copulas are sub-types of the AC skew-t copula; namely, the AC skew-normal copula ($\nu\rightarrow \infty$), the t-copula ($\deltavec=\bm{0}$), and the Gaussian copula
($\nu\rightarrow \infty$ and $\deltavec=\bm{0}$). The fourth copula is the GH skew-t copula, implemented
using the code and best fitting factor structure in~\cite{oh_patton_2021}; see Part~E.3 of the Web Appendix for implementation details. The fifth benchmark
model is an intraday Constant Conditional Correlation GARCH (CCC-GARCH) \citep{Bollerslev_1990} specified as in Section~\ref{sec:data}, but with $\varepsilon_{\tau,t}\sim N(0,1)$ and cross-asset correlation estimated using 
the sample correlation.
We treat the CCC-GARCH model as the baseline, and compute the cumulative difference
of each metric for each copula model and this baseline model. Figure~\ref{fig:cumsumdiff}
plots the differences over the validation period for (a)~LS and (b)~CRPS. The positive slope indicates
that the five copula models all out-perform the 
baseline model throughout the period, and the AC skew-t factor copula dominates the other factor copula models on both metrics.

\subsection{Asymmetric dependence}
If $\Delta_{\text{Major},i,j}(u)$ denotes asymmetric quantile dependence for
dimensions $i,j$, then we propose
\begin{equation}\label{eq:totalasym}
	\calT_{i,j} = \int_{\epsilon}^{0.5} \left| \Delta_{\text{Major},i,j}(u)\right| \mbox{d}u\,,
\end{equation}
as a measure of Total Asymmetric Dependence (TAD) in the major diagonal. We set $\epsilon=0.001$ and compute 
the integral over all quantiles numerically.
Asymmetry along the major diagonal is measured, rather than the minor diagonal, because
most equity pairs exhibit positive overall dependence. 
When dependence is symmetric $\calT_{i,j}=0$, while the maximum value of the TAD for any pair of variables in the AC skew-t copula can be computed as $\max_{\rho,\deltavec}\{\calT_{i,j}\} = 0.193$ with $\nu = 2$.  


\begin{figure}[htbp]
	\centering
	\includegraphics[width=\textwidth]{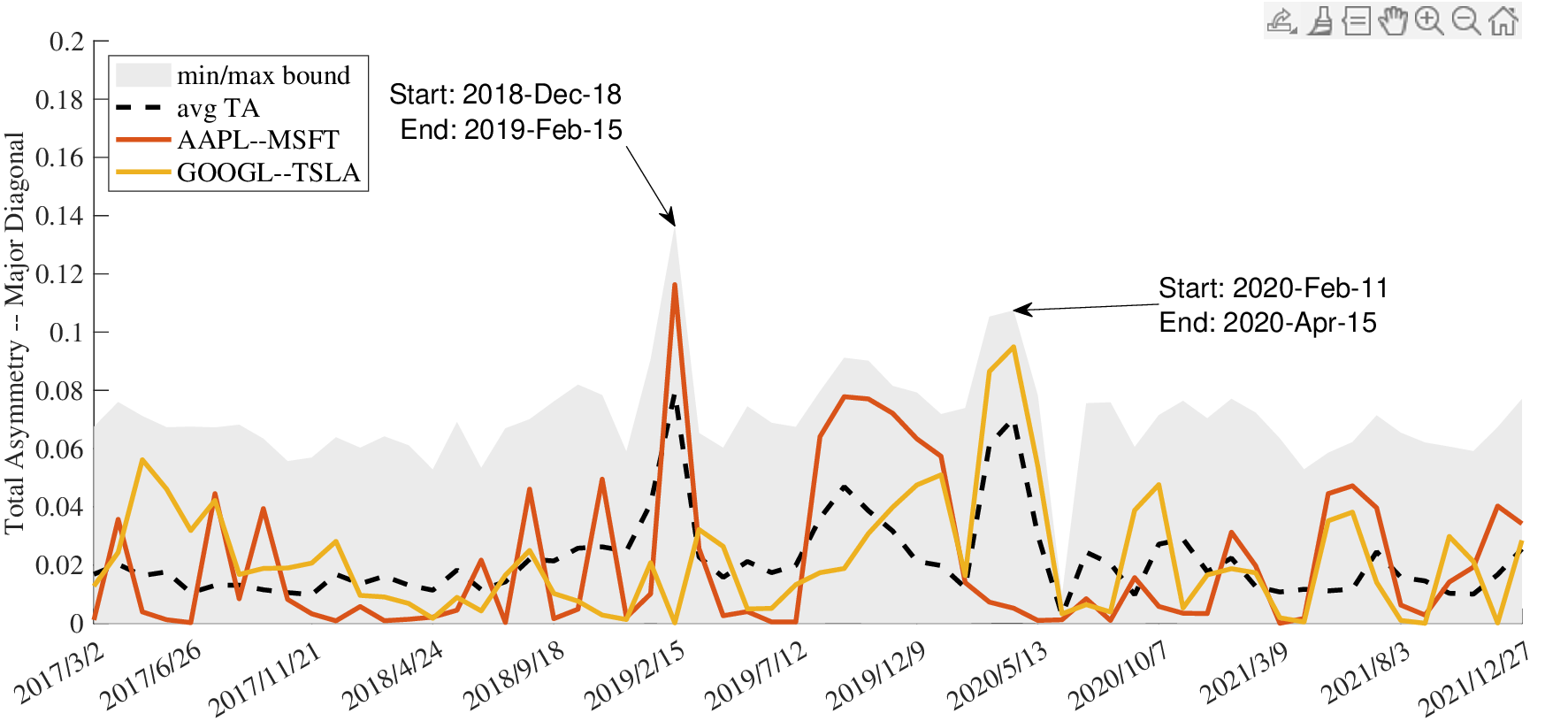} 
	\caption{Summary of the total asymmetric dependence metric values $\calT_{i,j}$ over the rolling windows in the five year period. The gray shaded area is the interval between $\min_{i,j}\{\calT_{i,j}\}$ and $\max_{i,j}\{\calT_{i,j}\}$, while the dashed black line depicts the mean value across all equity pairs.
		The values of $\calT_{i,j}$ for the pairs Apple--Microsoft and Google--Tesla
		are plotted in red and orange, respectively.}
	\label{fig:TA_major_ST_KG10_KB3}
\end{figure}

This metric is computed for all $93\times 92/2=4278$ pairs of equities and for
the AC skew-t copulas fitted at each
of the 60 windows. 
Figure~\ref{fig:TA_major_ST_KG10_KB3} summarizes the evolution of the $\calT_{i,j}$
values over the estimation windows as follows.
The gray shaded area is the interval between $\min_{i,j}\{\calT_{i,j}\}$ and $\max_{i,j}\{\calT_{i,j}\}$, while the dashed black line depicts the mean value across all equity pairs. The windows with the highest level of TAD are 18 Dec. 2018 to 15 Feb. 2019, and 11 Feb. 2020 to 15 Apr. 2020. 
The former window corresponds to an escalating trade war between the U.S. and China, 
while the second window corresponds to the height of the COVID-19 equity market crash. 
The $\calT_{i,j}$ values for the two pairs Apple--Microsoft and Google--Tesla
are also plotted, illustrating how the level of TAD can vary substantially over  
different equity pairs.

\begin{figure}[tbh]
	  \centering
	  \includegraphics[width=1\textwidth, height=0.6\textheight]{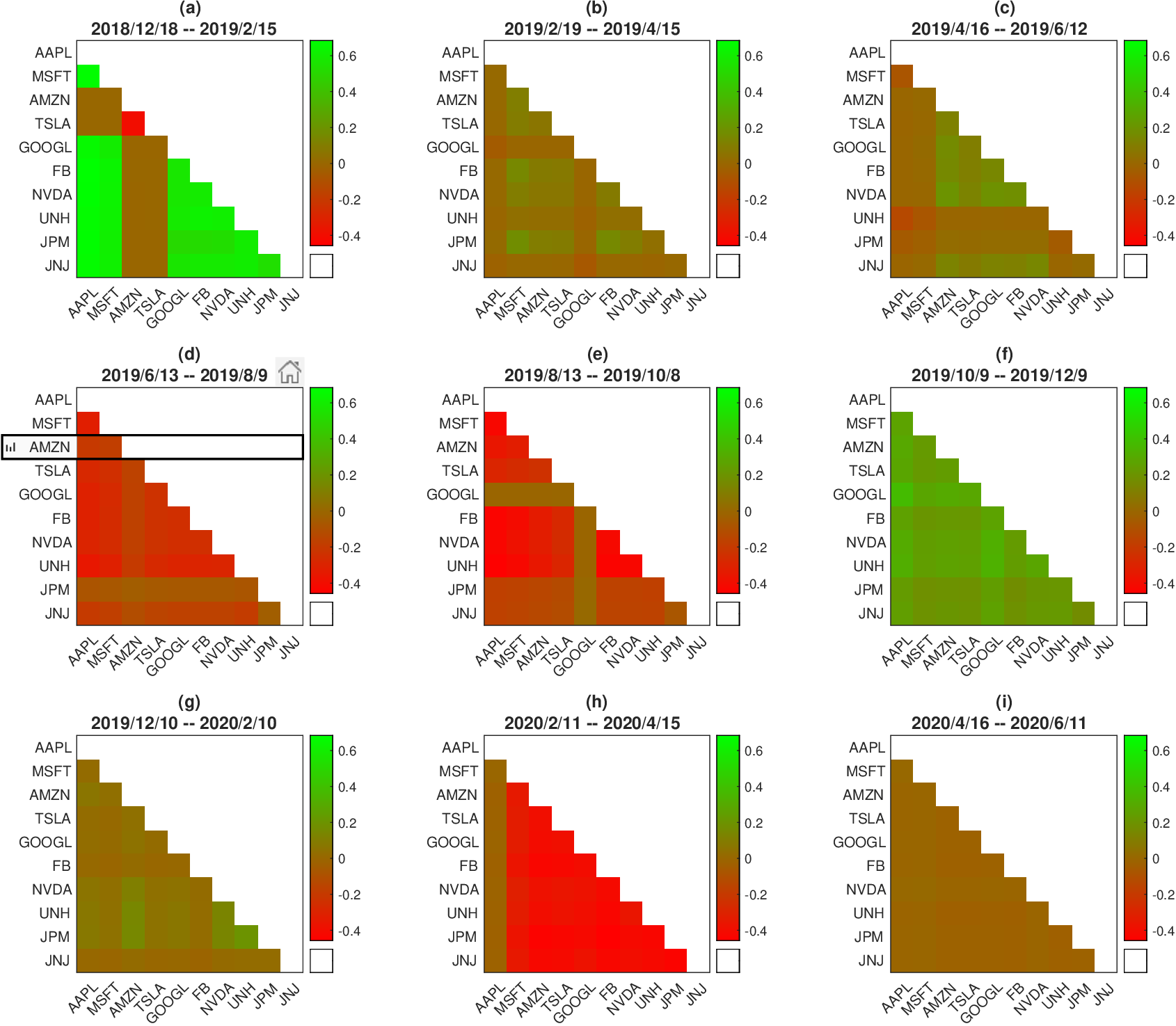} 
	  \caption{Heatmap of $\Delta_{\text{Major},i,j}(0.01)$ across the ten equity pairs $(i,j)$ with largest market capitalization between 18 December 2018 and 19 August 2020.}
	  \label{fig:heatmap_10d}
\end{figure}

Finally, to further highlight the heterogeneity in asymmetric quantile dependence,
Figure~\ref{fig:heatmap_10d} 
depicts $\Delta_{\text{Major},i,j}(0.01)$ over the final ten 
non-overlapping windows for the ten equities with the largest market capitalization. Positive values of $\Delta_{\text{Major},i,j}(0.01)$ (green) indicate
upside tail dependence, while negative values of $\Delta_{\text{Major},i,j}(0.01)$ (red) indicate downside tail dependence.
In particular, during the early stages of the COVID-19 crash in panel~(h),
the copula captures
extreme downside tail dependence across all equity pairs, 
with the exception of Apple. 


\subsection{Portfolio equity selection}
Dependence between equity returns is one of the major attributes of asset selection
strategies for investment portfolios. Traditionally, equities that are negatively correlated are preferred because they
achieve a superior efficient investment frontier.
More recently, consideration has been given to ``tail risk'' of extreme 
downside losses. For example, \cite{guidolin2008} and \cite{harvey2010} proposed portfolio selection based on skewness and/or kurtosis, \cite{giacometti2021} propose a portfolio allocation strategy with penalty terms based on tail risk measures, \cite{zhao2021} proposes a new portfolio optimization method using a bivariate peak-over-threshold approach; and \cite{bollerslev2022realized} made use of semi-beta risk measures, separating market upside and downside movements, for portfolio construction. 
We consider using pairwise asymmetric quantile dependencies from the skew-t copula to 
construct portfolios.
 
The following two strategies are considered to select five equity pairs from the 
4278 possible:
\begin{itemize}
\item {\em Upside Gains}: to maximize upside gains, equities that move together strongly in the upper tail are preferred, so that equity pairs with the largest values of $\Delta_{\text{Major}}(0.05)$ selected. 
\item {\em High Minor Tail Dependence}: to maximize negative tail dependence, equity pairs that maximize the sum of the minor diagonal tail dependencies, $\lambda_{UL}(0.05)+\lambda_{LR}(0.05)$, are selected.
\end{itemize}
The equities that make up the top five pairs according to both criteria 
are used to construct portfolios using market-value weights. 

Table \ref{tab:equitypairs} reports backtesting results from our equity selection strategies implemented
for the period from 3 Mar. 2017 to 25 Jan. 2022.
For comparison, we also include the market-value weighted portfolio with all 93 equities (i.e. the S\&P100 index corrected for sample selection), along with a strategy selecting the five equity pairs with 
the most negative Pearson's correlation, as our benchmarks. 
All strategies are re-balanced with the most current estimates and market values every 20 trading days. The high minor tail dependence strategy performs strongly,
even against the benchmark S\&P100 portfolio. It has the highest Sharpe ratio and 
lower downside risk as measured by the Value-at-Risk (VaR) at the 5\% level.\footnote{The risk-free rate used to calculate the Sharpe Ratio is the Federal Reserve Effective Rate, while higher values of VaR(5\%) indicate a decrease in downside risk at this quantile.}
While this study
does not incorporate  trading costs, it suggests that trading strategies
based on quantile dependencies from our AC skew-t copula model
have the potential to improve the risk profile---particularly downside risk---of
portfolios. Finally, we repeat the study using the GH skew-t copula, which performed poorly.

\begin{table}[H]
	\begin{center}
		\caption{Backtesting results of the four equity selection strategies.}
		\label{tab:equitypairs}
		\resizebox{1.0\textwidth}{!}{
			\begin{tabular}{lccccccccccc}
				\toprule
				\toprule
				& 93-Equities &       & Pearson's &       & \multicolumn{3}{c}{Upside Gain} &       & \multicolumn{3}{c}{High Minor Tail Dependence} \\
				&\multicolumn{1}{l}{(Benchmark}     &       &Correlation       &       & \multicolumn{3}{c}{(max $\Delta_{\text{Major}}(0.05)$)} &       & \multicolumn{3}{c}{(max $\lambda_{UL}(0.05)+\lambda_{LR}(0.05)$)} \\
				&\multicolumn{1}{r}{S\&P100)} &       & (Largest Negative) &       & AC Copula &       & GH Copula &       & AC Copula &       & GH Copula \\
				\cmidrule{2-2}\cmidrule{4-4}\cmidrule{6-6}\cmidrule{8-8}\cmidrule{10-10}\cmidrule{12-12}    Sharpe Ratio & 0.293 &       & 0.159 &       & 0.248 &       & 0.155 &       & 0.381 &       & 0.276 \\
				Return (p.a.) & 15.736\% &       & 12.695\% &       & 17.443\% &       & 10.606\% &       & 26.204\% &       & 18.123\% \\
				Stdev (p.a.) & 15.087\% &       & 22.449\% &       & 19.759\% &       & 19.205\% &       & 19.327\% &       & 18.425\% \\
				VaR(5\%) & -6.054\% &       & -8.064\% &       & -7.633\% &       & -8.247\% &       & -5.804\% &       & -7.303\% \\
				\bottomrule
				\bottomrule
			\end{tabular}
		}
	\end{center}
\caption*{The Sharpe ratio, annualized realized portfolio return and standard deviation, and the 5\% Value-at-Risk are reported.  ``93 Equities'' is the 
	S\&P 100 portfolio corrected for stock selection. The two asymmetric dependence-based stock selection strategies ``Upside Gain'' and ``High Minor Tail Dependence'' are implemented using estimates from both the AC and GH skew-t copulas.}
\end{table}
  
\section{Discussion}\label{sec:conc}
This paper makes three main contributions. First, 
we show that the AC skew-t copula can capture greater asymmetry in dependence
than two other types of skew-t copulas. This is important because capturing asymmetry is the sole
reason to adopt a skew-t copula over the t-copula. Second, we propose a VI method that can estimate the AC skew-t copula parameters for large panels of financial data. Third, our empirical work shows that the direction and extent of asymmetric
dependence can vary over both equity pair and time, and that exploiting this in portfolio formation can
improve investment performance. 
Below, we make some additional comments on the methodology.

While VI methods are increasingly employed in econometrics 
(see~\cite{LMDN2022,chan2022fast} and~\cite{gefang2023} for examples) for more complex models the choice of both target and approximating densities
is crucial. Standard VI methods cannot be used to approximate the posterior $p(\thetavec|\yvec)$
because it has a complex geometry.
However, 
the augmented posterior $p(\psivec|\yvec)$ is well approximated using~\eqref{eq:hybrid_q} and hybrid VI. While the SGA algorithm is widely used to solve variational optimization problems (for example, see~\cite{ranganath2014}) 
it is the efficient re-parameterization gradient at~\eqref{eq:repargrad} that makes our VI algorithm fast.

When estimating
any implicit
copula, it is necessary to compute the  quantile functions $F_{Z_j}^{-1}$, for $j=1,\ldots,d$, at every observation.
For the GH skew-t distribution this is difficult and large Monte Carlo 
samples (e.g. one million draws) are typically used to compute them to a necessary level of accuracy. 
One way to speed this computation is to set $\deltavec=\delta(1,1,\ldots,1)^\top$, however this restricts  variation
in pairwise asymmetric dependence, which is the key focus of our study. This restriction, plus the difficulty
in computing the MLE, is likely to contribute to its poor relative performance
in Section~\ref{sec:risk}. In contrast, 
for the AC skew-t distribution
$F_{Z_j}^{-1}$ can be computed with high accuracy numerically, and there is no
need to restrict $\deltavec$.  
Most previous factor copula studies use only one or two global factors, often
enriched with an industry or latent group-specific factor. 
The computational efficiency of our VI method allows for a larger number of global
factors to be used. This proves important in our empirical work, where 
we find 10 global factors improves performance,
compared to a smaller number. 

Finally, we conclude by mentioning two possible extensions of our work.
In our study 
of intraday returns we use a static copula with a rolling window of width 40 trading days. 
But for studies of lower frequency returns, allowing for time variation
in the AC skew-t copula can be achieved by following previous authors and 
adopting a dynamic specification. 
A second promising extension would be to consider generalizations of the AC skew-t distribution
underlying the implicit copula to provide greater flexibility in capturing
tail dependence. However, in both extensions identification of the copula parameters  and an effective estimation methodology would need to be considered.
\noindent
\appendix
\section{Quantile Dependence}\label{app:qdep}
For each skew-t copula, the pairwise quantile dependence metrics in Section~\ref{sec:properties}
are evaluated from their bivariate copula function $C(u_1,u_2)=F_{Z_1,Z_2} \left( F_{Z_1}^{-1}(u_1), F_{Z_2}^{-1}(u_2) \right)$. 
Computation of $F_{Z_j}^{-1}(u_j)$ is discussed in the manuscript for each distribution. The
function $F_{Z_1,Z_2}$ is obtained for the SDB skew-t distribution by 
numerical integration.
For the AC skew-t distribution it is computed as the trivariate 
student t distribution function
$
F_{Z_{\mbox{\tiny AC}}}\left( (z_1,z_2)^\top; \deltavec,\bar\Omega, \nu \right) = 2F_t\left((z_1,z_2,0)^\top; \Omega^*, \nu\right)\,,
$
where 
\begin{equation*}
	\Omega^* = \left(\begin{array}{ll} \bar\Omega & -\deltavec  \\ -\deltavec ^\top & 1 \end{array}\right) \,,
\end{equation*} 
which can be derived from~\eqref{eq:ac_jointnorm}. The GH skew-t distribution  function cannot be computed in closed form, and numerical integration of its joint density is difficult. Therefore, a kernel-based approximation based on one million Monte Carlo draws is employed. 
%

\section{Generative Representations for AC Skew-t}\label{app:gr}
We list below two different generative representations for $\bm{Z}_{\mbox{\tiny AC}}$
that can be derived from~\eqref{eq:ac_jointnorm}. 
\begin{itemize}
	\item[GR1:] The first is to generate from the marginal $L\sim t_1(0,1,\nu)$ constrained so that $L>0$, and then from the conditional $\bm{Z}_{\mbox{\tiny AC}}=(\bm{X}|L)\sim t_d\left(\bm{0},\frac{\nu+L^2}{\nu+1}(\bar \Omega -\deltavec \deltavec^\top),\nu+1\right)$. 
	\item[GR2:] The second uses a scale mixture of normals representation 
	for a t-distribution, where if
	$W\sim \mbox{Gamma}(\nu/2,\nu/2)$, then $(\bm{X}^\top,L)^\top=W^{-1/2}(\tilde{ \bm{X}}^\top,\tilde L)^\top$ with $(\tilde{\bm{X}}^\top,\tilde L)^\top \sim
	N_{d+1}(\bm{0},\Omega)$. Generating sequentially from $W$, $\tilde L$ and then from $\bm{Z}_{\mbox{\tiny AC}}=(\bm{X}|\tilde L,W)\sim N_d\left(W^{-1/2}\deltavec \tilde{L},W^{-1}(\bar \Omega -\deltavec \deltavec^\top)\right)$. as in Section~\ref{sec:ac_cop}, gives a generative representation.
%
\end{itemize}
 The extended likelihood in Section~\ref{sec:like} is based on generative
 representation GR2. 

\section{Gradient}\label{app:grads}
\begin{table}[H]
	\centering
	\caption{Computationally efficient closed form expression for the gradient $\nabla_\theta \log p(\thetavec,\ltildevec, \wvec|\yvec)$  }
 	\small
	  \begin{tabular}{ll}
	  \toprule
	  $
	  \begin{aligned}[t]
		&\text{Computing } \nabla_G \log p(\thetavec,\ltildevec, \wvec|\yvec)
		\\\hline
		\\&\nabla_G \log p(\thetavec,\ltildevec, \wvec|\yvec) 
		\\ &= \nabla_G \Big\{- \frac{n}{2} T_{G1} - \frac{1}{2} \sum_{i=1}^n(T_{G2i} + T_{G3i} + T_{G4i}) \Big\} \\ &+\nabla_{G} \log p(G)
		  \\& \nabla_G T_{G1} = -2 (V_2 \odot VV_1C_4  + V_1 C_4 V_1 ) G 
		  \\& \nabla_G T_{G2i} = -2 ( V_2 \odot VV_1 C_{7i} + V_1 C_{7i} V_1 ) G
		  \\& \nabla_G T_{G3i} = -2 ( V_2 \odot VV_1 C_{11i} + V_1 C_{11i} V_1 ) G
		  \\& \nabla_G T_{G4i} = -2 ( V_2 \odot VV_1 C_{15i} + V_1 C_{15i} V_1 ) G
		  \\& \nabla_{G} \log p(G) = -(a_1 + 1) \operatorname{sign}(G) \frac{1}{b_1 + |G|}
		  \\& C_0  =\bar\Omega^{-1} + \frac{\bar\Omega^{-1}\deltavec \deltavec^\top  \bar\Omega^{-1}}{1 - \deltavec^\top \bar\Omega^{-1}\deltavec}
		  \\& C_1 = 1+\alphavec^\top  \bar{\Omega} \alphavec
		  \\& C_2 = \alphavec\alphavec^\top  \bar\Omega
		  \\& C_3 = C_1^{-1} C_2  C_0 +  C_1^{-1} C_0  C_2^\top  - C_1^{-2} C_2C_0 C_2^\top 
		  \\& C_4 = C_0 - C_3
		  \\& C_{5i} = -w_i C_0 \yvec_i \yvec_i^\top  C_0
		  \\& C_{6i} = C_1^{-1} C_2 C_{5i} + C_1^{-1} C_{5i} C_2^\top  - C_1^{-2} C_2 C_{5i} C_2^\top 
		  \\& C_{7i} = C_{5i} - C_{6i}
		  \\& C_{8i} = - 2\tilde{l}_i w_i^{1/2} C_1 ^{-1/2} C_0 \yvec_i \alphavec^\top  + \tilde{l}_i w_i^{1/2} C_1 ^{-3/2} C_2 C_0 \yvec_i \alphavec^\top
		  \\& C_{9i} = 2\tilde{l}_i w_i^{1/2} C_0 \yvec_i \deltavec^\top  C_0
		  \\& C_{10i} = C_1^{-1} C_2 C_{9i} + C_1^{-1} C_{9i} C_2^\top  - C_1^{-2} C_2 C_{9i} C_2^\top
		  \\& C_{11i} = C_{8i} + C_{9i} - C_{10i}
		  \\& C_{12i} = 2\tilde{l}_i^2 C_1^{(-1/2)} \alphavec \deltavec^\top C_0 - \tilde{l}_i^2 C_1^{-3/2} \alphavec \deltavec^\top  C_0 C_2^\top
		  \\& C_{13i} = - \tilde{l}_i^2 C_0 \deltavec \deltavec^\top  C_0
		  \\& C_{14i} = C_1^{-1} C_2 C_{13i} + C_1^{-1} C_{13i}^\top C_2^\top  - C_1^{-2} C_2 C_{13i} C_2^\top
		  \\& C_{15i} = C_{12i} + C_{13i} - C_{14i}
		  \\& V_1 = \diag(V)^{-1/2} \quad \& \quad V_2 = \diag(V)^{-3/2}
		  \\
		  \\\hline
		  &\text{Computing } \nabla_{\widetilde{G}} \log p(\thetavec,\ltildevec, \wvec|\yvec)
		  \\\hline
		  \\&\nabla_{\widetilde{G}} \log p(\thetavec,\ltildevec, \wvec|\yvec) = \nabla_{G} \log p(\thetavec,\ltildevec, \wvec|\yvec) \times \exp(\widetilde{G}) + 1
	  \end{aligned}
	  $ &  		
	  $
	  \begin{aligned}[t]
		&\text{Computing } \nabla_\alphavec \log p(\thetavec,\ltildevec, \wvec|\yvec)
		\\\hline
		\\&\nabla_\alphavec \log p(\thetavec,\ltildevec, \wvec|\yvec) \\ &= \nabla_\alphavec \Big\{- \frac{n}{2} T_{\alphavec1} - \frac{1}{2} \sum_{i=1}^n(T_{\alphavec2i} + T_{\alphavec3i} + T_{\alphavec4i}) \Big\} \\&+\nabla_{\alphavec} \log p(\alphavec)
 		\\& \nabla_\alphavec T_{\alphavec 1} = 2C_1^{-3/2} \bar\Omega \alphavec \alphavec^\top \bar\Omega C_0 \deltavec - 2 C_1^{-1/2} \bar\Omega C_0 \deltavec
 		\\& \nabla_\alphavec T_{\alphavec 2i} = C_1^{-2} \bar\Omega\alphavec\alphavec^\top \bar\Omega (C_{5i} +C_{5i}^\top ) \bar\Omega\alphavec \\& - C_1^{-1} \bar\Omega (C_{5i} + C_{5i}^\top ) \bar\Omega \alphavec
 		\\& \nabla_\alphavec T_{\alpha3i} = C_{16i} + C_1^{-1} \bar\Omega (C_{17i} + C_{17i}^\top ) \bar\Omega \alphavec \\& - C_1^{-2} \bar\Omega\alphavec\alphavec^\top \bar\Omega (C_{17i} + C_{17i}^\top )  \bar\Omega\alphavec
 		\\& \nabla_\alphavec T_{\alphavec 4i} = 2C_{18i} + C_1^{-1} \bar\Omega (C_{19i} + C_{19i}^\top ) \bar\Omega \alphavec \\& - C_1^{-2} \bar\Omega\alphavec\alphavec^\top \bar\Omega (C_{19i} + C_{19i}^\top ) \bar\Omega \alphavec
 		\\& \nabla_{\alpha} \log p(\alpha) = -\frac{1}{\sigma^2} \sum_i \alpha_i 
 		\\& C_{16i} = - 2\tilde{l}_i w_i^{1/2} C_1^{-1/2}\bar\Omega C_0 \yvec_i \\& + 2\tilde{l}_i w_i^{1/2} C_1^{-3/2} \alphavec^\top \bar\Omega C_0 \yvec_i  \bar\Omega \alphavec
 		\\& C_{17i} = - 2\tilde{l}_i w_i^{1/2} C_0 \yvec_i \deltavec^\top  C_0
 		\\& C_{18i} = \tilde{l}_i^2 C_1^{-1/2} \bar\Omega C_0 \deltavec - \tilde{l}_i^2 C_1^{-3/2} \bar\Omega \alphavec \alphavec^\top  \bar\Omega C_0 \deltavec
 		\\& C_{19i} = \tilde{l}_i^2 C_0 \deltavec \deltavec^\top  C_0 
		\\
		\\
		\\
		\\
		\\\hline
		&\text{Computing } \nabla_{\tilde{\nu}} \log p(\thetavec,\ltildevec, \wvec|\yvec)
		\\\hline
		\\&\nabla_{\tilde{\nu}} \log p(\thetavec,\ltildevec, \wvec|\yvec) 
		\\& = \nabla_{\nu} \Big\{\calL( \yvec, \ltildevec, \wvec | \thetavec)  + \log(p(\nu)) \Big\} \times \exp(\tilde{\nu}) + 1     
		\\& \nabla_{\nu} \calL( \yvec, \ltildevec, \wvec | \thetavec) = \frac{1}{2} \sum^n_{i=1}\log(w_i) 
		\\& + n\Big(\frac{1}{2}\log(\frac{\nu}{2}) + \frac{1}{2} - \frac{1}{2}\psi_0(\frac{\nu}{2})  \Big) - \frac{1}{2}\sum_{i=1}^n w_i
		\\& \nabla_{{\nu}} \log(p({\nu}))
		= (a_2 - 1)\frac{1}{\nu} - b_2
 		\end{aligned}
 		$ 
		\\ 
	  \bottomrule
	  \end{tabular}%
	  \caption*{Note: `$\odot$' denotes the Hadamard (i.e. element-wise) product, and the four gradients are computed by evaluating the terms sequentially from the bottom upwards. Their derivation is found in Part B of the Web Appendix. MATLAB routines to evaluate the gradients are provided. }
	  \label{tab:grad}%
\end{table}%

\section{MCMC Scheme}\label{app:mcmcscheme}
Algorithm~\ref{alg:mcmc} is an MCMC sampling scheme to evaluate
the augmented posterior of the AC skew-t factor copula parameters $\thetavec$.
Steps~1 and~2 are also used in the hybrid VI Algorithm~\ref{alg:hvi}.

\begin{algorithm}
	\caption{MCMC Scheme for AC Skew-\textit{t} Copula} \label{alg:mcmc}
	\begin{algorithmic}
	\State Step 0. Initialize feasible values for $ \thetavec$,  $\ltildevec$ and $\wvec$
	\State Step 1. Generate from $p(\ltildevec|\thetavec, \wvec, \yvec)=\prod_{i=1}^n p(\tilde{l}_i| \thetavec, w_i, \yvec_i)$ 
    \State Step 2. Generate from $p(\wvec|\thetavec, \ltildevec, \yvec)=\prod_{i=1}^n p(w_i| \thetavec, \tilde{l}_i, \yvec_i)$ 
	\State Step 3. Generate from $p(\alpha_j|\{\thetavec\backslash \alpha_j\},\ltildevec,\wvec,\yvec)$ for $j=1,\ldots,d$.
	\State Step 4. Generate from $p(\tilde{g}_{ij}|\{\thetavec\backslash \tilde{g}_{ij}\},\ltildevec,\wvec,\yvec)$ for non-zero
	elements $\tilde{g}_{ij}$ of $\widetilde{G}$.
	\State Step 5. Generate from $p(\nu|\{\thetavec\backslash \nu\},\ltildevec,\wvec,\yvec)$ 
	\end{algorithmic}
\end{algorithm}
\noindent The conditional posteriors at Steps~3, 4 and~5 can be derived from~\eqref{eq:augpost}, but are unrecognizable, so we employ adaptive random walk Metropolis-Hastings schemes to generate each element.

To derive the posteriors at Steps~1 and 2, note that from~\eqref{eq:elike1} and~\eqref{eq:elike2}, 
\begin{eqnarray*}
p(\ltildevec,\wvec|\thetavec,\yvec) &\propto &p(\ltildevec,\wvec,\yvec|\thetavec) 
\propto \prod_{i=1}^n p(\zvec_i|\tilde{l}_i,w_i,\thetavec)p(\tilde{l}_i|w_i)p(w_i|\nu)\\
&= &\prod_{i=1}^n \phi_d(\zvec_i;\muvec_{z,i},\Sigma_{z,i})\phi_1(\tilde{l}_i;0,1)\mathds{1}(\tilde{l}_i>0) p(w_i|\nu)\,,
\end{eqnarray*}
with $\muvec_{z,i}=\deltavec \tilde{l}_i w_i^{-1/2}$, $\Sigma_{z,i}=w_i^{-1}(\bar\Omega - \deltavec\deltavec^\top)$ and $p(w_i|\nu)$ is a Gamma$(\nu/2,\nu/2)$ density.

 From the above, at Step~1 the conditional posterior $p(\ltildevec|\wvec,\thetavec,\yvec)= 
\prod_{i=1}^n p(\tilde{l}_i|w_i,\thetavec,\yvec)$, where
$\tilde{l}_i|\thetavec, w_i, \yvec_i \sim N_{+}(A^{-1}B_i, A^{-1}) $, with $A = 1 + \deltavec^\top (\bar\Omega - \deltavec\deltavec^\top)^{-1} \deltavec, B_i = w_i^{1/2} \deltavec^\top(\bar\Omega - \deltavec\deltavec^\top)^{-1}\zvec_i$ and $N_+$
denotes a univariate normal distribution constrained to positive values.
Similarly, at Step~2 the conditional posterior 
$p(\wvec|\ltildevec,\thetavec,\yvec)=\prod_{i=1}^n p(w_i|\tilde{l}_i,\thetavec,\yvec)$, with
\begin{equation*}
    p(w_i|\thetavec, \tilde{l}_i, \yvec) \propto w_i^{\frac{d+\nu}{2}-1} \exp \left\{ -\frac{1}{2} \left( w_i \zvec_i^\top (\bar\Omega - \deltavec\deltavec^\top)^{-1} \zvec_i - 2 \tilde{l}_i w_i^{1/2} \deltavec^\top (\bar\Omega - \deltavec\deltavec^\top)^{-1} \zvec_i + w_i\nu \right) \right\}\,.
\end{equation*}
An adaptive random walk Metropolis-Hastings step is used to draw from this posterior.

In Step~(b) of Algorithm~\ref{alg:hvi}, a draw of $\ltildevec,\wvec$  is obtained by repeatedly drawing from the conditionals 25 times. 
We stress this is fast because all demanding computations do not involve $\lvec$ and $\wvec$,
so that they only
need to be computed once per SGA step. 
We found 25 draws to be adequate, which is consistent with the findings in~\cite{Loaiza-Maya_Smith_Nott_Danaher_2021} for other models.

\newpage

\singlespacing
\bibliography{references}


\newpage
\onehalfspacing
\newpage
\noindent
\setcounter{page}{1}
\begin{center}
	{\bf \Large{Online Appendix for ``Large Skew-t Copula Models and Asymmetric Dependence in Intraday Equity Returns''}}
\end{center}

\vspace{10pt}

\setcounter{figure}{0}
\setcounter{table}{0}
\setcounter{section}{0}
\setcounter{equation}{0}
\setcounter{algorithm}{1}
\renewcommand{\thetable}{A\arabic{table}}
\renewcommand{\thefigure}{A\arabic{figure}}
\renewcommand{\thealgorithm}{\Alph{section}\arabic{algorithm}}
\renewcommand{\thesection}{Part~\Alph{section}}
\renewcommand{\theequation}{A\arabic{equation}}
\noindent
This Online Appendix has five parts:

 
\begin{itemize}
    \item[] {\bf Part~A}: Supporting information for Section 2.
    \begin{itemize}
    	\item[] A.1: Density for the GH skew-t distribution
    	\item[] A.2: Plot of the AC, SDB and GH copula densities in Section~2.
    \end{itemize}
	\item[] {\bf Part~B}: Gradient for the AC skew-t copula with representation GR2.
	\begin{itemize}
		\item[] B.1: Priors
		\item[] B.2: Logarithm of the augmented posterior
		\item[] B.3--B.5: Derivation of the gradients in Table~1 of the paper
	\end{itemize}
	\item[] {\bf Part~C}: Additional details for the simulation study in Section 4.
	\begin{itemize}
		\item[] C.1: Data generating processes
		\item[] C.2: Additional empirical results
		\item[] C.3: Comparison of different generative representations
		\item[] C.4: Estimation accuracy
\end{itemize}
	\item[] {\bf Part~D}: Additional details for Section 5.
	\begin{itemize}
		\item[] D.1: Estimates of the SDB and GH skew-t copulas
		\item[] D.2: Geometry of the posterior and its impact on VI
	\end{itemize}
	\item[] {\bf Part~E}: Additional details for Section 6.
\begin{itemize}
	\item[] E.1: Generating from the predictive distribution of portfolio returns
	\item[] E.2: Quantile dependence of AAPL-MSFT and GOOGL-TSLA
	\item[] E.3: Comparison with the GH skew-t copula
\end{itemize}
\end{itemize}

\newpage
\section{Supporting Information for Section 2}
\subsection{Density for the GH skew-t distribution}
The joint density of $\bm{Z}_{\mbox{\tiny GH}}$ is
\begin{equation*}
	\begin{aligned}
		f_{Z_{\mbox{\tiny GH}}}(\zvec;\bar{\Omega},\deltavec,\nu)&={\cal A}\frac{ K_{(\nu+d) / 2}\left(\sqrt{\left(\nu+\zvec^{\prime} \bar{\Omega}^{-1}\zvec\right) \deltavec^{\prime} \bar{\Omega}^{-1} \deltavec}\right) \exp \left(\zvec^{\prime} \bar{\Omega}^{-1} \deltavec\right)}{\left.\left(\nu+\zvec^{\prime} \bar{\Omega}^{-1}\zvec\right) \deltavec^{\prime} \bar{\Omega}^{-1} \deltavec\right)^{-\frac{\nu+d}{4}}\left(1+\frac{\zvec^{\prime} \bar{\Omega}^{-1}\zvec}{\nu}\right)^{\frac{\nu+d}{2}}}\,,\mbox{ with}
		\\ 
		{\cal A}&=\frac{2^{\frac{2-\nu-d}{2}}}{\Gamma(\nu / 2)(\pi \nu)^{d / 2} |\bar{\Omega}|^\frac{1}{2}}\,.
	\end{aligned}
\end{equation*}
The function $K_\beta(s)$ is the modified Bessel function of the second kind with index $\beta$.

\subsection{Plot of the AC, SDB, GH copula densities in Section 2}
\begin{figure}[htbp]
	\centering
	\includegraphics[width=0.9\textwidth]{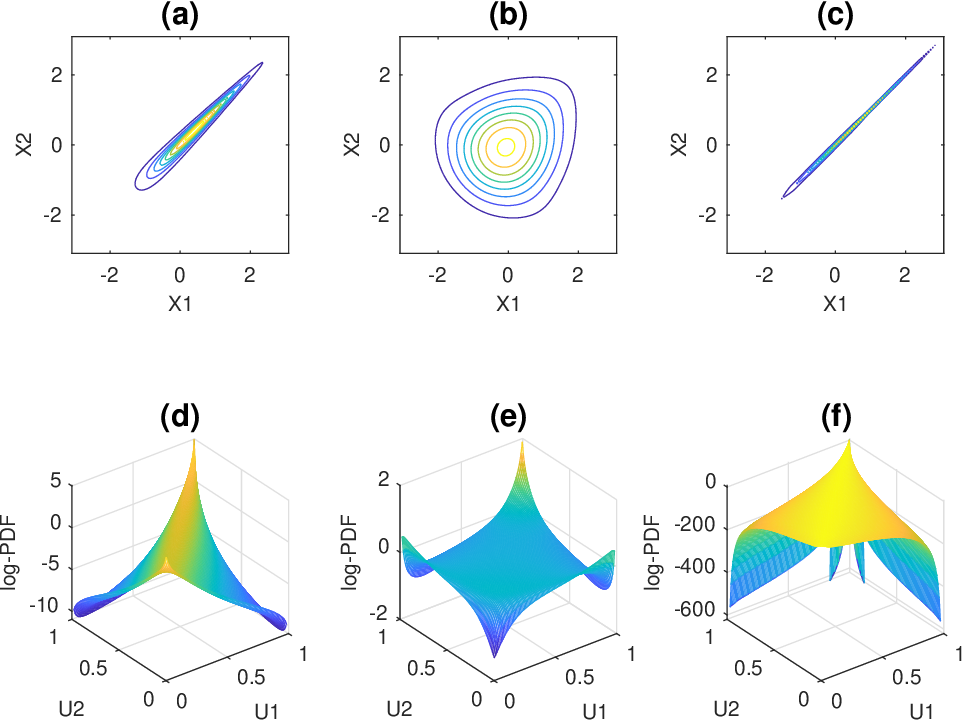}
	\caption{The first row plots density contours of three bivariate skew-{\em t} copula models with $N(0,1)$ marginals and maximal asymmetric dependence $\Delta(0.01)$ for (a)~AC, (b)~SDB, and (c)~GH.
		The second row plots the logarithm of the corresponding copula densities.}
	\label{fig:STcopulaPDF_wth_Margins}
\end{figure}

Figure~\ref{fig:STcopulaPDF_wth_Margins} plots the bivariate copula model densities with $N(0,1)$ marginals and copula parameters which give
the maximal asymmetric dependence at the 1\% quantile (i.e. maximum
value of $\Delta(0,1)$.)
\newpage

\section{Gradient for the AC skew-t copula with representation GR2}
\noindent 
In this part of the Web Appendix we give further details on the VI methodology for 
the AC skew-t copula when using the generative representation GR2 with $\psivec\equiv\{\thetavec,\tilde \lvec,\wvec\}$. We list the priors employed and then the log augmented posterior for this case. 
We derive the gradient $\nabla_\lambda \log h(\psivec)$ for this case, the result of which is given 
in Table~2 of the manuscript. This derivation uses the trace operator to express 
gradients as directional derivatives, which provides a computationally efficient expression.

\subsection{Priors}
The parameters $\thetavec = \{ \mbox{vech}(\widetilde{G}), \alphavec, \nu \}$ are 
transformed to unconstrained real values because the VI algorithm is applicable in this case.
Thus we defined $\mbox{vech}(\widetilde{G}) = \{G_{p,k}, \widetilde{G}_{k,k}\}, p<k$, to represent off-diagonal and diagonal values of $\mbox{vech}(G)$ respectively, $\widetilde{G}_{k,k} = \log(\Diag(G))$ denotes the logarithms of leading diagonal values in $P \times K$ matrix $G$ with $K \ll P$. And $\tilde{\nu} = \log(\nu - 2)$ to promise VA inferenced $\nu > 2$. 

The prior distribution is defined as $p(\thetavec) = p(G_{P,K}) p(\widetilde{G}_{K,K}) p(\alphavec) p(\tilde{\nu})$ with 

\begin{equation*}
    \begin{split}
        (1) \ &p(G_{p,k}) = f_{\mbox{\tiny GDP}}(G_{p,k}; a_1, b_1), p<k \quad
        (2) \ p(\widetilde{G}_{k,k}) = f_{\mbox{\tiny GDP}}(\exp(\widetilde{G}_{k, k}); a_1, b_1) \exp(\widetilde{G}_{k, k}) \\
        (3) \ &p(\alphavec) = \prod_{p=1}^{P}f_N(\alpha_p; 0, \sigma^2) \quad
        (4) \ p(\tilde{\nu}) = f_{\mbox{\tiny Gamma}}(\exp(\tilde{\nu}); a_2, b_2) \exp(\tilde{\nu})
    \end{split}
\end{equation*}

\subsection{Logarithm of the Augmented Posterior}
Below is the logarithm of the augmented posterior density for the generative representation GR2. 
\begin{equation*}
    \begin{aligned}
        \log p(\thetavec, \ltildevec,\wvec | \yvec) & = (\frac{d+v}{2}-1)\sum^n_{i=1}\log(w_i) - \frac{n}{2}\log\bigg(\det\Big(\bar\Omega-\deltavec\deltavec'\Big)\bigg) 
        \\& \quad + n\bigg(\frac{\nu}{2}\log(\frac{\nu}{2}) - \log(\Gamma(\frac{\nu}{2}))\bigg)
        \\& \quad  - \frac{1}{2}\sum^n_{i=1} \bigg\{w_i \yvec_i' \Big(\bar\Omega-\deltavec \deltavec'\Big)^{-1}\yvec_i - 2\tilde{l}_i w_i^{1/2} \deltavec'\Big(\bar\Omega-\deltavec \deltavec'\Big)^{-1} \yvec_i 
        \\& \quad + \tilde{l}_i^2 \deltavec'\Big(\bar\Omega-\deltavec \deltavec'\Big)^{-1} \deltavec + w_i\nu + \tilde{l}_i^2  \bigg\} + \log p(\thetavec)
        \\& = \calL( \yvec, \ltildevec, \wvec | \thetavec) + \log p(\thetavec)
    \end{aligned}
\end{equation*}
, where $\bar\Omega = V_1^{-1/2}VV_1^{-1/2}, V = GG^\top + D, V_1 = \diag(V)^{-1/2}$ and $\deltavec = (1+\alphavec^{\top} \bar{\Omega} \alphavec)^{-1 / 2} \bar{\Omega} \alphavec $

\subsection{Computing $\nabla_{G_{p,k}} \log p(\thetavec, \ltildevec,\wvec | \yvec)$, $\nabla_{G_{k,k}} \log p(\thetavec, \ltildevec,\wvec | \yvec)$}

First we note that in expressing these gradients we adopt the following notation. If $f(\xvec)$ is a 
scalar-valued function of a column vector $\xvec$, then
$\nabla_x g(\xvec)=\frac{\partial g}{\partial \xvec}^\top$ which is a column vector.
If the vector-valued function $g(\xvec)\in \mathbb{R}^d \times 1$ and $\xvec \in \mathbb{R}^n \times 1$, then $\frac{\partial g}{\partial \xvec}$ is a $d\times n$ matrix with $(i,j)$ element $\frac{\partial g_i}{\partial x_j}$.

\begin{alignat*}{2}
    & \nabla_{G_{p,k}} \log p(\thetavec, \ltildevec,\wvec | \yvec) =\nabla_{G_{p,k}} \Big\{\calL( \yvec, \ltildevec, \wvec | \thetavec) + \log{p(G_{p,k})}\Big\}, && \text{if } k < p, 
    \\
    & \nabla_{\widetilde{G}_{k,k}} \log p(\thetavec, \ltildevec,\wvec | \yvec) =\nabla_{G_{k,k}} \Big\{\calL( \yvec, \ltildevec, \wvec | \thetavec) + \log{p(G_{k,k})}\Big\} \times \exp(\widetilde{G}_{k,k}) + 1
\end{alignat*}

\begin{equation*}
    \begin{aligned}
        \nabla_G \calL( \yvec, \ltildevec, \wvec | \thetavec) &= \nabla_G \Bigg\{ - \frac{n}{2}\log\bigg(\det\Big(\bar\Omega-\deltavec\deltavec'\Big)\bigg)  - \frac{1}{2}\sum^n_{i=1} \bigg(w_i \yvec_i' \Big(\bar\Omega-\deltavec \deltavec'\Big)^{-1}\yvec_i  
        \\& \quad - 2\tilde{l}_i w_i^{1/2} \deltavec'\Big(\bar\Omega-\deltavec \deltavec'\Big)^{-1} \yvec_i + \tilde{l}_i^2 \deltavec'\Big(\bar\Omega-\deltavec \deltavec'\Big)^{-1} \deltavec \bigg)\Bigg\}
        \\&= \nabla_G \Big\{- \frac{n}{2} T_{G1} - \frac{1}{2} \sum_{i=1}^n(T_{G2i} + T_{G3i} + T_{G4i}) \Big\}
    \end{aligned}
\end{equation*}
with $\deltavec = (1+\alphavec^{\top} \bar{\Omega} \alphavec)^{-1 / 2} \bar{\Omega} \alphavec$ and $\bar\Omega = V_1VV_1, V = GG^\top + D, V_1 = \diag(V)^{-1/2} $


\subsubsection{Computing $\nabla_G T_{G1}$}

\begin{equation*}
    \begin{aligned}
        T_{G1} &= \log(\det(\bar\Omega-\deltavec\deltavec^\top ))
    \end{aligned}
\end{equation*}

\begin{equation*}
    \begin{aligned}
        \dd T_{G1} & = \dd\log(\det(\bar\Omega-\deltavec\deltavec^\top ))
        \\&= \tr((\bar\Omega-\deltavec\deltavec^\top )^{-1} \dd(\bar\Omega-\deltavec\deltavec^\top ))
        \\&=  \tr( C_0 \dd{\bar\Omega} ) - \tr( C_0 \dd{(\deltavec\deltavec^\top )}) 
        \\& = M_1 - M_2
    \end{aligned}
\end{equation*}


By Woodbury formula and Sherman–Morrison formula, we have
\begin{equation*}
    \begin{aligned}
        \bar\Omega &= V_1VV_1
        \\& = V_1(GG^\top  + D)V_1
        \\& = V_1GG^\top V_1 + V_1DV_1
        \\& = G_1G_1^\top  + D_1
    \end{aligned}
\end{equation*}

\begin{equation*}
    \begin{aligned}
        \bar\Omega^{-1} &= (D_1 + G_1G_1^\top )^{-1}
        \\& = D_1^{-1} - D_1^{-1} G_1 (I_K + G_1^\top   D_1^{-1} G_1)^{-1} G_1^\top  D_1^{-1} 
        \\ C_0 & = (\bar\Omega-\deltavec\deltavec^\top )^{-1}
        \\&= \bar\Omega^{-1} + \frac{\bar\Omega^{-1}\deltavec \deltavec^\top  \bar\Omega^{-1}}{1 - \deltavec^\top \bar\Omega^{-1}\deltavec}
    \end{aligned}
\end{equation*}

\begin{equation*}
    \begin{aligned}
        M_1 &= \tr( C_0 \dd{\bar\Omega} )
        \\
        M_2 &= \tr( C_0 \dd{(\deltavec\deltavec^\top )}) 
       \\&= \tr \Big( C_0 \dd ((1+\alphavec^\top  \bar{\Omega} \alphavec)^{-1} \bar\Omega \alphavec\alphavec^\top  \bar\Omega) \Big)
       \\&= \tr \Big( C_0 \dd (C_1^{-1} \bar\Omega \alphavec\alphavec^\top  \bar\Omega) \Big)
       \\&=  \tr \Big(C_1^{-1} \alphavec\alphavec^\top  \bar\Omega  C_0  \dd{ \bar\Omega } \Big) + \tr \Big( C_1^{-1}C_0  \bar\Omega \alphavec\alphavec^\top  \dd{\bar\Omega} \Big) 
       \\& \quad + \tr \Big( - C_1^{-2}  \alphavec\alphavec^\top \bar\Omega C_0 \bar\Omega\alphavec\alphavec^\top  \dd \bar\Omega \Big)
       \\& = \tr \Big( (C_1^{-1} C_2  C_0 +  C_1^{-1} C_0  C_2^\top  - C_1^{-2} C_2C_0 C_2^\top ) \dd{ \bar\Omega } \Big) 
       \\& = \tr \Big( C_3 \dd{ \bar\Omega } \Big) 
    \end{aligned}
\end{equation*}

\begin{equation*}
    \begin{aligned}
        M_1 - M_2 &= \tr( (C_0 - C_3) \dd{\bar\Omega} )
        \\&= \tr( C_4 \dd{\bar\Omega} )
        \\& = \tr( VV_1 C_4 \dd{V_1}) + \tr( C_4 V_1V\dd{V_1}) + \tr(V_1 C_4 V_1 \dd{V})
        \\& = M_{1-1} + M_{1-2} + M_{1-3}
    \end{aligned}
\end{equation*}
where $C_1 = 1+\alphavec^\top  \bar{\Omega} \alphavec$, $C_2 = \alphavec\alphavec^\top  \bar\Omega $, $C_3 = C_1^{-1} C_2  C_0 +  C_1^{-1} C_0  C_2^\top  - C_1^{-2} C_2C_0 C_2^\top  $, $C_4 = C_0 - C_3$

\begin{equation*}
    \begin{aligned}
        M_{1-1} & = \tr( VV_1 C_4 \dd{V_1})
        \\& = \tr( - \frac{1}{2} V^{-3/2}  \odot (VV_1 C_4) \dd{V})
        \\& = \tr( - \frac{1}{2} G ^\top ( V_2 \odot (VV_1 C_4) ) \dd{G}) 
        \\& \quad + \tr( - \frac{1}{2} (V_2  \odot (VV_1 C_4) ) G \dd{G^\top })
        \\& = \tr(- \frac{1}{2}( V_2 \odot (VV_1 C_4 + (VV_1 C_4)^\top ) ) G \dd{G^\top })
    \end{aligned}
\end{equation*}

\begin{equation*}
    \begin{aligned}
        M_{1-2} & = \tr(C_4 V_1V \dd{V_1})
        \\& = \tr(- \frac{1}{2}( V_2 \odot (C_4 V_1V + (C_4 V_1V)^\top) ) G \dd{G^\top })
    \end{aligned}
\end{equation*}

\begin{equation*}
    \begin{aligned}
        M_{1-3} & = \tr(V_1 C_4 V_1 \dd{V})
        \\& = \tr(V_1 C_4 V_1 \dd{GG^\top })
        \\& = \tr(G^\top  V_1 C_4 V_1 \dd{G}) + \tr( V_1 C_4 V_1 G \dd{G^\top })
        \\& = \tr( (V_1 C_4 V_1 + (V_1 C_4 V_1)^\top) G \dd{G^\top })
    \end{aligned}
\end{equation*}

, where $V_2 = \diag(V)^{-3/2}$.

From above we could derive the gradient of $T_{G1}$
\begin{equation*}
    \begin{aligned}
        \nabla_G T_{G1} &= -2 (V_2 \odot VV_1C_4  + V_1 C_4 V_1 ) G
    \end{aligned}
\end{equation*}

\subsubsection{$\nabla_G T_{G2i}$}

\begin{equation*}
    T_{G2i} = w_i \yvec_i^\top  \Big(\bar\Omega-\deltavec \deltavec^\top \Big)^{-1}\yvec_i 
\end{equation*}

\begin{equation*}
    \begin{aligned}
        \dd T_{G2i} &= \tr \bigg(-w_i(\bar\Omega-\deltavec \deltavec^\top )^{-1} \yvec_i \yvec_i^\top  (\bar\Omega-\deltavec \deltavec^\top )^{-1} \dd{(\bar\Omega-\deltavec \deltavec^\top )}\bigg)
        \\& = \tr \Big( C_{5i} \dd{\bar\Omega}\Big) - \tr \Big( C_{5i} \dd{\deltavec \deltavec^\top }\Big) 
        \\& = M_3 - M_4
    \end{aligned}
\end{equation*}
where $C_{5i} = -w_i(\bar\Omega-\deltavec \deltavec^\top )^{-1} \yvec_i \yvec_i^\top  (\bar\Omega-\deltavec \deltavec^\top )^{-1} = -w_i C_0 \yvec_i \yvec_i^\top  C_0$

\begin{equation*}
    \begin{aligned}
        M_3 & = \tr ( C_{5i} \dd{\bar\Omega})
        \\
        M_4 &= \tr \Big( C_{5i} \dd{\deltavec \deltavec^\top }\Big) 
        \\&= \tr \Big(C_1^{-1} C_2 C_{5i} + C_1^{-1} C_{5i} C_2^\top  - C_1^{-2} C_2 C_{5i} C_2^\top  \dd{\bar\Omega} \Big) 
        \\& = \tr( C_{6i} \dd{\bar\Omega} )
    \end{aligned}
\end{equation*}

\begin{equation*}
    \begin{aligned}
        M_3 - M_4 &= \tr( (C_{5i} - C_{6i}) \dd{\bar\Omega} )
        \\&= \tr( C_{7i} \dd{\bar\Omega} )
    \end{aligned}
\end{equation*}
where $C_{6i} = C_1^{-1} C_2 C_{5i} + C_1^{-1} C_{5i} C_2^\top  - C_1^{-2} C_2 C_{5i} C_2^\top $, $C_{7i} = C_{5i} - C_{6i} $


From above we could derive the gradient of $T_{G2i}$


\begin{equation*}
    \begin{aligned}
        \nabla_G T_{G2i} &= -2 ( V_2 \odot VV_1 C_{7i} + V_1 C_{7i} V_1 ) G
    \end{aligned}
\end{equation*}
\\

\subsubsection{Computing $\nabla_G T_{G3i}$}

\begin{equation*}
    T_{G3i} = - 2\tilde{l}_i w_i^{1/2} \deltavec^\top \Big(\bar\Omega-\deltavec \deltavec^\top \Big)^{-1} \yvec_i 
\end{equation*}
\begin{equation*}
    \begin{aligned}
        \dd T_{G3i} & = \tr \Big( - 2\tilde{l}_i w_i^{1/2} (\bar\Omega-\deltavec \deltavec^\top )^{-1} \yvec_i  \dd \deltavec^\top  \Big) + \tr \Big( - 2\tilde{l}_i w_i^{1/2} \yvec_i \deltavec^\top   \dd (\bar\Omega-\deltavec \deltavec^\top )^{-1} \Big) 
        \\& = \tr \Big( - 2\tilde{l}_i w_i^{1/2} (1+\alphavec^\top \bar\Omega\alphavec) ^{-1/2} (\bar\Omega-\deltavec \deltavec^\top )^{-1} \yvec_i \alphavec^\top   \dd \bar\Omega \Big)
        \\& \quad + \tr \Big( \tilde{l}_i w_i^{1/2} (1+\alphavec^\top \bar\Omega\alphavec) ^{-3/2} \alphavec \alphavec^\top  \bar\Omega (\bar\Omega-\deltavec \deltavec^\top )^{-1} \yvec_i \alphavec^\top  \dd \bar\Omega \Big)
        \\& \quad + \tr \Big(  2\tilde{l}_i w_i^{1/2} (\bar\Omega-\deltavec \deltavec^\top )^{-1}\yvec_i \deltavec^\top  (\bar\Omega-\deltavec \deltavec^\top )^{-1}   \dd \bar\Omega \Big) 
        \\& \quad - \tr \Big(  2\tilde{l}_i w_i^{1/2} (\bar\Omega-\deltavec \deltavec^\top )^{-1}\yvec_i \deltavec^\top  (\bar\Omega-\deltavec \deltavec^\top )^{-1}   \dd \deltavec \deltavec^\top  \Big) 
        \\& = \tr((C_{8i} + C_{9i}) \dd \bar\Omega) - \tr(C_{9i} \dd \deltavec\deltavec^\top )
        \\& = M_5 - M_6
    \end{aligned}
\end{equation*}
where $C_{8i} = - 2\tilde{l}_i w_i^{1/2} C_1 ^{-1/2} C_0 \yvec_i \alphavec^\top  + \tilde{l}_i w_i^{1/2} C_1 ^{-3/2} C_2 C_0 \yvec_i \alphavec^\top  $, $C_{9i} = 2\tilde{l}_i w_i^{1/2} C_0 \yvec_i \deltavec^\top  C_0$.

\begin{equation*}
    \begin{aligned}
        M_5 & = \tr((C_{8i} + C_{9i}) \dd \bar\Omega)
        \\
        M_6 &= \tr(C_{9i} \dd \deltavec\deltavec^\top )
        \\&= \tr \Big((C_1^{-1} C_2 C_{9i} + C_1^{-1} C_{9i} C_2^\top  - C_1^{-2} C_2 C_{9i} C_2^\top ) \dd{\bar\Omega} \Big) 
        \\&= \tr(C_{10i} \dd \bar\Omega)          
    \end{aligned}
\end{equation*}

\begin{equation*}
    \begin{aligned}
        M_5 - M_6 &= \tr((C_{8i} + C_{9i} - C_{10i}) \dd \bar\Omega)
        \\& = \tr(C_{11i} \dd \bar\Omega)
    \end{aligned}
\end{equation*}
where $C_{10i} = C_1^{-1} C_2 C_{9i} + C_1^{-1} C_{9i} C_2^\top  - C_1^{-2} C_2 C_{9i} C_2^\top $, $C_{11i} = C_{8i} + C_{9i} - C_{10i} $


\begin{equation*}
    \begin{aligned}
        \nabla_G T_{G3i} &= -2 ( V_2 \odot VV_1 C_{11i} + V_1 C_{11i} V_1 ) G
    \end{aligned}
\end{equation*}
\\

\subsubsection{Computing $\nabla_G T_{G4i}$}

\begin{equation*}
    T_{G4i} = \tilde{l}_i^2\deltavec^\top (\bar\Omega-\deltavec \deltavec^\top )^{-1} \deltavec
\end{equation*}

\begin{equation*}
    \begin{aligned}
        \dd T_{G4i} & =  \tr \Big(2\tilde{l}_i^2 (\bar\Omega-\deltavec \deltavec^\top )^{-1} \deltavec \dd \deltavec^\top  \Big) + \tr \Big(\tilde{l}_i^2 \deltavec \deltavec^\top  \dd (\bar\Omega-\deltavec \deltavec^\top )^{-1} \Big) 
        \\& = \tr \Big(\tilde{l}_i^2 C_1^{(-1/2)} C_0 \deltavec \alphavec^\top  \dd \bar\Omega \Big) + \tr \Big( - \frac{1}{2} \tilde{l}_i^2 C_1^{-3/2} \alphavec \alphavec^\top  \bar\Omega C_0 \deltavec \alphavec^\top  \dd \bar\Omega \Big)
        \\&  + \tr \Big(- \tilde{l}_i^2 C_0 \deltavec \deltavec^\top  C_0 \dd \bar\Omega \Big) - \tr \Big(- \tilde{l}_i^2 C_0 \deltavec C_0 \dd \deltavec \deltavec^\top \Big) 
        \\& = \tr((C_{12i} + C_{13i}) \dd \bar\Omega) - \tr(C_{13i} \dd \deltavec\deltavec^\top )
        \\& = M_7 - M_8
    \end{aligned}
\end{equation*}
where $C_{12i} = 2\tilde{l}_i^2 C_1^{(-1/2)} \alphavec \deltavec^\top C_0 - \tilde{l}_i^2 C_1^{-3/2} \alphavec \deltavec^\top  C_0 C_2^\top$, $C_{13i} = - \tilde{l}_i^2 C_0 \deltavec \deltavec^\top  C_0$.
\\

\begin{equation*}
    \begin{aligned}
        M_7 &= \tr( ( C_{12i} + C_{13i} ) \dd \bar\Omega)
        \\
        M_8 &= \tr(C_{13i} \dd \deltavec\deltavec^\top )
        \\&= \tr \Big((C_1^{-1} C_2 C_{13i} + C_1^{-1} C_{13i} C_2^\top  - C_1^{-2} C_2 C_{13i} C_2^\top ) \dd{\bar\Omega} \Big) 
        \\&= \tr(C_{14i} \dd \bar\Omega) 
    \end{aligned}
\end{equation*} 

\begin{equation*}
    \begin{aligned}
        M_7 - M_8 &= \tr( ( C_{12i} + C_{13i} - C_{14i}) \dd \bar\Omega) 
        \\&= \tr( C_{15i}  \dd \bar\Omega) 
    \end{aligned}
\end{equation*}
where $C_{14i} = C_1^{-1} C_2 C_{13i} + C_1^{-1} C_{13i}^\top C_2^\top  - C_1^{-2} C_2 C_{13i} C_2^\top $, $C_{15i} = C_{12i} + C_{13i} - C_{14i}$


\begin{equation*}
    \begin{aligned}
        \nabla_G T_{G4i} &= -2 ( V_2 \odot VV_1 C_{15i} + V_1 C_{15i} V_1 ) G
    \end{aligned}
\end{equation*}
\\

\subsubsection{Computing $\nabla_{G_{p,k}} \log p(G_{p,k})$}
\begin{equation*}
    \begin{aligned}
        \pdv{}{G_{p,k}} \log(p(G)) &= \pdv{}{G_{p,k}} \log(f_{\mbox{\tiny GDP}}(G_{p,k}, a_1, b_1) )
        \\&= \pdv{}{G_{p,k}} \bigg(-(a_1+1) \log(1 + \frac{|G_{p,k}|}{b_1})\bigg)
        \\&= -(a_1 + 1) \operatorname{sign}(G_{p,k}) \frac{1}{b_1 + |G_{p, k}|} 
    \end{aligned}
\end{equation*}

\subsection{Computing $\nabla_\alphavec \log p(\thetavec, \ltildevec,\wvec | \yvec)$}
\begin{equation*}
    \begin{aligned}
        \nabla_\alphavec \calL( \yvec, \ltildevec, \wvec | \thetavec) & = \nabla_\alphavec \bigg(- \frac{n}{2}\log\bigg(\det\Big(\bar\Omega-\deltavec\deltavec^\top \Big)\bigg) 
        \\& \quad - \frac{1}{2}\sum^n_{i=1} \bigg(w_i \yvec_i^\top  \Big(\bar\Omega-\deltavec \deltavec^\top \Big)^{-1}\yvec_i - 2\tilde{l}_i w_i^{1/2} \deltavec^\top \Big(\bar\Omega-\deltavec \deltavec^\top \Big)^{-1} \yvec_i 
        \\& \quad + \tilde{l}_i^2 \deltavec^\top \Big(\bar\Omega-\deltavec \deltavec^\top \Big)^{-1} \deltavec \bigg)\bigg) 
        \\& = \nabla_\alphavec \Big (- \frac{n}{2} T_{\alphavec 1} - \frac{1}{2}\sum_{i=1}^n(T_{\alpha2i} + T_{\alpha3i} + T_{\alpha4i}) \Big) 
    \end{aligned}
\end{equation*}

\subsubsection{Computing $\nabla_\alphavec T_{\alphavec 1} $}
\begin{equation*}
    T_{\alphavec 1} = \log(\det(\bar\Omega-\deltavec\deltavec^\top ))
\end{equation*}

\begin{equation*}
    \begin{aligned}
        \dd T_{\alphavec 1} &= \tr ((\bar\Omega-\deltavec\deltavec^\top )^{-1} \dd (\bar\Omega-\deltavec\deltavec^\top ))
        \\& = \tr ( - C_0 \dd (\deltavec\deltavec^\top ))
        \\& = \tr ( - 2  C_0 \deltavec \dd \deltavec^\top)
        \\& = \tr ( - 2 C_0 \deltavec \dd (1+\alphavec^\top  \bar\Omega \alphavec)^{(-1/2)} \alphavec^\top \bar\Omega )
        \\& = \tr (-2 C_1^{-1/2} \bar\Omega C_0 \deltavec  \dd \alphavec^\top + 
        C_1^{-3/2} \bar\Omega \alphavec (\alphavec^\top \bar\Omega C_0 \deltavec) \dd \alphavec^\top + 
        C_1^{-3/2}  (\alphavec^\top \bar\Omega C_0 \deltavec)\alphavec^\top \bar\Omega \dd \alphavec )
        \\&= \tr (-2 C_1^{-1/2} \bar\Omega C_0 \deltavec  \dd \alphavec^\top + 
        2C_1^{-3/2} \bar\Omega \alphavec (\alphavec^\top \bar\Omega C_0 \deltavec) \dd \alphavec^\top)
    \end{aligned}
\end{equation*}

\begin{equation*}
    \nabla_\alphavec T_{\alphavec 1} = 2C_1^{-3/2} \bar\Omega \alphavec (\alphavec^\top \bar\Omega C_0 \deltavec) - 2 C_1^{-1/2} \bar\Omega C_0 \deltavec
\end{equation*}
\\

\subsubsection{Computing $\nabla_\alphavec T_{\alphavec 2i} $}
\begin{equation*}
    T_{\alphavec 2i} = w_i \yvec_i^\top  \Big(\bar\Omega-\deltavec \deltavec^\top \Big)^{-1}\yvec_i
\end{equation*}

\begin{equation*}
    \begin{aligned}
        \dd T_{\alphavec 2i} &= \tr(w_i \yvec_i^\top  \dd (\bar\Omega-\deltavec \deltavec^\top )^{-1}\yvec_i )
        \\&= \tr(- w_i C_0 \yvec_i \yvec_i^\top  C_0 \dd (\bar\Omega-\deltavec \deltavec^\top ) )
        \\&= \tr( w_i C_0\yvec_i \yvec_i^\top  C_0 \dd (\deltavec \deltavec^\top ) )
        \\&= \tr( 2 w_i C_0\yvec_i \yvec_i^\top C_0 \deltavec \dd \deltavec^\top  )
        \\& = \tr( 2 C_{5i} \deltavec \dd (1+\alphavec^\top \bar\Omega\alphavec)^{-1/2} \alphavec^\top \bar\Omega)
        \\& = \tr( 2 C_1^{-1/2} \bar\Omega C_{5i} \deltavec \dd \alphavec^\top ) + \tr(- C_1^{-3/2} C_{5i} \bar\Omega \alphavec \alphavec^\top  \bar\Omega \dd \alphavec) + \tr(- C_1^{-3/2} \bar\Omega \alphavec C_{5i} \bar\Omega \alphavec \dd \alphavec^\top ) 
    \end{aligned}
\end{equation*}



\begin{equation*}
    \begin{aligned}
        \nabla_\alphavec T_{\alphavec 2i} &= C_1^{-2} \bar\Omega\alphavec\alphavec^\top \bar\Omega (C_{5i} +C_{5i}^\top ) \bar\Omega\alphavec - C_1^{-1} \bar\Omega (C_{5i} + C_{5i}^\top ) \bar\Omega \alphavec 
    \end{aligned}
\end{equation*}
\\

\subsubsection{ Computing $\nabla_\alphavec T_{\alphavec 3i} $ }
\begin{equation*}
    T_{\alphavec 3i} = - 2\tilde{l}_i w_i^{1/2} \deltavec^\top \Big(\bar\Omega-\deltavec \deltavec^\top \Big)^{-1} \yvec_i 
\end{equation*}

\begin{equation*}
    \begin{aligned}
        \dd T_{\alphavec 3i} &= \tr( - 2\tilde{l}_i w_i^{1/2} (\bar\Omega-\deltavec \deltavec^\top )^{-1} \yvec_i \dd \deltavec^\top ) 
        \\& \quad + \tr( 2\tilde{l}_i w_i^{1/2} (\bar\Omega-\deltavec \deltavec^\top )^{-1} \yvec_i \deltavec^\top  (\bar\Omega-\deltavec \deltavec^\top )^{-1}\dd (\bar\Omega-\deltavec \deltavec^\top ) )
        \\&= M_9 + M_{10}
        \\
        M_9 &= \tr( - 2\tilde{l}_i w_i^{1/2} (\bar\Omega-\deltavec \deltavec^\top )^{-1} \yvec_i \dd ((1+\alphavec^\top \bar\Omega \alphavec)^{-1/2} \alphavec^\top  \bar\Omega ))
        \\&= \tr( - 2\tilde{l}_i w_i^{1/2} C_1^{-1/2}\bar\Omega C_0 \yvec_i \dd \alphavec^\top  ) + \tr ( \tilde{l}_i w_i^{1/2}  C_1^{-3/2} \bar\Omega \alphavec \alphavec^\top \bar\Omega C_0 \yvec_i \dd \alphavec^\top)
        \\& \quad + \tr ( \tilde{l}_i w_i^{1/2}  C_1^{-3/2} \alphavec^\top \bar\Omega  C_0 \yvec_i \alphavec^\top \bar\Omega \dd \alphavec)
        \\& = \tr(C_{16i} \dd \alphavec^\top )
        \\
        M_{10} &= \tr( 2\tilde{l}_i w_i^{1/2} (\bar\Omega-\deltavec \deltavec^\top )^{-1} \yvec_i \deltavec^\top  (\bar\Omega-\deltavec \deltavec^\top )^{-1}\dd (\bar\Omega-\deltavec \deltavec^\top ) )
        \\&= \tr(- 2\tilde{l}_i w_i^{1/2} (\bar\Omega-\deltavec \deltavec^\top )^{-1} \yvec_i \deltavec^\top  (\bar\Omega-\deltavec \deltavec^\top )^{-1}\dd (\deltavec \deltavec^\top ) )
        \\& = \tr( C_{17i}\dd ((1+\alphavec^\top \bar\Omega\alphavec)^{-1}\bar\Omega\alphavec\alphavec^\top \bar\Omega) )
        \\&= \tr (C_1^{-1} \alphavec^\top  \bar\Omega C_{17i} \bar\Omega \dd \alphavec) + \tr( C_1^{-1} \bar\Omega C_{17i} \bar\Omega \alphavec \dd\alphavec^\top ) 
        \\& \quad + \tr(- C_1^{-2} \bar\Omega\alphavec\alphavec^\top \bar\Omega C_{17i}  \bar\Omega\alphavec \dd \alphavec^\top ) + \tr( - C_1^{-2} \alphavec^\top  \bar\Omega C_{17i}  \bar\Omega\alphavec\alphavec^\top \bar\Omega \dd \alphavec )
    \end{aligned}
\end{equation*}
where $C_{16i} = - 2\tilde{l}_i w_i^{1/2} C_1^{-1/2}\bar\Omega C_0 \yvec_i + 2\tilde{l}_i w_i^{1/2} C_1^{-3/2} \alphavec^\top \bar\Omega C_0 \yvec_i  \bar\Omega \alphavec$, $C_{17i} = - 2\tilde{l}_i w_i^{1/2} C_0 \yvec_i \deltavec^\top  C_0$

\begin{equation*}
    \begin{aligned}
        \nabla_\alphavec T_{\alphavec 3i} &= C_{16i} + C_1^{-1} \bar\Omega (C_{17i} + C_{17i}^\top ) \bar\Omega \alphavec - C_1^{-2} \bar\Omega\alphavec\alphavec^\top \bar\Omega (C_{17i} + C_{17i}^\top )  \bar\Omega\alphavec
    \end{aligned}
\end{equation*}
\\

\subsubsection{ Computing $\nabla_\alphavec T_{\alphavec 4} $ }
\begin{equation*}
    T_{\alphavec 4} = \tilde{l}_i^2 \deltavec^\top (\bar\Omega-\deltavec \deltavec^\top )^{-1} \deltavec
\end{equation*}

\begin{equation*}
    \begin{aligned}
        \dd T_{\alphavec 4} &= \tr(\tilde{l}_i^2 (\bar\Omega-\deltavec \deltavec^\top )^{-1} \deltavec \dd \deltavec^\top ) + \tr(  \tilde{l}_i^2 \deltavec^\top (\bar\Omega-\deltavec \deltavec^\top )^{-1} \dd \deltavec) 
        \\& + \tr( - \tilde{l}_i^2 (\bar\Omega-\deltavec \deltavec^\top )^{-1} \deltavec \deltavec^\top (\bar\Omega-\deltavec \deltavec^\top )^{-1} \dd (\bar\Omega-\deltavec \deltavec^\top )) 
        \\& = 2 M_{11} + M_{12}
    \end{aligned}
\end{equation*}

\begin{equation*}
    \begin{aligned}
        M_{11} &= \tr(\tilde{l}_i^2 (\bar\Omega-\deltavec \deltavec^\top )^{-1} \deltavec \dd \deltavec^\top ) 
        \\&= \tr(\tilde{l}_i^2 C_0 \deltavec \dd ((1+\alphavec^\top  \bar\Omega \alphavec)^{(-1/2)}\alphavec^\top  \bar\Omega)
        \\& = \tr(\tilde{l}_i^2 C_1^{-1/2} \bar\Omega C_0 \deltavec \dd \alphavec^\top  ) \\& \quad + \tr(-\frac{1}{2} \tilde{l}_i^2 C_1^{-3/2} \bar\Omega \alphavec \alphavec^\top  \bar\Omega C_0 \deltavec  \dd \alphavec^\top ) + \tr(-\frac{1}{2} \tilde{l}_i^2 C_1^{-3/2} \alphavec^\top  \bar\Omega C_0 \deltavec \alphavec^\top  \bar\Omega \dd \alphavec) 
        \\& = \tr(C_{18i} \dd \alphavec^\top)
        \\
        M_{12} &= \tr( \tilde{l}_i^2 C_0 \deltavec \deltavec^\top  C_0\dd (\deltavec \deltavec^\top ) )
        \\& = \tr( C_{19i}\dd ((1+\alphavec^\top \bar\Omega\alphavec)^{-1}\bar\Omega\alphavec\alphavec^\top \bar\Omega) )
        \\&= \Big( \tr (C_1^{-1} \alphavec^\top  \bar\Omega C_{19i} \bar\Omega \dd \alphavec) + \tr( C_1^{-1} \bar\Omega C_{19i} \bar\Omega \alphavec \dd\alphavec^\top ) 
        \\& \quad + \tr(- C_1^{-2} \bar\Omega\alphavec\alphavec^\top \bar\Omega C_{19i}  \bar\Omega\alphavec \dd \alphavec^\top ) + \tr( - C_1^{-2} \alphavec^\top  \bar\Omega C_{19i}  \bar\Omega\alphavec\alphavec^\top \bar\Omega \dd \alphavec ) \Big)
    \end{aligned}
\end{equation*}
where $C_{18i} = \tilde{l}_i^2 C_1^{-1/2} \bar\Omega C_0 \deltavec - \tilde{l}_i^2 C_1^{-3/2} \bar\Omega \alphavec \alphavec^\top  \bar\Omega C_0 \deltavec  , C_{19i} = \tilde{l}_i^2 C_0 \deltavec \deltavec^\top  C_0$

\begin{equation*}
    \begin{aligned}
    \nabla_\alphavec T_{\alphavec 4i} = 2C_{18i} + C_1^{-1} \bar\Omega (C_{19i} + C_{19i}^\top ) \bar\Omega \alphavec - C_1^{-2} \bar\Omega\alphavec\alphavec^\top \bar\Omega (C_{19i} + C_{19i}^\top ) \bar\Omega \alphavec
    \end{aligned}
\end{equation*}
\\

\subsubsection{ Computing $ \nabla_{\alphavec} \log p(\alphavec) $ }

\begin{equation*}
    \begin{aligned}
        \nabla_{\alpha_i} \log p(\alpha_i) &\propto \nabla_{\alpha_i}(-\frac{1}{2\sigma^2} \alpha_i^2 )
        \\&\propto -\frac{1}{\sigma^2} \alpha_i 
    \end{aligned}
\end{equation*}
\\

\subsection{ Computing $\nabla_{\tilde{\nu}} \log p(\thetavec, \ltildevec,\wvec | \yvec)$ }

\begin{equation*}
    \begin{aligned}
        \nabla_{\tilde{\nu}} \log g(x, \thetavec, L)  &= \nabla_{\tilde{\nu}} \Big\{\calL( \yvec, \ltildevec, \wvec | \thetavec) + \log(p(\nu)) + \log(\frac{d\nu}{d\tilde{\nu}})\Big\}
        \\& = \nabla_{\tilde{\nu}} \Big\{\calL( \yvec, \ltildevec, \wvec | \thetavec) + \log(p(\nu)) + \tilde{\nu} + \log(2) \Big\}
        \\& = \nabla_{\nu} \Big\{\calL( \yvec, \ltildevec, \wvec | \thetavec) + \log(p(\nu)) \Big\} \frac{d\nu}{d\tilde{\nu}}+ \frac{d}{d\tilde{\nu}}(\tilde{\nu} + \log(2))
        \\&= \nabla_{\nu} \Big\{\calL( \yvec, \ltildevec, \wvec | \thetavec) + \log(p(\nu)) \Big\} \times \exp(\tilde{\nu}) + 1
    \end{aligned}
\end{equation*}

\begin{equation*}
    \begin{aligned}
        \nabla_{\nu} \calL( \yvec, \ltildevec, \wvec | \thetavec) &= \nabla_\nu \bigg\{\frac{v}{2} \sum^n_{i=1}\log(w_i) + n\bigg(\frac{\nu}{2}\log(\frac{\nu}{2}) - \log(\Gamma(\frac{\nu}{2}))\bigg) - \frac{1}{2}\sum_{i=1}^n w_i\nu \bigg\}
        \\&= \frac{1}{2} \sum^n_{i=1}\log(w_i) + n\Big(\frac{1}{2}\log(\frac{\nu}{2}) + \frac{1}{2} - \frac{1}{2}\psi_0(\frac{\nu}{2})  \Big) - \frac{1}{2}\sum_{i=1}^n w_i
    \end{aligned}
\end{equation*}
, where $\psi_0(t) = \diff{\log(\Gamma(t))}{(t)}$ is the digamma function

\begin{equation*}
    \begin{aligned}
        \nabla_{{\nu}} \log(p({\nu}))
        &= \nabla_{{\nu}} \log \Big(f_{Gamma}(\nu), a_2, b_2) \Big)
        \\&= \nabla_{\nu} \Big( (a_2-1) \log(\nu) - b_2 \nu\Big)
        \\&= (a_2 - 1)\frac{1}{\nu} - b_2 
    \end{aligned}
\end{equation*}

\newpage
\section{Additional Details for the Simulation}
This web appendix provides additional details for the simulation in Section~4.
\subsection{DGPs}
In Case~1 the dimension $d=5$ and number of factors $k=1$, with parameters obtained by fitting the skew-t copula with $\nu=10$ to 
$n=1040$ observations of 15 minute returns in the period between 2019-Sep-11 and 2019-Nov-6. The parameter values
are:
\begin{equation*}
	\begin{aligned}
		\alphavec &= [6.7933, -0.6313, -0.0269, 0.2665, -1.0225]^\top\\
		G &= [0.7526, 0.6048, 3.1338, 2.5151, 0.7016]^\top\\
		\nu & = 10
	\end{aligned}
\end{equation*} 

In Case~2 the dimension $d=30$ and number of factors $k=5$, with parameters (reported below) obtained by fitting the skew-t copula with $\nu=10$ to 
$n=1040$ observations of 15 minute returns in the period between 2019-Sep-11 and 2019-Nov-6.

\begin{table}[H]
    \centering
      \begin{tabular}{ccccccccccc}
      $\alphavec$ =  & [0.872 & -0.066 & 0.369 & -5.555 & -0.324 & -0.313 & 0.270 & 0.147 & -0.216 & 0.367 \\
            & 0.431 & 0.357 & -0.049 & 0.387 & -0.089 & 1.930 & 0.659 & -0.358 & -0.040 & -0.265 \\
            & 0.326 & 0.346 & -0.450 & 0.271 & -0.786 & 0.070 & -0.098 & -0.447 & -0.004 & $-0.414]^\top$ \\
      \end{tabular}%
    \label{tab:addlabel}%
  \end{table}%
$G = $ 
$\left[ \begin{smallmatrix}
    & 1831.606 & 0.000 & 0.000 & 0.000 & 0.000 \\
            & 0.158 & 0.495 & 0.000 & 0.000 & 0.000 \\
            & 0.198 & 0.411 & 0.013 & 0.000 & 0.000 \\
            & 0.128 & -0.023 & -0.437 & 0.511 & 0.000 \\
            & -0.037 & 1.595 & -23.314 & 8.952 & 5.215 \\
            & 6.425 & -11.706 & -17.094 & 21.695 & -41.596 \\
            & 0.062 & 0.380 & -0.127 & 0.132 & -0.057 \\
            & 1.055 & 16.057 & -23.007 & -27.291 & -10.508 \\
            & 0.119 & 0.188 & -0.263 & 0.138 & -0.203 \\
            & 0.248 & 0.389 & -0.168 & 0.062 & -0.170 \\
            & 0.281 & 0.351 & -0.116 & 0.167 & -0.103 \\
            & 0.151 & 0.161 & -0.293 & 0.232 & -0.512 \\
            & 6.331 & -11.536 & -16.839 & 21.378 & -40.998 \\
            & 0.006 & 0.328 & -0.121 & 0.031 & -0.024 \\
            & 0.029 & 0.555 & -0.431 & -0.165 & -0.186 \\
            & 0.249 & 0.436 & -0.133 & 0.087 & -0.075 \\
            & 0.067 & 0.494 & -0.322 & -0.146 & -0.222 \\
            & 0.114 & 0.459 & -0.149 & 0.087 & -0.199 \\
            & 0.115 & 0.551 & -0.182 & 0.131 & -0.140 \\
            & 0.019 & 0.549 & -0.424 & 0.165 & -0.104 \\
            & 1.050 & 15.973 & -22.878 & -27.142 & -10.458 \\
            & 0.102 & 0.428 & -0.261 & 0.097 & -0.120 \\
            & -0.124 & 1.530 & -23.476 & 8.823 & 3.646 \\
            & 0.005 & 0.507 & -0.284 & -0.045 & -0.136 \\
            & 82.161 & -0.001 & 0.000 & -0.001 & -0.001 \\
            & 0.203 & 0.311 & -0.154 & 0.220 & -0.143 \\
            & 0.203 & 0.445 & -0.214 & 0.122 & -0.155 \\
            & 0.014 & 0.464 & -0.299 & 0.035 & -0.089 \\
            & 0.253 & 0.048 & -0.071 & 0.074 & -0.161 \\
            & 6.351 & -11.576 & -16.895 & 21.442 & -41.113
\end{smallmatrix} \right]$ 
\\

$\nu = 10$ 

\subsection{Additional Empirical Results}
\begin{figure}[htbp]
	\centering
	\includegraphics[width=0.7\textwidth]{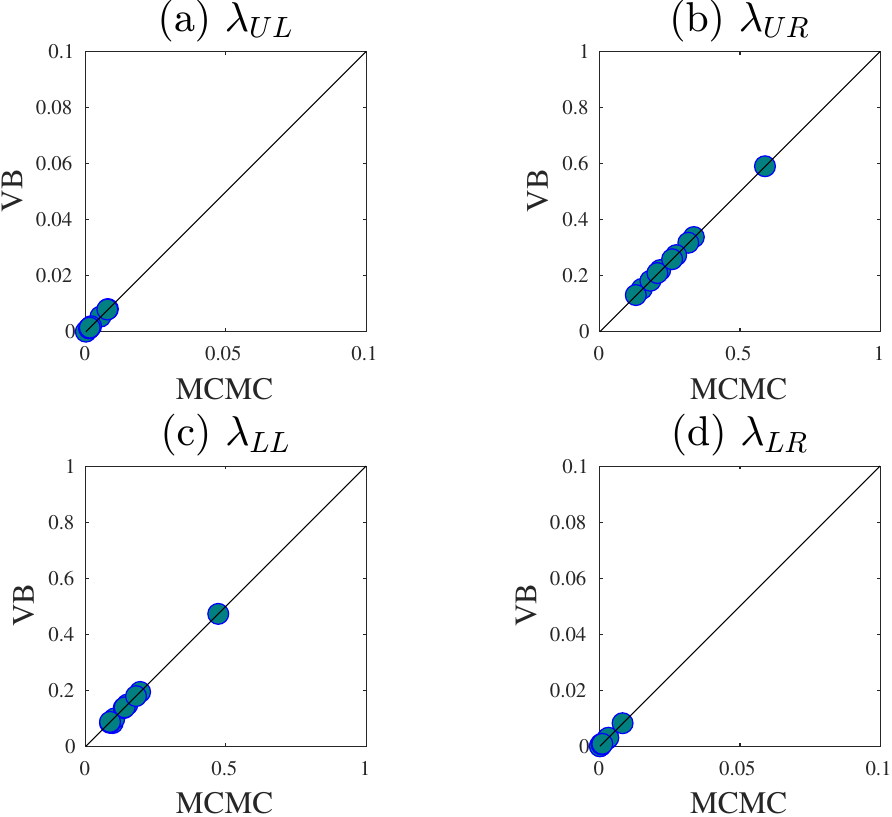}
	\caption{For Case~1, each panel plots the posterior means of the pairwise quantile dependences (at the 1\% quantile level). These are computed exactly using MCMC (horizontal axis) and approximately using VI (vertical axis). Each panel gives results for a different direction. The closer the scatters are to 
		the 45 degree line, the more accurate the variational posterior mean.}
	\label{fig:mcmc_vs_vb_qd_01_mean_5D}
\end{figure}

Figure~\ref{fig:mcmc_vs_vb_qd_01_mean_5D} plots the marginal posterior means
of the four quantile pairwise dependence metrics at the 1\% level for Case~1. On the horizontal
axis is the exact posterior mean computed using MCMC, and on the vertical axis is the
approximate posterior mean computed using VI. Each point in a scatterplot corresponds
to a specific pairwise dependence.

\begin{figure}[htbp]
    \centering
    \includegraphics[width=0.8\textwidth]{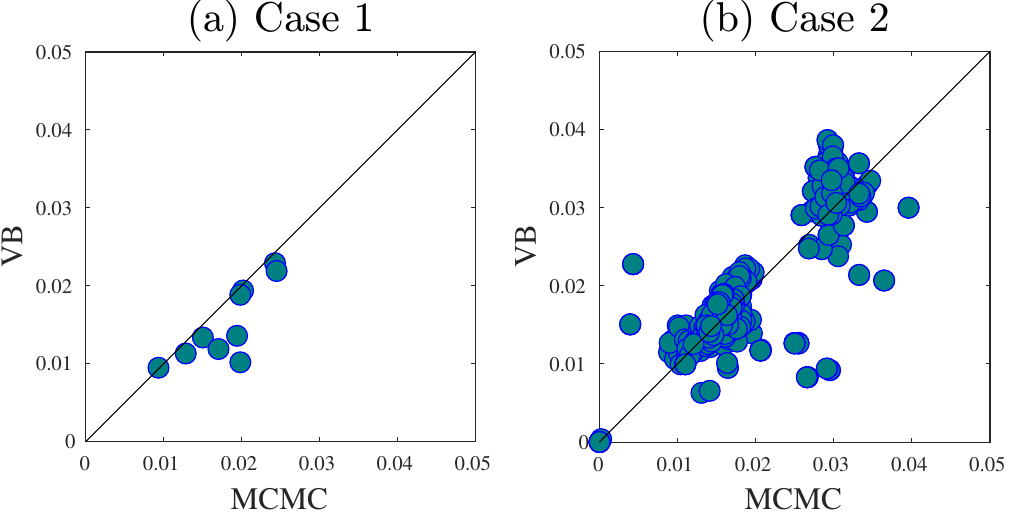}
    \caption{Comparison of the estimates of the posterior standard deviation of Spearman correlation computed exactly using MCMC and approximately using VI for (a) Case 1 where $d=5$, and (b) Case 2 where $d=30$.}
    \label{fig:mcmc_vs_vb_spearmanrho_std}
\end{figure}

Figure~\ref{fig:mcmc_vs_vb_spearmanrho_std} plots the standard deviation of the marginal posterior (i.e. computed using MCMC) and its variational approximation (i.e. computed using VI) for 
the pairwise Spearman correlations. 
It is well-known that variational approximations
are less well-calibrated to higher order moments, and this is apparent for some of the pairwise Spearman correlations. The accuracy of the variational posterior standard deviations 
can be further improved by considering more complex variational approximations, although the trade-off typically involves greater computation.

\subsection{Comparison of different generative representations}
\begin{figure}[htbp]
    \centering
    \includegraphics[width=0.8\textwidth]{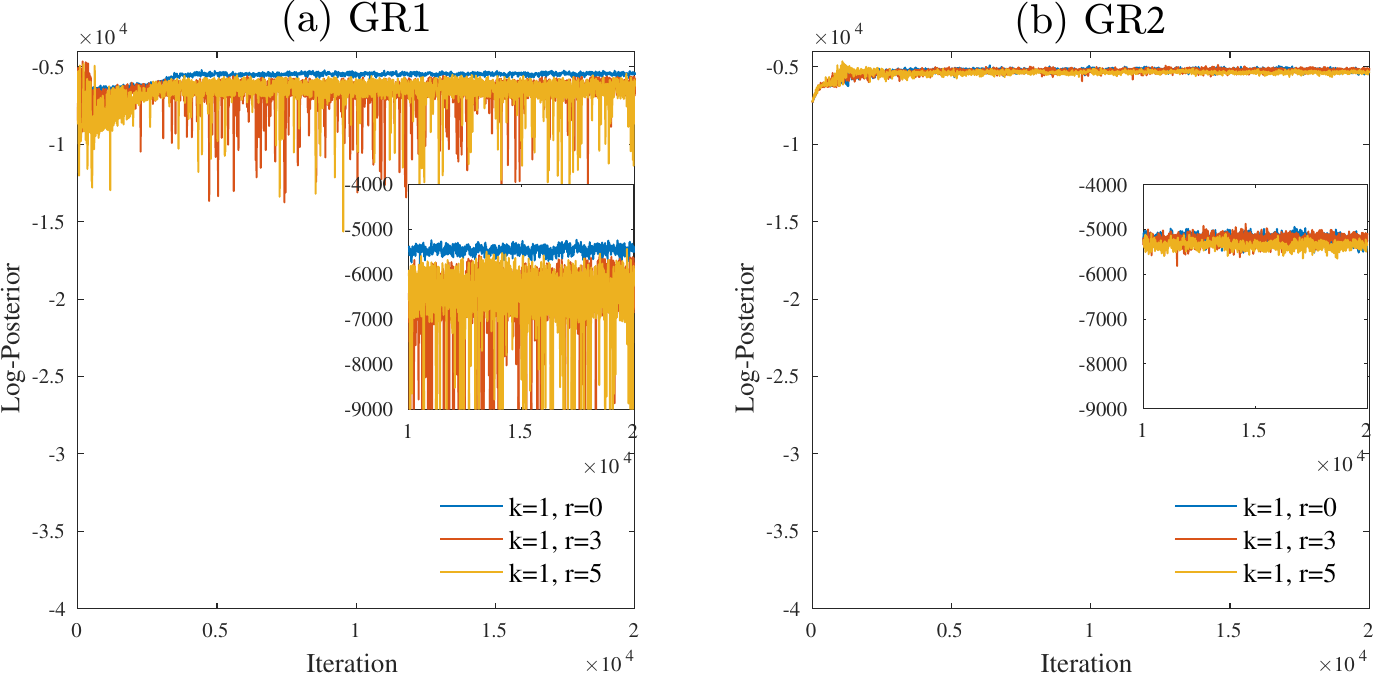}
    \caption{Plots of log-posterior against step number 
	    	are used to monitor the convergence of the SGA algorithm for  
	    	the Case~1 simulation. Panel~(a) is where generative representation GR1 is
	    used in the hybrid VI algorithm, and panel~(b) is where GR2 is used (which is our 
	    recommended algorithm).}
    \label{fig:grplot}
\end{figure}

We find the generative representation GR2 to provide an augmented posterior
that is most effective to estimate using our hybrid VI method, in comparison to GR1.
To illustrate, Figure~\ref{fig:grplot} plots the logarithm of the posterior
density against SGA step for both VI algorithms applied to Case~1. Results are given
for the Gaussian factor
approximation $q_\lambda^0$ using $r=0, 3$ and $5$
factors, and the optimization converges faster by this metric for GR2. Similar
results (unreported) were found for Case~2 and in the analysis of the financial data.
    
\FloatBarrier

\subsection{Estimation accuracy}
\begin{figure}[htbp]
	\centering
	\includegraphics[width=0.6\textwidth]{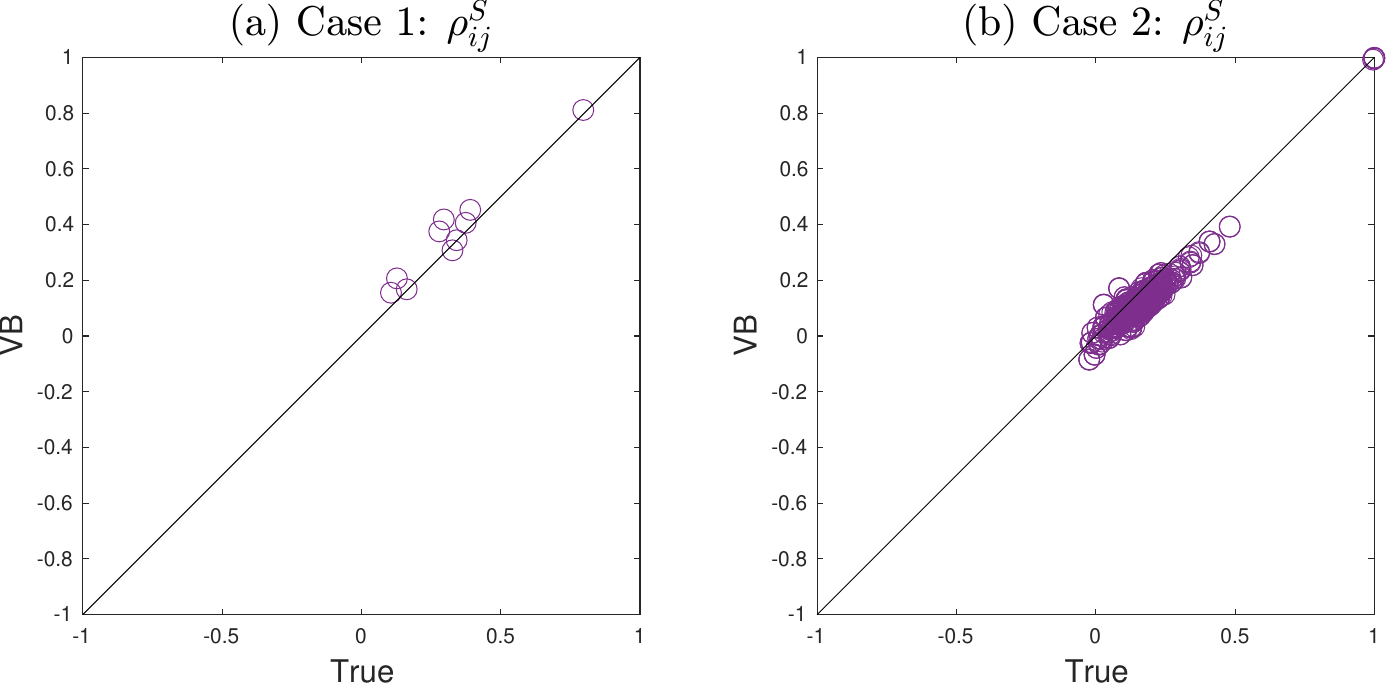}
	\caption{This plot is a reproduction of Figure~2 in the manuscript, but where the variational posterior means are plotted against the true values of $\rho_{i,j}^S$ in the DGP, instead of the exact posterior means evaluated using MCMC. See the original figure caption for further details.}
	\label{afig:vit1}
\end{figure}

\begin{figure}[htbp]
	\centering
	\includegraphics[width=0.6\textwidth]{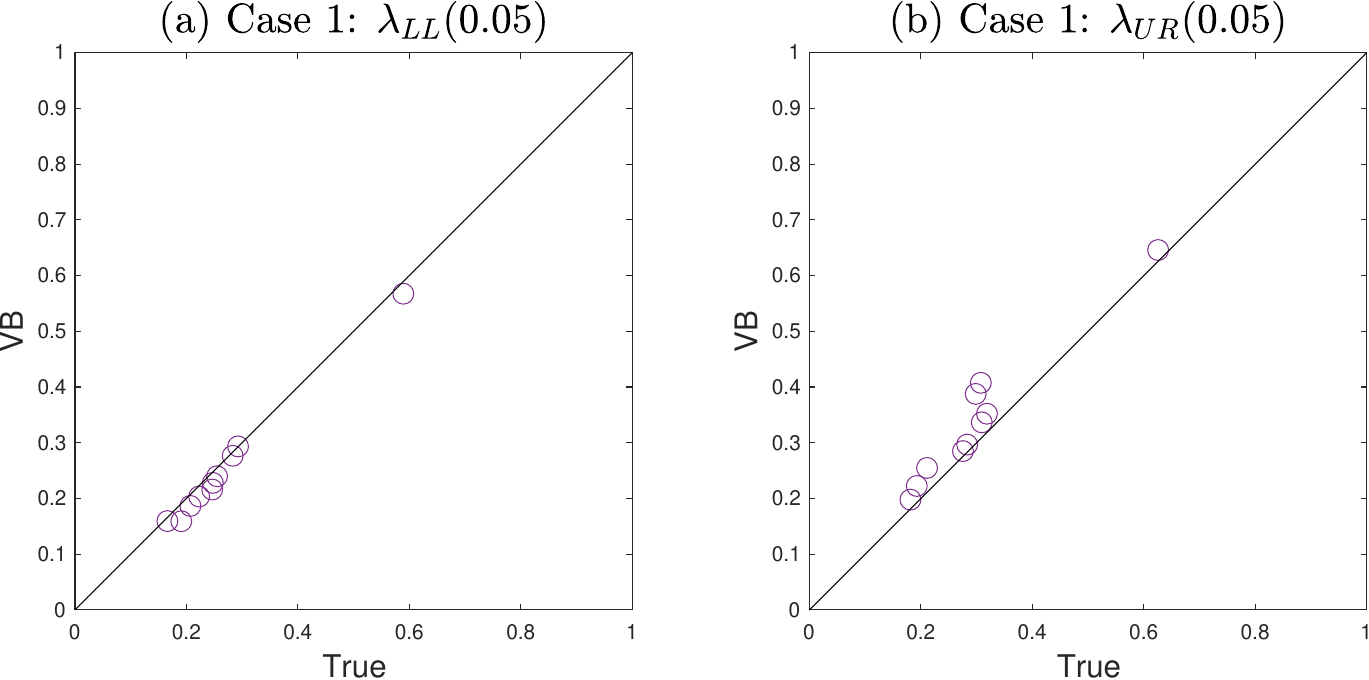}
	\caption{This plot is a reproduction of Figure~3 in the manuscript, but where the variational posterior means are plotted against the true values of $\lambda_{LL}(0.05)$ and $\lambda_{UR}(0.05)$ for Case~1 in the DGP, instead of the exact posterior means evaluated using MCMC. See the original figure caption for further details.}
	\label{afig:vit2}
\end{figure}

\begin{figure}[htbp]
	\centering
	\includegraphics[width=0.6\textwidth]{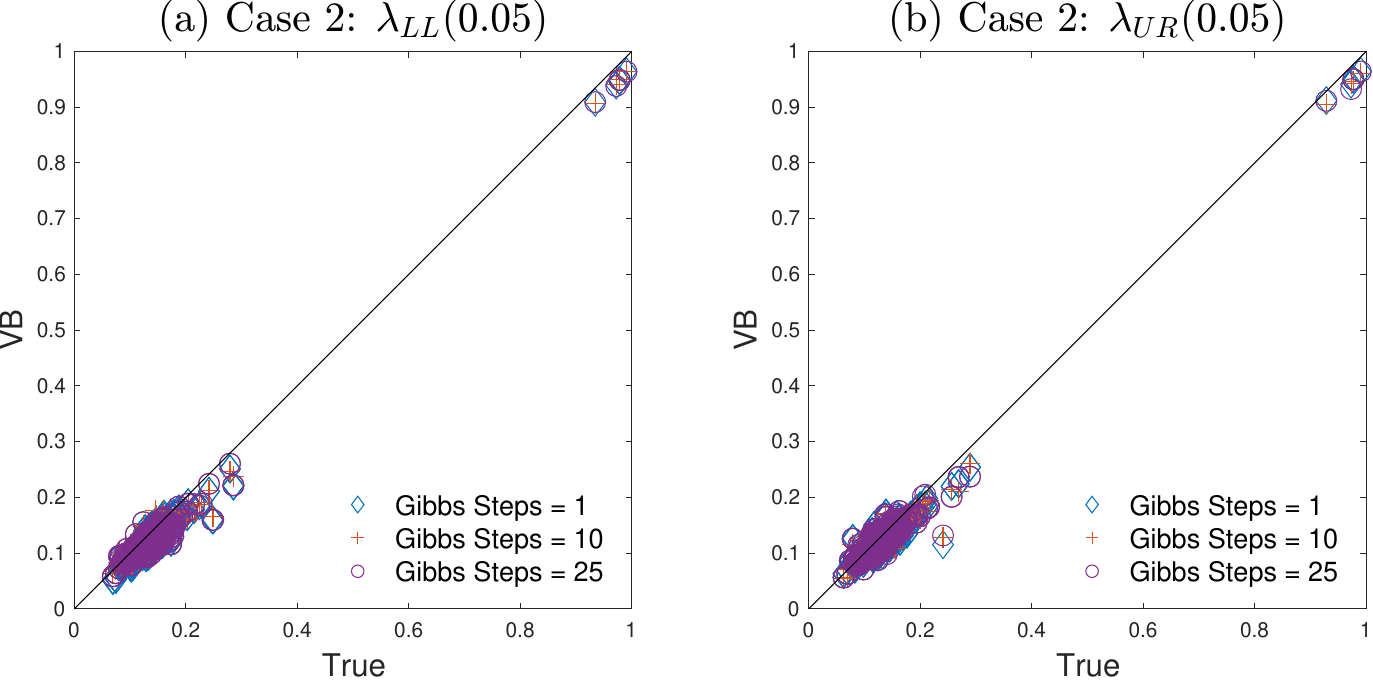}
	\caption{This plot is a reproduction of Figure~3 in the manuscript, but where the variational posterior means are plotted against the true values of $\lambda_{LL}(0.05)$ and $\lambda_{UR}(0.05)$ for Case~2 in the DGP, instead of the exact posterior means evaluated using MCMC. See the original figure caption for further details.}
	\label{afig:vit3}
\end{figure}

To show that the variational posterior mean is a good estimator of the 
true DGP, we reproduce Figures 3, 4 and~5 in the manuscript, but where we plot the variational posterior means against the true values. Figure~\ref{afig:vit1} does
so for the Spearman correlations of both DGPs, while
Figures~\ref{afig:vit2} and~\ref{afig:vit3} do so for quantile dependencies in both 
cases. This demonstrates the accuracy of the Bayesian posterior mean as an estimator.
\FloatBarrier

\newpage 
\section{Additional Details for Section 5}\label{app:D}
\subsection{Estimates of the SDB and GH skew-t copulas}
Below are  estimates of the dependence metrics and parameters for the SDB and GH skew-t copulas
fit to the $d=3$ dimensional example in Section 5.

\begin{table}[H]
	\begin{center}
	\caption{Estimated SDB Skew-t Copula for the BAC-JPM-VIX Example}
	\resizebox{0.95\textwidth}{!}{
        \begin{tabular}{ccccccccccc}
            \toprule
            \toprule
                  &       & \multicolumn{9}{c}{Low Volatility Period: From 24 October 2007 to 22 December 2017} \\
        \cmidrule{3-11}          &       & \multicolumn{2}{c}{Correlation} &       & \multicolumn{2}{c}{Tail Asymmetry} &       & \multicolumn{3}{c}{$\thetavec$} \\
        \cmidrule{3-4}\cmidrule{6-7}\cmidrule{9-11}    Pair  &       & Spearman & Kendall &       & $\Delta_\text{Major}(0.05)$ & $\Delta_\text{Minor}(0.05)$ &       & $\rho$ & $\deltavec^\top$ & $\nu$ \\
            \multirow{2}[0]{*}{BAC -- VIX} &       & -0.345 & -0.235 &       & 0.000 & 0.000 &       & -0.361 & -0.009, 0.004 &  \\
                  &       & (0.023) & (0.016) &       & (0.002) & (0.007) &       & (0.023) & (0.019, 0.024) &  \\
            \multirow{2}[0]{*}{JPM -- VIX} &       & -0.363 & -0.248 &       & 0.000 & 0.000 &       & -0.379 & -0.006,0.004 & 31.061 \\
                  &       & (0.021) & (0.015) &       & (0.002) & (0.007) &       & (0.021) & (0.018, 0.024) & (9.115) \\
            \multirow{2}[1]{*}{JPM -- BAC} &       & 0.788 & 0.594 &       & 0.000 & 0.000 &       & 0.804 & -0.006, -0.009 &  \\
                  &       & (0.010) & (0.010) &       & (0.008) & (0.000) &       & (0.010) & (0.018, 0.019) &  \\
            \midrule
            \midrule
                  &       & \multicolumn{9}{c}{High Volatility Period: From 11 February 2020 to 15 April 2020} \\
        \cmidrule{3-11}          &       & \multicolumn{2}{c}{Correlation} &       & \multicolumn{2}{c}{Tail Asymmetry} &       & \multicolumn{3}{c}{$\thetavec$} \\
        \cmidrule{3-4}\cmidrule{6-7}\cmidrule{9-11}    Pair  &       & Spearman & Kendall &       & $\Delta_\text{Major}(0.05)$ & $\Delta_\text{Minor}(0.05)$ &       & $\rho$ & $\deltavec^\top$ & $\nu$ \\
            \multirow{2}[0]{*}{BAC -- VIX} &       & -0.606 & -0.432 &       & 0.000 & 0.000 &       & -0.628 & 0.028, 0.020 &  \\
                  &       & (0.016) & (0.013) &       & (0.001) & (0.009) &       & (0.016) & (0.017, 0.023) &  \\
            \multirow{2}[0]{*}{JPM -- VIX} &       & -0.607 & -0.432 &       & 0.000 & 0.000 &       & -0.628 & -0.002, 0.020 & 20.518 \\
                  &       & (0.016) & (0.013) &       & (0.001) & (0.008) &       & (0.016) & (0.017, 0.023) & (3.401) \\
            \multirow{2}[1]{*}{JPM -- BAC} &       & 0.851 & 0.664 &       & 0.000 & 0.000 &       & 0.864 & -0.002, 0.020 &  \\
                  &       & (0.007) & (0.008) &       & (0.008) & (0.000) &       & (0.007) & (0.017, 0.017) &  \\
            \bottomrule
            \bottomrule
            \end{tabular}%
	}
	\end{center}
	Note: These results were obtained using MCMC as outlined in~\cite{smith_gan_kohn_2010}, but where
	the matrix $\bar{\Omega}$ was modelled using a factor model with $k=1$ factors as in Section~2.4. The posterior mean of the dependence metrics and parameter values are reported, along with the posterior 
	standard deviations in parentheses below. 
	\label{tab:addlabel}%
\end{table}%

\begin{table}[H]
	\begin{center}
	\caption{Estimated GH Skew-t Copula for the BAC-JPM-VIX Example}
	\resizebox{0.95\textwidth}{!}{
        \begin{tabular}{ccccccccccr}
            \toprule
            \toprule
                  &       & \multicolumn{9}{c}{Low Volatility Period: From 24 October 2007 to 22 December 2017} \\
        \cmidrule{3-11}          &       & \multicolumn{2}{c}{Correlation} &       & \multicolumn{2}{c}{Tail Asymmetry} &       & \multicolumn{3}{c}{$\thetavec$} \\
        \cmidrule{3-4}\cmidrule{6-7}\cmidrule{9-11}    Pair  &       & Spearman & Kendall &       & $\Delta_\text{Major}(0.05)$ & $\Delta_\text{Minor}(0.05)$ &       & $\rho$ & $\delta$ & \multicolumn{1}{c}{$\nu$} \\
            \multicolumn{1}{l}{BAC -- VIX} &       & -0.342 & -0.234 &       & -0.027 & 0.000 &       & -0.388 &       &  \\
            \multicolumn{1}{l}{JPM -- VIX} &       & -0.365 & -0.250 &       & 0.001 & 0.000 &       & -0.413 & -0.413 & \multicolumn{1}{c}{19.554} \\
            \multicolumn{1}{l}{JPM -- BAC} &       & 0.788 & 0.595 &       & -0.046 & 0.000 &       & 0.802 &       &  \\
            \midrule
            \midrule
                  &       & \multicolumn{9}{c}{High Volatility Period: From 11 February 2020 to 15 April 2020} \\
        \cmidrule{3-11}          &       & \multicolumn{2}{c}{Correlation} &       & \multicolumn{2}{c}{Tail Asymmetry} &       & \multicolumn{3}{c}{$\thetavec$} \\
        \cmidrule{3-4}\cmidrule{6-7}\cmidrule{9-11}    Pair  &       & Spearman & Kendall &       & $\Delta_\text{Major}(0.05)$ & $\Delta_\text{Minor}(0.05)$ &       & $\rho$ & $\delta$ & \multicolumn{1}{c}{$\nu$} \\
            BAC -- VIX &       & -0.617 & -0.441 &       & -0.002 & 0.000 &       & -0.655 &       &  \\
            JPM -- VIX &       & -0.614 & -0.438 &       & 0.012 & 0.000 &       & -0.652 & -0.306 & \multicolumn{1}{c}{21.206} \\
            JPM -- BAC &       & 0.853 & 0.666 &       & -0.029 & 0.000 &       & 0.865 &       &  \\
            \bottomrule
            \bottomrule
            \end{tabular}%
	}
	\end{center}
	Note: These results were obtained using code provided by~\cite{oh_patton_2021} for fitting
	a static GH skew-t copula using the EM algorithm. Because the example features negative pairwise dependence we removed the positive dependence constraint in that code. This implementation imposes the constraint that $\deltavec=(1,1,1)\delta$. The factor representation for matrix $\bar{\Omega}$ has one global factor, along with three group factors with one allocated to each dimension.
	\label{tab:addlabel}%
\end{table}%

\subsection{Geometry of the posterior and its impact on VI}
\subsubsection{The posterior distribution}
\begin{figure}[htbp]
	\begin{center}
		\caption{Marginals of the posterior $p(\thetavec|\yvec)$ for the high volatility data in Section~5}	
		\label{afig:geometry}
	\includegraphics[width=0.8\textwidth]{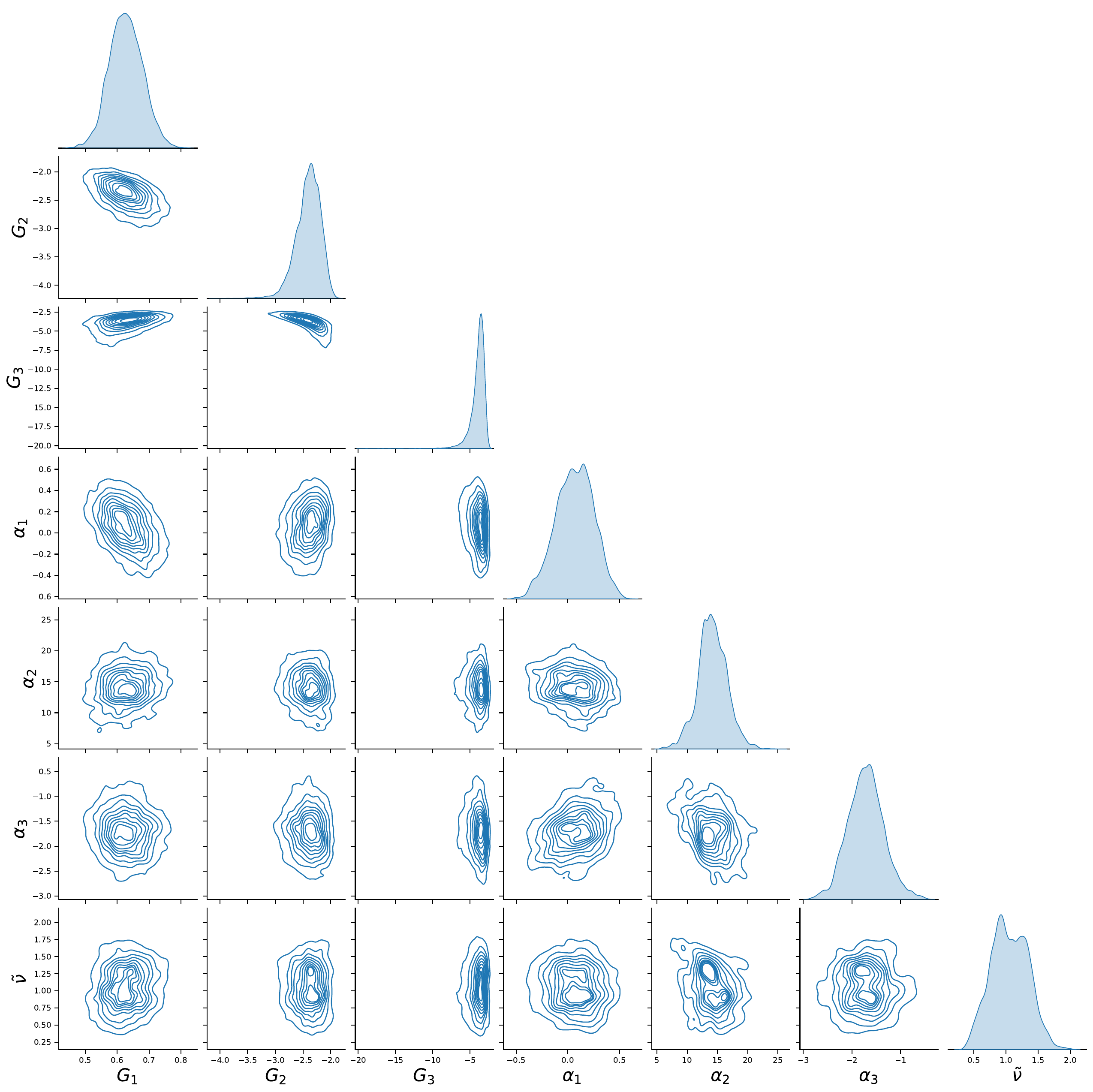}
	\end{center}
Note: Univariate marginals are given in the leading diagonal panels, and bivariate marginals are given in the off-diagonal panels. These are evaluated using univariate
and bivariate kernel density estimates from draws from the posterior distribution 
obtained using MCMC. 
\end{figure}

The posterior of the AC skew-t copula parameters $\thetavec$ has a complex geometry that we illustrate
here using the fit to the high volatility period data in Section~5. In this 
example, $d=3$ and there is $k=1$ factor, so that the loading matrix
\begin{equation*}
	G=(g_{11},g_{21},g_{31})^\top\\,
\end{equation*} 
and the transformed degrees of freedom parameter
$\tilde{\nu}=\log(\nu-2)$. Therefore,
$\thetavec=\{g_{11},g_{21},g_{31},\alpha_1,\alpha_2,\alpha_3,\tilde{\nu}\}$. With 
the vague proper priors in Section~3.3, the posterior
is dominated by the likelihood. Figure~\ref{afig:geometry} plots the univariate and bivariate marginals
of this posterior distribution, evaluated exactly using MCMC applied to the GR2 generative representation. Even in this low dimension, the complex geometry of the posterior is visible, so that it is a hard target 
distribution to approximate directly using SGA, or even traverse directly using MCMC. This motivates
the use of the more tractable augmented posterior based on the generative representation.

\subsubsection{Impact on variational inference}
\begin{figure}[htbp]
	\begin{center}
		\caption{Effectiveness of SGA applied to three target distributions}
		\label{afig:3dlogp}
	\includegraphics[width=0.7\textwidth]{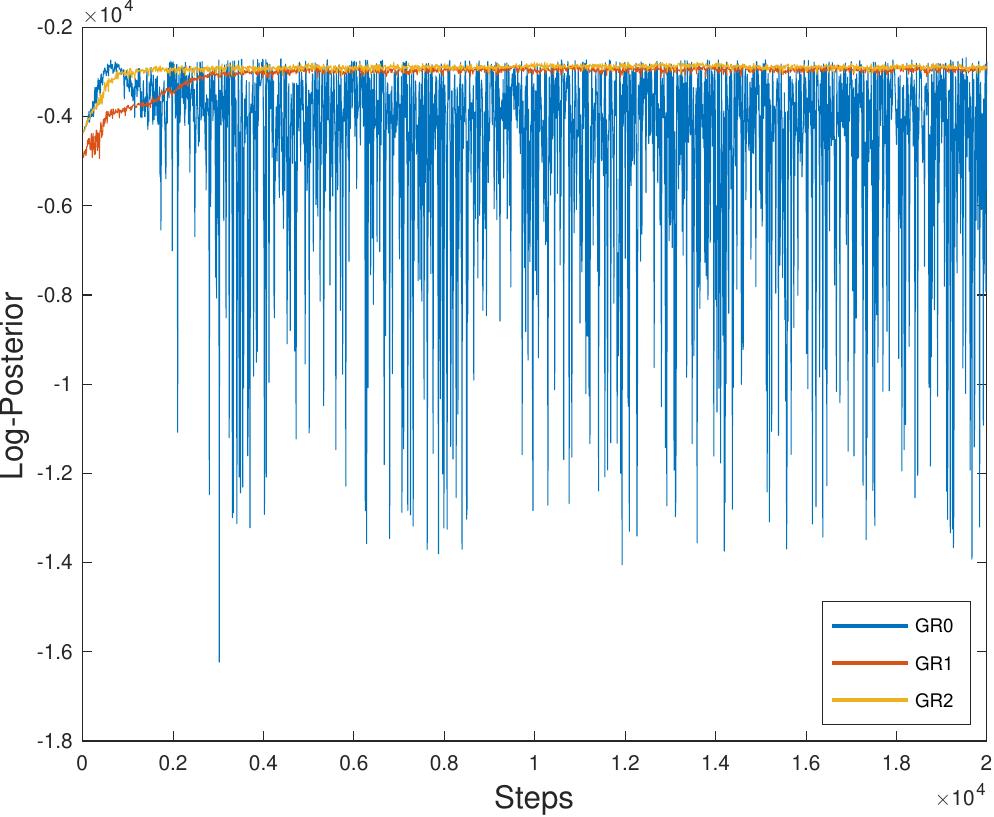}
	\end{center}
	Note: the log-posterior density is evaluated at the means of VA's at each step of the SGA algorithm. Faster and more stable convergence is observed for GR2 (yellow line). In contrast, SGA struggles when applied directly to the posterior GR0 (blue line) as a target distribution. The data used was the high volatility period in Section~5.
\end{figure}

The complex geometry of the posterior makes the variational optimization
a difficult task. We solve this problem by proposing the use of the generative representation GR2 as the target posterior. To illustrate the improvement this provides, we apply our VI approach to three different target distributions for the high volatility period data. 
These target distributions are:
\begin{itemize} 
\item[(i)] GR0: The posterior which is not augmented with any latent variables, and is displayed in Figure~\ref{afig:geometry}.
\item[(ii)] GR1: The augmented posterior using the GR1 generative representation in Appendix~B of the paper.
\item[(iii)] GR2: The augmented posterior using the GR2 generative representation in Appendix~B of the paper. This is our recommended target distribution.
\end{itemize}
Figure~\ref{afig:3dlogp} plots values of the log posterior density against
step number of the SGA optimization for all three target distributions. This allows
us to monitor the effectiveness of the optimization. SGA struggles to converge for the  
GR0 and GR1 target distributions, whereas GR2 converges quickly. In our experience, the difficulties of applying SGA to the GR0 and GR1 target distributions only multiply for higher dimensional copulas. In contrast, we find that applying SGA to the target distribution GR2 is stable, which is demonstrated  in our empirical work
in Section~6.

\newpage
\section{Additional Details for Section 6}\label{app:E}

\subsection{Generating the Predictive Distribution of Portfolio Return} \label{app:E1}

Algorithm 5.1 outlines the steps necessary to generate draws from the posterior predictive return distribution for the density forecast evaluation study in Section \ref{sec:density}.

\begin{cvbox}
	\noindent\rule{\textwidth}{1pt}
	\textbf{Algorithm 5.1: Generating from the Predictive Distribution of Portfolio Return }  \\
	For $m=1,...,M$, do \\
	
	\textit{Step 1.} Sample skew-t model parameter set $\theta$ from fitted VB parameters, $\theta \sim q(\lambda^*)$ \\
\textit{Step 2.} Generate copula data $\vect{\widetilde{U}} \in \mathbb{R}^{d}$ from fitted skew-t copula model $C_{\mbox{\tiny skew-t}}(\vect{U} = \{U_{1},...,U_{d}\}; \theta )$. \\
	\textit{Step 3.} Transform $j$th marginal predicted return from corresponding sampled copula data $\tilde{z}_{t+1,i}^{(j)} = F^{-1}_{\mbox{\tiny stud-t}}(\widetilde{U}_{j};\nu_j)$ for $ j = 1,...,d$.\\
	\textit{Step 4.} Then convert it to raw predictive returns $r_{t+1,i}^{(j)}$, with GARCH forecasting $\hat{\sigma}_{t+1,i}$ for $\tau = 1,...,N , j = 1,...,d$, as per the model given in Section \ref{eq:rti}.
	$$r_{\tau,t+1}^{(j)} = (h_{t} s_\tau)^{1/2} \hat{\sigma}_{\tau,t+1} \tilde{z}_{\tau,t+1}^{(j)}$$ 
	\textit{Step 5.} Compute weights $\omega_j$ from daily shares outstanding $S_{t}^{(j)} $ and last price $P_{\tau,t}^{(j)} $ for each asset. $K_{j} = \sum_{t=1}^T S_{t}^{(j)} P_{\tau,t}^{(j)}$, $\omega_j = K_{j} / \sum_{j=1}^d K_{j}$. These daily market value weighted portfolio compositions are calculated based on data from the Center for Research in Security Prices (CRSP), The University of Chicago Booth School of Business. \\
	\textit{Step 6.} Compute predicted portfolio return 
	$$R_{\tau,t+1}^{[m]} = \sum_{j=1}^d \omega_j r_{,\tau,t+1}^{(j)}$$.\\
	\noindent\rule{\textwidth}{1pt}
\end{cvbox}
	
The $M$ draws of $(R_{\tau,t+1}^{[1]},...,R_{\tau,t+1}^{[M]})$ represents the draw from the posterior portfolio return distribution. These predictive distributions are evaluated using density forecast evaluation metrics in Section \ref{sec:density}.

\subsection{Quantile Dependence of AAPL--MSFT \& GOOGL--TSLA}\label{app:E2}
\begin{figure}[H]
    \centering
    \includegraphics[width=\textwidth]{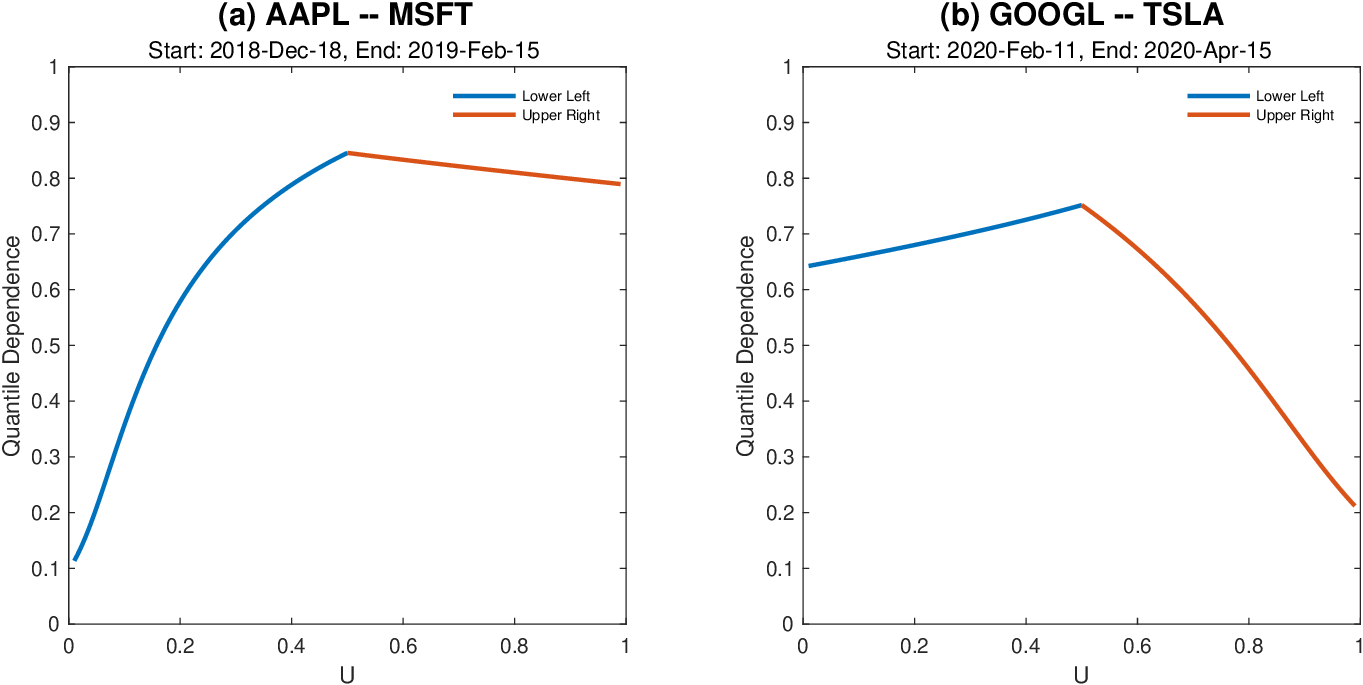} 
    \caption{Panel (a) and (b) depict the quantile dependence measure $\lambda(u)$ along the major diagonals for the peak asymmetry period for Apple vs Microsoft and Google vs Tesla, respectively. }
    \label{fig:TA_major_2Pairs}
  \end{figure}

Figure~\ref{fig:TA_major_2Pairs} gives the quantile dependence plots between AAPL -- MSFT and GOOGL--TSLA from the fitted AC skew-t copula, 
calculated at the peak asymmetry period for each respective pair.

\subsection{Comparison with the GH skew-t copula}
We repeat the prediction and investment studies in Sections~6 using the GH skew-t factor copula. 
The same marginals were used as for the AC skew-t copula. To estimate
the GH skew-t copula the EM algorithm is used, as implemented in 
the routine ``EM\_static\_Gcop.m'' provided by~\cite{oh_patton_2021} with one global factor plus 
heterogeneous loadings for 21 groups.  This code determines group allocation and parameter estimation
iteratively. 

The optimal number of groups identified by~\cite{oh_patton_2021} was 21 for their analysis of daily returns on the 110 stocks that were ever
part of the S\&P100 between 2010 and 2019. These authors found that their results were fairly robust to between 10 and 30 groups. The main difference with our study is that we employ intraday returns on (mostly) the 
same stocks, so that it is likely the optimum number of groups differs somewhat from 21. We tried a lower number of groups (10) and the prediction results were slightly less accurate, but (similar to these authors)
we found the results relatively unchanged between 15 and 25 factors. Therefore, we simply present the
results with  same number of groups as the original study for simplicity.

Table~4 in the manuscript gives the equivalent investment performance using the quantile dependence
based strategies in Section~6, but using the quantile dependence estimates from the GH skew-t copula. 
The relatively poor risk-adjusted performance for the GH skew-t copula is likely due to poor estimates of the quantile dependencies, and
is consistent with the weaker one-step-ahead forecasts using this copula observed in Figure~7 in the manuscript.

\end{document}